# In silico tool for identification of colorectal cancer from cell-free DNA biomarkers

Kartavya Mathur[1,2], Shipra Jain[1], Nisha Bajiya[1], Nishant Kumar[1], Gajendra P. S. Raghava *[1]

1. Department of Computational Biology, Indraprastha Institute of Information Technology, Okhla Phase 3, New Delhi-110020, India.

2. School of Biotechnology, Gautam Buddha University, Gautam Buddha Nagar, Greater Noida, Uttar Pradesh, 201308

**Mailing Address of Authors**

Kartavya Mathur (KM): 221pbg002@gbu.ac.in    ORCID ID: https://orcid.org/0000-0003-3563-5666

Shipra Jain: shipra@iiitd.ac.in    ORCID ID: https://orcid.org/0000-0002-7045-5188

Nishant Kumar (NK): nishantk@iiitd.ac.in    ORCID ID: https://orcid.org/0000-0001-7781-9602

Nisha Bajiya (NB): nishab@iiitd.ac.in    ORCID ID: https://orcid.org/0000-0002-5075-5386

Gajendra P. S. Raghava (GPSR): raghava@iiitd.ac.in    ORCID ID: https://orcid.org/0000-0002-8902-2876

**\*Corresponding Author**

Prof. Gajendra P. S. Raghava

Head and Professor

Department of Computational Biology

Indraprastha Institute of Information Technology, Delhi

Okhla Industrial Estate, Phase III (Near Govind Puri Metro Station)

New Delhi, India – 110020 Office: A-302 (R&D Block)

Phone: 011-26907444

Email: raghava@iiitd.ac.in

Website: http://webs.iiitd.edu.in/raghava/

# 1. Abstract

Colorectal cancer (CRC) remains a major global health concern, with early detection being pivotal for improving patient outcomes. In this study, we leveraged high-throughput methylation profiling of cell-free DNA (cfDNA) to identify and validate diagnostic biomarkers for CRC. The GSE124600 study data were downloaded from the Gene Expression Omnibus (GEO), as the discovery cohort, comprising 142 CRC and 132 normal cfDNA methylation profiles obtained via MCTA-seq. After preprocessing and filtering, 97,863 CpG sites were retained for further analysis. Differential methylation analysis using statistical tests identified 30,791 CpG sites as significantly altered in CRC samples ($p < 0.05$). Univariate scoring enabled the selection of top-ranking features, which were further refined using multiple feature selection algorithms, including Recursive Feature Elimination (RFE), Sequential Feature Selection (SFS), and SVC-L1. Various machine learning (ML) models—such as Logistic Regression, Support Vector Machines, Random Forest, and Multi-layer Perceptron—were trained and tested using independent validation datasets. The best performance was achieved with an MLP model trained on 25 features selected by RFE, reaching an AUROC of 0.89 and MCC of 0.78 on validation data. Additionally, a deep learning-based convolutional neural network (CNN) achieved an AUROC of 0.78. Functional annotation of the most predictive CpG sites identified several genes involved in key cellular processes, some of which were validated for differential expression in CRC using the GEPIA2 platform. Our study highlights the potential of cfDNA methylation markers combined with ML/DL models for non-invasive and accurate CRC detection, paving the way for clinically relevant diagnostic tools.

**Author's Biography**

1. Kartavya Mathur is currently studying as an M.Sc. student at Gautam Buddha University, Greater Noida, Uttar Pradesh, India. He is currently working as an Intern on a Project position at the Department of Computational Biology, Indraprastha Institute of Information Technology (IIIT), New Delhi, India.
2. Shipra Jain is currently working as Ph.D. in Computational Biology from the Department of Computational Biology, Indraprastha Institute of Information Technology, New Delhi, India.

3. Nisha Bajiya is currently working as Ph.D. in Computational Biology from the Department of Computational Biology, Indraprastha Institute of Information Technology, New Delhi, India.
4. Nishant Kumar is currently working as Ph.D. in Computational Biology from the Department of Computational Biology, Indraprastha Institute of Information Technology, New Delhi, India.
5. Gajendra P. S. Raghava is currently working as Professor and Head of the Department of Computational Biology, Indraprastha Institute of Information Technology, New Delhi, India.

**Abbrevations**

**AB:** AdaBoost

**AUROC**: Area Under the Receiver Operating Characteristics

**Acc**: Accuracy

**CEA**: Carcinoembryonic Antigen

**CNN**: Convolutional Neural Network

**CRC**: Colorectal carcinogenesis

**DT**: Decision Tree

**ET**: Extra Trees Classifier

**GB**: Gradient Boosting

**GEO**: Gene Expression Omnibus

**KN**: K-Nearest Neighbor

**LR**: Logistic Regression

**MCC**: Matthews Correlation Coefficient

**MLP**: Multi-Layer Perceptron

**NB**: Naïve Bayes

**RF**: Random Forest

**RFE**: Recursive Feature Elimination

**Sens**: Sensitivity

**SFS**: Sequential Feature Selector

**Spec**: Specificity

**SVC**: Support Vector Classifier

## 2. Introduction

Colorectal carcinogenesis (CRC), arising due to abnormal growth of cells lining the colon and rectum regions of the large intestine, has become one of the most frequent malignancies globally [1,2]. According to the GLOBOCAN 2022 survey by Global Cancer Observatory (https://gco.iarc.fr/en), colorectal cancer is the third most common carcinoma globally in terms of incidence, with an Age-Standardized Rate (ASR) of 18.4 per 100,000 and mortality and ranks second most deadliest among all cancer types with 5.7 million prevalent cases in a span of 5-years [3]. According to cancer statistics published in the Cancer Journal for Clinicians, the United States alone accounts for nearly 152,810 cases of CRC, with an estimated death of nearly 53,010 [4]. Among the highest colorectal cancer incidence countries in 2022, India ranks sixth in terms of incidence (n = 70,038) and mortality (n = 40,993) (https://www.wcrf.org/preventing-cancer/cancer-statistics/colorectal-cancer-statistics/). These statistics emphasize the importance of early detection of CRC to reduce morbidity and mortality. Early recognition of CRC is challenging as there are very limited clinical symptoms, therefore, the majority of cases are identified at progressive stages when radical action is no longer a viable option.

Colorectal cancer (CRC) testing involves a spectrum of screening, diagnostic, and biomarker evaluations available in the market [5,6]. Standard screening methods include Fecal Occult Blood Tests (FOBT) and the Fecal Immunochemical Test (FIT), which detect hidden blood in stool but may miss polyps or cancers that do not bleed consistently [7–9]. Stool DNA tests detect genetic level changes but are costly and less accessible in small labs [10,11]. Colonoscopy is the cornerstone examination recommended by medical practitioners, which enables direct visualization, biopsy, and polyp removal, but it is an invasive procedure and carries risks like perforation of the intestine [12–14]. CT colonography and flexible sigmoidoscopy are less invasive but cannot detect small polyps developed in the initial stage and may require follow-up colonoscopy in case of abnormalities [15–18]. Blood tests such as Carcinoembryonic Antigen (CEA) and Carbohydrate Antigen (CA-19) are also widely used for monitoring but lack sensitivity and specificity for early detection [19–22]. CEA sensitivity in CRC varies from 65 to 74%, while CA19-9 sensitivity ranges between 26% and 48%. However, together, they provide an increased specificity and sensitivity [23]. A recent study found that patients with CEA or CA19-9 levels exceeding 200 had a lower 5-year overall survival (OS) than those below 200 or within the normal range. Additionally, individuals with elevated levels of both markers experienced a significantly shorter 5-year OS [24]. Recently, detecting methylated SEPT9 (mSEPT9) in blood has gained attention as a valuable tool for CRC screening. An enhanced mSEPT9 test demonstrated an overall recall of 90% and a specificity of 88% [25,26]. Despite the significant advancements made in analyzing genomic, transcriptomic and epigenomic profiles of colorectal cancer patients, we still lack specific biomarkers.

With recent advancements in genomic and transcriptomic profiling, liquid biopsies have emerged as a minimally invasive alternative and have appeared promising for detecting circulating biomarkers in body fluids, especially blood. Epigenetic modifications, including methylation and histone alterations, and non-coding RNAs like long noncoding RNAs (lncRNAs) and microRNAs (miRNAs) expressed during early carcinogenesis have been suggested as highly specific indicators of cancer presence and progression [27]. DNA methylation, in particular, plays an essential role in modulating gene expression and, therefore, is dysregulated in cancer cells [28]. Aberrant methylation patterns in circulating tumor DNA (ctDNA) from liquid biopsies (or cell-free DNA) can be used to monitor disease states as well as distinguish between cancerous and normal samples [29,30]. Molecular and genetic testing, including KRAS, NRAS, and BRAF mutations and Microsatellite Instability (MSI), shows highly specific personalized treatment but may not be universally accessible and can have high costs [31–33]. Tailored testing strategies are the need of the hour, which can optimize CRC prevention, early detection, and proper disease management.

Over the years, machine learning (ML) has emerged as a powerful tool in cancer diagnostics. ML techniques can analyze high-dimensional data generated by liquid biopsy methods, allowing for identifying patterns and genetic alterations that traditional methods might overlook. ML-based in-silico prediction tools can detect cancer before symptoms and reduce the risk of invasive procedures like repeated biopsies. These techniques hold immense promise in improving diagnostic accuracy, personalizing treatment strategies, and reducing the reliance on invasive procedures. In the present study, we aimed to develop a robust ML classifier capable of distinguishing between normal and colorectal cancer liquid biopsy samples using epigenomic signatures, specifically focusing on DNA methylation profiles. We hypothesized that ML models trained on these methylation patterns could achieve high sensitivity and specificity in differentiating cancerous from non-cancerous samples, potentially paving the way for a non-invasive and reliable diagnostic tool for CRC screening. Our approach involved the integration of key steps, including the extraction and processing of DNA methylation data, the selection of informative features, and the training of ML models. We evaluated the performance of various classifiers, optimized their parameters, and validated the results on an independent validation dataset. By leveraging the power of epigenomics and ML, this study seeks to contribute to developing next-generation diagnostic tools for colorectal cancer, focusing on early detection and improved patient outcomes.

## 3. Methodology

### 3.1. Data Acquisition

We explored the methylation profiles of Colorectal Cancer (CRC) samples from Gene Expression Omnibus (GEO) using a pre-designed keyword search strategy: "((cell free DNA[All Fields] OR cfDNA[All Fields] OR circulating tumor DNA[All Fields] OR ctDNA[All Fields]) AND (("colorectal neoplasms"[MeSH Terms] OR colorectal cancer[All Fields]) OR CRC[All Fields] OR (colorectal[All Fields] AND ("adenocarcinoma"[MeSH Terms] OR adenocarcinoma[All Fields])) OR ("colorectal neoplasms"[MeSH Terms] OR colorectal carcinoma[All Fields]))) AND "Homo sapiens"[porgn]". The accessions retrieved from the keywords were further filtered based on publication year, availability of processed dataset, and cell-free DNA obtained from blood specimens. The filtered datasets thus obtained were further grouped based on their methodology and ranked according to sample size. An in-house R script was used to download the processed datasets. The datasets were reassessed for sample size, balanced dataset, number of methylation sites and availability of proper annotations. Based on these parameters, GSE124600 was selected as the discovery cohort. We obtained 142 cancerous (CRC) and 132 control (Normal) sample methylation profiles from this dataset. The stage distribution of the cancerous sample of these patients is displayed in Figure 1. The data originated from the methylated CpG tandem amplification (MCTA) sequencing technique, which provided methylated alleles per million mapped reads as a measure for the quantification of gene expression using methylation profiling through high-throughput sequencing.

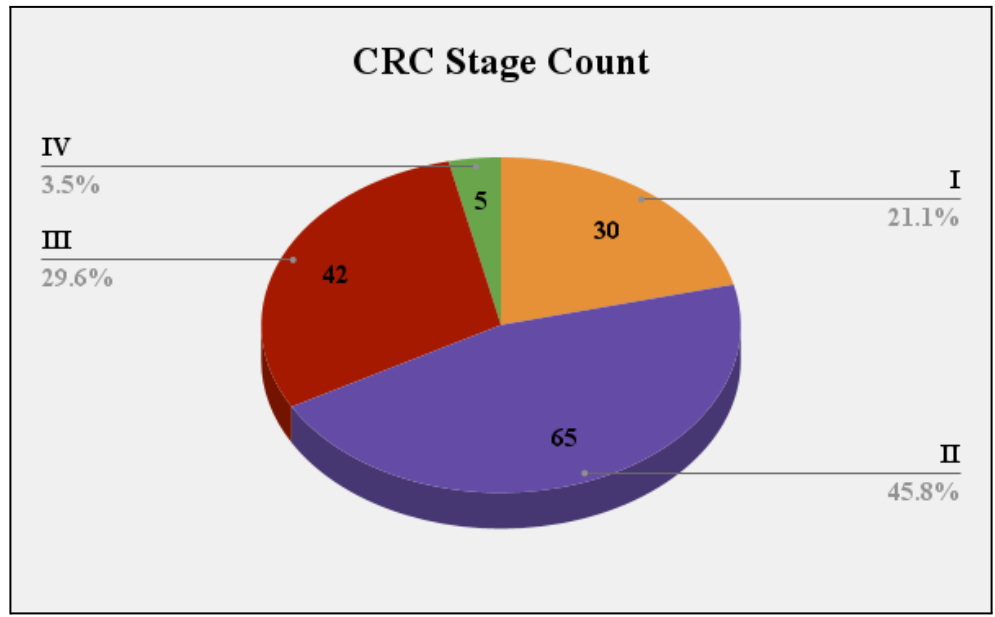

Figure 1: Pie chart showing stage distribution of CRC samples.

## 3.2. Data Preprocessing

We identified approximately 214,490 CpG methylated sites across the dataset for the cancer and control groups, which were examined for any missing values. The missing values were replaced with '0'. Basic statistical measures were calculated for each methylation site, including mean, standard deviation, and variance. The mean difference was computed to differentiate between cancer and normal samples. Around 54.38% of the sites showed a mean difference of 0 between the two groups, which were excluded from the study using a custom Python script. As a result, the number of CpG sites was reduced to 97,863 for further analysis.

## 3.3. Identification of differentially methylated CpG sites

We utilized the previously estimated mean methylation scores for each site in cancer and normal samples to identify differentially expressed methylation sites. Following this, a two-tailed independent Student's t-test was performed using a custom Python script to evaluate the statistical significance of the filtered methylation sites based on the mean values of CRC and normal samples. 30791 methylation sites out of 97863 with a confidence level greater than 95% ($p < 0.05$) were considered significant. We performed further analysis on these 30791 differentially methylated sites for the identification of diagnostic biomarker methylation sites, which can screen CRC and control groups with precision.

### 3.4. Feature Engineering and Feature Selection

Using the univariate score of 30791 sites, methylation sites were differentiated as positively correlated (U > 0) and negatively correlated (U < 0) among CRC and control samples. Upon that, Area Under Receiver Operating Characteristics (AUROC) based feature engineering was implemented, and single-feature-based models were developed to discriminate between cancerous and non-cancerous samples. The single-feature-based (or threshold-based) model was developed wherein the features with a probability score above the threshold are assigned to the normal class if found upregulated in normal samples, or it is assigned to a colorectal cancer class if upregulated in cancerous samples. Sorting was done based on maximum AUROC, and the top 100 methylation sites for both positive and negative correlation were selected. From the selected 100 methylation sites, sets of 25, 15, and 5 best-performing methylation sites were created for positively and negatively correlated sites to identify the minimal features, which can screen for either group. In addition to this, we have also implemented popular feature selection techniques separately on each positively and negatively correlated sets, like Recursive Feature Selection (RFE), Sequential Feature Selection and SVC-LI-based methods from the scikit-learn package.

### 3.5. Development of ML and DL based Models

In this study, for training our machine learning based classifier, we deployed eleven different algorithms, including Support Vector Classifier (SVC), Decision Tree (DT), Random Forest (RF), K-Nearest Neighbor (KNN), Logistic Regression (LR), Naïve Bayes (NB), XGBoost, AdaBoost (AB), Gradient Boosting (GB), Extra Trees Classifier (ET), and Multi-Layer Perceptron (MLP). A deep learning-based Convolutional Neural Network (CNN) was also implemented to develop a colorectal cancer diagnostic model.

### 3.6. Performance Evaluation

In this study, both internal validation via k-fold cross-validation and external validation using an independent dataset were executed to evaluate the models' performance. As depicted in Figure 2, the dataset of a total of 274 samples was randomly split in an 80:20 ratios, where 80% of the data

219 samples (103 healthy and 116 CRC) is allocated for training called as training dataset, and 20% data of 55 samples (29 healthy and 26 CRC) is used for validation called as validation dataset or independent dataset. The training dataset was used to build and evaluate models through a 5-fold cross-validation technique. Model performance on the testing dataset is called internal validation, where parameters are optimized to achieve the best results. To prevent over-optimization, the final model from internal validation (5-fold cross-validation) was further tested on an independent dataset that was not involved in training or testing. Additionally, in the present study, a cross-platform validation was also conducted using data from a different colorectal cancer study, GSE149438.

Both threshold-dependent and threshold-independent parameters were employed to measure the performance of models. In the case of threshold-dependent parameters, we computed sensitivity, specificity, accuracy and Matthew's correlation coefficient (MCC) using the following equations.

$$Sensitivity = \frac{TP}{TP+FN} * 100 \qquad (1)$$

$$Specificity = \frac{TN}{TN+FP} * 100 \qquad (2)$$

$$Accuracy = \frac{TP+TN}{TP+FP+TN+FN} * 100 \qquad (3)$$

$$MCC = \frac{(TP*TN) - (FP*FN)}{\sqrt{(TP+FP)(TP+FN)(TN+FP)(TN+FN)}} \qquad (4)$$

Where FP is false positive, FN is false negative, TP is true positive, and TN is true negative.

For threshold-independent measures, we used the standard parameter Area Under the Receiver Operating Characteristic curve (AUROC). The AUROC curve is generated by plotting sensitivity or true positive rate against the false positive rate (1-specificity) at various thresholds. Finally, the area under the ROC curve is calculated to compute a single AUROC parameter. AUROC with CI (Confidence Interval) were computed using the scikit-learn metrics module.

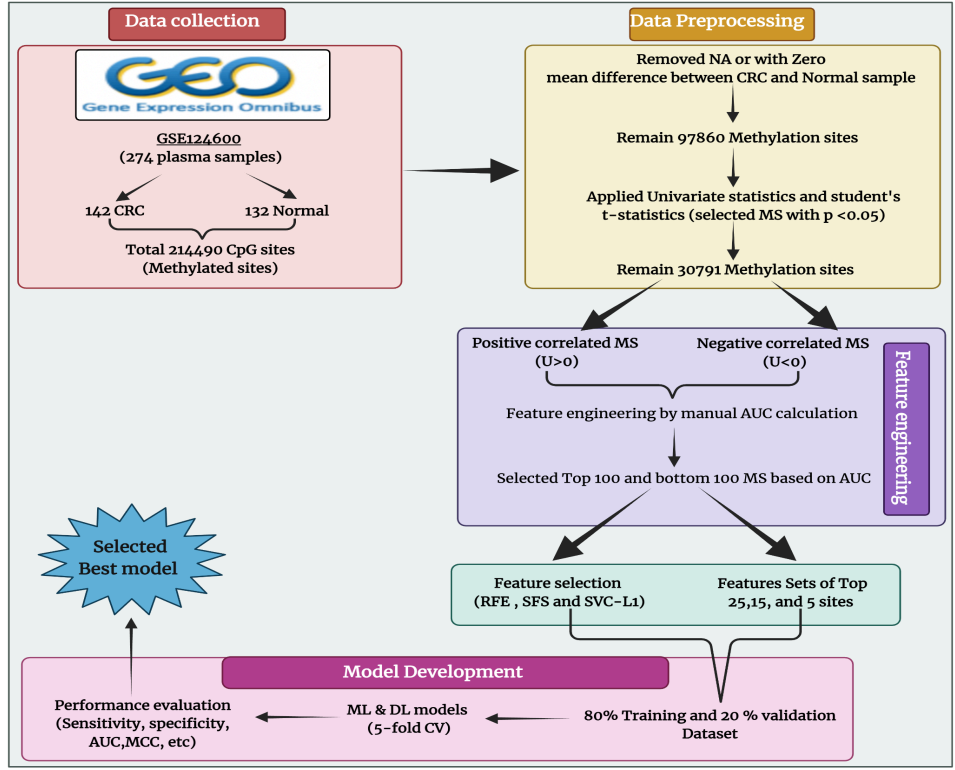

**Figure 2: Illustration of the complete study pipeline, providing a detailed stepwise description.**

### 3.7. Functional Annotation and Gene Ontology Analysis

Gene enrichment analysis was conducted using gProfiler to determine the biological significance of genes associated with signature CpG sites. This tool employs the Fisher exact test to compute gene enrichment scores and applies a correction to derive a Z-score. Further to this, the biological functions of relevant genes were retrieved from the GeneCards database for extrapolation. In addition, we have also utilized GEPIA2 tool [34] to validate the expression of our genes in colon and rectum cancers.

### 4. Results and Discussion

### 4.1. Preliminary data analysis

### 4.1.1 Differentially methylated CpG sites:

In the present study, we performed preliminary data statistical analysis by calculating the mean difference of various methylated CpG sites among CRC and control groups. Firstly, we designated a unique ID to each methylated CpG site, which can be referred to in Supplementary Table 1. Then, we calculated several statistical parameters, such as mean, variance, and standard deviation for each ID for both the cancer and control groups. As evident from Table 1, we observed the highest absolute mean difference among CRC and normal groups is 346.43 for 60881 ID. The statistical analysis parameters for 30791 IDs can be referred to in Supplementary Table 2.

**Table 1: List of top 10 significant differential methylation sites based on the highest absolute mean difference.**

| CpG site ID | Mean_CRC | Mean_NOR | Abs_Mean_Diff |
|---:|---:|---:|---:|
| 60881 | 761.17 | 1107.60 | 346.43 |
| 117819 | 694.09 | 963.57 | 269.49 |
| 60882 | 624.16 | 891.25 | 267.08 |
| 117820 | 650.18 | 888.81 | 238.62 |
| 60883 | 522.78 | 742.58 | 219.80 |
| 60884 | 394.99 | 571.43 | 176.45 |
| 117848 | 453.43 | 628.26 | 174.83 |
| 60885 | 328.94 | 474.48 | 145.55 |
| 117829 | 356.05 | 488.69 | 132.64 |
| 117830 | 348.69 | 473.21 | 124.51 |

### 4.1.2. Univariate analysis

In this study, we screened 30791 significant methylation sites out of 97863 with a confidence level greater than 95% ($p < 0.05$) using their univariate score. Based on univariate statistics, we divided our 30791 significantly enriched methylation sites into positively correlated ($U > 0$) and negatively correlated ($U < 0$) sets and selected the top 100 positively and negatively correlated sites. The top 5 positively and negatively correlated methylated sites are provided in Table 2, while the complete score of each site can be found in Supplementary Table 2. We identified that most of our significantly enriched methylation sites were associated with essential protein-coding genes and non-coding transcripts, which were potentially linked to various cell signaling, developmental, or gene regulatory pathways. The list of the top 5 positively and negatively correlated methylation sites between colorectal cancer and normal samples is depicted in Table 2.

**Table 2: List of the top 5 positively and negatively correlated methylation sites between colorectal cancer and normal samples.**

| Top 5 positively correlated sites | | | | Top 5 negatively correlated sites | | | |
|---|---|---|---|---|---|---|---|
| Site ID | Univariate score | t-test score | p-value | Site ID | Univariate score | t-test score | p-value |
| 155365 | 0.46 | 5.77 | 2.17E-08 | 75318 | -0.37 | -6.09 | 3.80E-09 |
| 155366 | 0.46 | 5.74 | 2.57E-08 | 75319 | -0.37 | -6.11 | 3.39E-09 |
| 162112 | 0.46 | 6.08 | 4.02E-09 | 7576 | -0.37 | -5.99 | 6.53E-09 |
| 162111 | 0.45 | 6.02 | 5.64E-09 | 7575 | -0.37 | -5.99 | 6.64E-09 |
| 197621 | 0.45 | 5.32 | 2.18E-07 | 7574 | -0.39 | -6.29 | 1.26E-09 |

## 4.2. Performance of Machine learning models

### 4.2.1. Single feature-based ML performance:

In the present study, we developed single-feature-based machine learning models based on the mean score of two groups, i.e. cancer and control groups. In order to implement a single feature-based model, we choose the top 100 positively and 100 negatively correlated methylated sites of 30791 sites. The performance parameters for 200 sites can be referred in Supplementary Table 3. We reported the highest AUROC as 0.75 for positively correlated site "181489" and 0.67 for negatively correlated site "3658", respectively.

### 4.2.2. ML model on significant methylation sites:

In this study, we have also developed a machine learning model based on all 30791 significant methylation sites to classify between the CRC and control groups. As depicted in Supplementary Table 4, we have achieved balanced parameters using an SVC classifier with the highest AUROC of 0.71 over the training and validation datasets, respectively, with an MCC score of 0.46 over validation.

### 4.2.3. Feature selection based on Univariate analysis:

To develop a robust prediction method, we have selected the top 25, 15, and 5 positive and negatively correlated methylated sites based on the univariate analysis score. The selected list of top 25, 15 and 5 positive and negative correlated methylated sites can be referred to in Supplementary Table 5.1. Using this approach, we reported the highest AUROC score as 0.88 using a Logistic Regression-based classifier over the top 15 positively correlated methylation sites. Whereas evident from Table 3, we reported the highest AUROC as 0.74 using an Extra tree-based classifier developed using the top 25 negatively correlated methylation sites. The complete results over the training and validation datasets can be referred in Supplementary Table 5.2.

**Table 3: Performance of ML-based models over validation dataset developed using top 25, 15 and 5 positive and negative correlated methylation sites based on univariate score.**

| Top Positively correlated methylated sites | | | | | | | |
|---|---|---|---|---|---|---|---|
| Name | Features | Thresh | Sens | Spec | Acc | AUROC | MCC |
| LR | 25 | 0.42 | 83.87 | 79.17 | 81.82 | 0.87 | 0.63 |
| LR | 15 | 0.42 | 83.87 | 79.17 | 81.82 | 0.88 | 0.63 |
| KN | 5 | 0.60 | 67.74 | 79.17 | 72.73 | 0.80 | 0.47 |
| Top Negatively correlated methylated sites | | | | | | | |
| Name | Features | Thresh | Sens | Spec | Acc | AUROC | MCC |
| ET | 25 | 0.50 | 83.87 | 54.17 | 70.91 | 0.74 | 0.40 |
| RF | 15 | 0.48 | 80.65 | 58.33 | 70.91 | 0.69 | 0.40 |
| ET | 5 | 0.52 | 67.74 | 58.33 | 63.64 | 0.64 | 0.26 |

# Thresh, Threshold; Sens, Sensitivity; Spec, Specificity; Acc, Accuracy; AUROC, Area Under Receiver Operating Characteristics; MCC, Matthews Correlation Coefficient; LR, Logistic Regression; KN, K-Nearest Neighbor; ET, Extra Trees Classifier; RF, Random Forest

### 4.2.4. Model developed on selected features by feature selection methods

In addition to this, we have implemented several well-known feature selection algorithms such as Recursive feature selection, Sequential feature selection, and SVC-L1 over the top 100 positively and negatively correlated methylation sites. Again, the top 25, 15, 10 and 5 best features identified as the feature selection results were deployed to train machine learning models. In Table 4, we have reported the best-performing model over the validation dataset using RFE, SFS, and SVC-L1-based feature selection techniques. As evident from Table 4, we observed the MLP-based model outperformed all and reported the highest AUROC as 0.89 and MCC as 0.78 with 25 methylated sites selected using the RFE feature selection technique. In contrast, SVC SVC-based ML model developed using 4 methylated sites extracted using the SFS selection algorithm reported the highest AUROC as 0.76 and MCC score as 0.53.

**Table 4: Performance of ML-based models over validation dataset developed using positive and negative correlated methylation sites selected based on feature selection techniques like RFE, SFS, and SVC-L1.**

| Top Positively correlated methylated sites | | | | | | | | |
|---|---|---|---|---|---|---|---|---|
| Feature selection | Name | Features | Sens | Spec | Acc | AUROC | MCC | Kappa |
| RFE | MLP | 25 | 93.10 | 84.62 | 89.09 | 0.89 | 0.78 | 0.78 |
| SFS | ET | 10 | 89.66 | 76.92 | 83.64 | 0.83 | 0.67 | 0.67 |
| SVC-L1 | SVC | 9 | 93.10 | 65.38 | 80.00 | 0.79 | 0.61 | 0.59 |
| Top Negatively correlated methylated sites | | | | | | | | |
| Feature selection | Name | Features | Sens | Spec | Acc | AUROC | MCC | Kappa |
| RFE | NB | 25 | 48.28 | 96.15 | 70.91 | 0.72 | 0.50 | 0.43 |
| SFS | SVC | 4 | 75.86 | 76.92 | 76.36 | 0.76 | 0.53 | 0.53 |
| SVC-L1 | NB | 50 | 55.17 | 92.31 | 72.73 | 0.74 | 0.51 | 0.46 |

#Sens, Sensitivity; Spec, Specificity; Acc, Accuracy; AUROC, Area Under the Curve; MCC, Matthews Correlation Coefficient; SVC, Support Vector Classifier; MLP, Multi-Layer Perceptron; NB, Naïve Bayes;

The list of selected positive and negative correlated methylated sites and their ML model performance score for RFE, SFS and SCV-L1 feature selection techniques can be referred to in Supplementary Tables 6, 7, and 8, respectively.

### 4.2.5. Deep learning model performance:

In the present study, we have also deployed a deep learning (DL) based model using CNN to investigate the discriminative ability between CRC and normal samples. The performance parameters of the CNN-based model among both groups are depicted in Table 5 for the training and validation sets. As shown, we observed the highest AUROC score of 0.87 and 0.78 on the training and validation datasets, respectively.

**Table 5: Performance of DL-based model (CNN) of training and validation sets on 30,479 methylation sites.**

| Model | Pos | Neg | Spec | Sens | AUROC | Acc | Kappa | MCC |
|---|---|---|---|---|---|---|---|---|
| Train | 116 | 103 | 89.32 | 84.48 | 0.87 | 86.76 | 0.735 | 0.737 |
| Validation | 26 | 29 | 79.31 | 76.92 | 0.78 | 78.18 | 0.562 | 0.562 |

#Pos, Positive; Neg, Negative; Sens, Sensitivity; Spec, Specificity; Acc, Accuracy; AUROC, Area Under the Curve; MCC, Matthews Correlation Coefficient

### 4.2.6 Functional annotation and gene ontology analysis

The methylation sites for our best-performing model derived from RFE selection were associated with ten protein-coding genes and three intronic regions. The protein-coding genes included T-cell leukemia homeobox (TLX2), Ellis van syndrome gene (EVC), SLIT homolog 3 (SLIT3), Sidekick cell adhesion molecule (SDK1), Brain-specific angiogenesis inhibitor 1 (BAI1), Growth Differentiation Factor 6 (GDF6), Tight Junction Protein 2 (TJP2), complement component 1,q subcomponent like protein (C1QL1), Fibroblast growth factor 4 (FGF4), and hydroxysteroid (17-beta) dehydrogenase 12 (HSD17B12). The function of all these genes is depicted in Table 6. Further, when performing gene enrichment analysis for these genes, we could not identify any tumor-specific pathways and biological/molecular processes associated with our gene set. We identified three molecular functions, including heparin-binding, glycosaminoglycan binding, and sulfur compound binding, as well as two biological processes, including neuron differentiation and generation of neurons.

**Table 6: The functional annotation of methylation sites linked with the best-performing model.**

| S.No. | Chromosome | Start | End | Strand | Gene | Function | References |
|---|---|---|---|---|---|---|---|
| 1 | chr2 | 74742649 | 74742650 | - | T-cell Leukemia homeobox (TLX2) | ❖ The homeobox-containing (HOX11 family) transcription factor family is majorly involved in cell differentiation and ALK1 signalling.<br>❖ Crucial for the development of the enteric nervous system.<br>❖ Loss-of-function of TLX2 is often linked with the tumorigenesis of gastrointestinal stromal tumors.<br>❖ Involved in the dilatation of the colon. | TLX2 Gene - GeneCards \| TLX2 Protein \| TLX2 Antibody [35] |
| 2 | chr4 | 147559358 | 147559359 | - | Intronic | Not Applicable | - |

| 3 | chr4 | 5713305 | 5713306 | + | Ellis van Creveld Syndrome (EVS) | ❖ A leucine zipper and transmembrane containing protein that positively regulates ciliary Hedgehog (Hh) signalling.<br>❖ Crucially involved in endochondral growth as well as skeletal development. | EVC Gene - GeneCards \| EVC Protein \| EVC Antibody [36] |
|---|---|---|---|---|---|---|---|
| 4 | chr4 | 5713227 | 5713228 | - | Ellis van Creveld Syndrome (EVS) | | |
| 5 | chr5 | 168727827 | 168727828 | + | SLIT homolog 3 (SLIT3) | ❖ Molecular guidance cues through cell migration by interacting with roundabout homolog receptors.<br>❖ Crucial part of SLIT-ROBO signalling and Netrin-signalling.<br>❖ Involved in soft palate malignancy, breast benign neoplasms, as well as mitochondrial complex deficiency.<br>❖ It has been suggested to influence thermogenesis and sympathetic activity.<br>❖ Shown to be inactivated in colorectal cancer. | SLIT3 Gene - GeneCards \| SLIT3 Protein \| SLIT3 Antibody) [37–39] |
| 6 | chr7 | 3341559 | 3341560 | + | Sidekick cell adhesion molecule (SDK1) | ❖ Single-pass transmembrane proteins have extracellular regions composed of an immunoglobulin domain, a fibronectin type III domain, and a short intracellular postsynaptic motif.<br>❖ Involved in cell junction organization and influences lamina-specific synaptic connections in the retina.<br>❖ Majorly involved in hypoparathyroidism.<br>❖ SDK1 mutations have been reported in malignant mesothelioma, adrenocortical carcinoma, lung cancer, as well as in gastric cancer. | SDK1 Gene - GeneCards \| SDK1 Protein \| SDK1 Antibody [40,41] |
| 7 | chr7 | 3341561 | 3341562 | + | Sidekick cell adhesion molecule (SDK1) | | |
| 8 | chr8 | 143533657 | 143533658 | + | Brain-specific angiogenesis inhibitor 1 (BAI1) (ADGRB1) | ❖ BAI1, a member of adhesion G protein-coupled receptors, mediates cell-cell and cell-matrix interactions.<br>❖ Loss may lead to deficits in neural development.<br>❖ Responsible for maintaining a balance between initiating and inhibiting new blood vessel formation. | ADGRB1 Gene - GeneCards \| AGRB1 Protein \| AGRB1 Antibody [42–44] |

| | | | | | | ❖ Suggested to be transcriptionally regulated by p53 to inhibit angiogenesis in glioblastoma.<br>❖ Key regulator in immune system processes, phagocytosis, besides cell adhesion.<br>❖ Crucial in gastric and medulloblastoma oncogenesis due to its role as an angiogenesis inhibitor.<br>❖ Reduced expression in colorectal cancer. | |
|---|---|---|---|---|---|---|---|
| 9 | chr8 | 97171925 | 97171926 | - | Growth differentiation factor (GDF6) | ❖ Responsible for encoding a key ligand for TGF-beta superfamily proteins to recruit and activate SMAD family transcription factors.<br>❖ Controls proliferation and cellular differentiation in retina/bone formation.<br>❖ Participates in various signalling pathways like GPCR, ERK signalling, and CREB, besides the GSK3 signalling pathway.<br>❖ Regulates apoptosis.<br>❖ Reduced expression of GDF6 in colorectal cancer tissues has been noted. | GDF6 Gene - GeneCards \| GDF6 Protein \| GDF6 Antibody [45–47] |
| 10 | chr9 | 71789398 | 71789399 | + | Tight junction protein 2 (TJP2) | ❖ Membrane-associated guanylate kinase homologs.<br>❖ Responsible for membrane-associated guanylate kinase activity and functions as a component of the tight junction and adherens junction barrier in epithelial and endothelial cells.<br>❖ Transcriptional repression of c-myc, programmed cell death, blood-brain-barrier and immune cell transmigration. | TJP2 Gene - GeneCards \| ZO2 Protein \| ZO2 Antibody) [48] |
| 11 | chr10 | 110226202 | 110226203 | - | Intronic | Not Applicable | - |
| 12 | chr10 | 23462279 | 23462280 | - | Intronic | | |

| 13 | chr11 | 43602856 | 43602857 | + | Hydroxysteroid (17-beta) dehydrogenase 12 (HSD17B12) | ❖ Mostly expressed in ovarian tissues and is responsible for converting estrone to estradiol.<br>❖ Involved in fatty acid biosynthesis and metabolism, especially fatty acid elongation.<br>❖ Responsible for oxidoreductase activity and collagen binding.<br>❖ Increased risk of death and progression of CRC. | HSD17B12 Gene - GeneCards \| DHB12 Protein \| DHB12 Antibody [49–51] |
|---|---|---|---|---|---|---|---|
| 14 | chr11 | 69589193 | 69589194 | - | Fibroblast growth factor 4 (FGF4) | ❖ Responsible for mitogenic as well as cell survival activities, including embryonic development, cell growth, morphogenesis, tissue repair, tumor growth and invasion.<br>❖ Essential in breast cancer-related pathways, MAPK signalling, focal adhesion through PI3K-Akt-mTOR signalling, etc.<br>❖ Associated with cardiac valve leaflet formation and limb development<br>❖ Dysregulated in haematological cancers, glioma susceptibility, embryonal carcinoma, etc. | FGF4 Gene - GeneCards \| FGF4 Protein \| FGF4 Antibody [52,53] |
| 15 | chr17 | 43045319 | 43045320 | + | Complement component 1, q subcomponent-like 1 (C1QL1) | ❖ C1Q complement family characterized by a C-terminal globular gC1Q signature domain involved in the formation of hetero- or homo-trimer.<br>❖ Signaling receptor binding activity.<br>❖ Interacts with BAI3 and controls the formation and maintenance of synapses.<br>❖ Overexpression of C1QL1 has been noted in CRC | C1QL1 Gene - GeneCards \| C1QRF Protein \| C1QRF Antibody [54,55] |

The literature and previous evidence show that hypermethylation of CpG islands can silence genes, i.e., suppress the expression of specific genes. The selected methylation sites based on positive correlation (high methylation value in cancerous samples and low methylation value in normal samples) suggested hypermethylation of these CpG regions. To correlate whether hypermethylation of our signature CpG methylation sites is associated with dysregulation of this gene expression, we explored gene expression patterns of the signature genes in the GEPIA2 server (http://gepia2.cancer-pku.cn/#index), which is trained on gene expression patterns of patients submitted on the TCGA database. Table 7 highlighted differentially expressed genes specific for Colon Adenocarcinoma (COAD) and Rectum Adenocarcinoma (READ). Out of 10, 7 (EVC, SLIT3, SDK1, ADGRB1, TJP2, HSD17B12, and C1QL1) genes were significantly expressed in colon and rectum adenocarcinoma. From these seven genes, 5 (EVC, SLIT3, SDK1, ADGRB1, and C1QL1) were predicted to be downregulated, suggesting that increased methylation of these genes in cancer samples could have caused their reduced expression. Two genes (HSD17B12 and TJP2) were overexpressed in both colon and rectum cancers, suggesting that they might have been hypomethylated, paving the way for transcriptional activity. However, more information is required to clearly deduce the correlation between methylation patterns and gene expression of these CpG sites.

**Table 7: CpG Signature Genes with differential gene expression from the GEPIA2 database.**

| S. No. | Gene | Colon Cancer (COAD) | | Rectum Cancer (READ) | |
|---|---|---|---|---|---|
| | | Significance ($p_{adj} < 0.05$) | Regulation (log2FoldChange) | Significance ($p_{adj} < 0.05$) | Regulation (log2FoldChange) |
| 1 | T-cell Leukemia homeobox (TLX2) | - | - | - | - |
| 2 | Ellis van Creveld Syndrome (EVC) | Yes, 2.55e-33 | Downregulated (-1.619) | Yes, 1.98e-15 | Downregulated (-1.752) |
| 3 | SLIT homolog 3 (SLIT3) | Yes, 9.59e-51 | Downregulated (-2.358) | Yes, 2.09e-22 | Downregulated (-2.305) |
| 4 | Sidekick cell adhesion molecule (SDK1) | Yes, 1.65e-39 | Downregulated (-1.638) | Yes, 1.18e-20 | Downregulated (-1.733) |
| 5 | Brain-specific angiogenesis inhibitor 1 (BAI1) / ADGRB1 | Yes, 9.03e-67 | Downregulated (-1.006) | Yes, 4.64e-38 | Downregulated (-1.116) |
| 6 | Growth differentiation factor (GDF6) | - | - | - | - |
| 7 | Tight junction protein 2 (TJP2) | Yes, 3.52e-48 | Upregulated (1.229) | Yes, 1.45e-25 | Upregulated (1.364) |
| 8 | Hydroxysteroid (17-beta) dehydrogenase 12 (HSD17B12) | Yes, 4.12e-71 | Upregulated (1.422) | Yes, 1.40e-33 | Upregulated (1.508) |
| 9 | Fibroblast growth factor 4 (FGF4) | - | - | - | - |

| 10 | Complement component 1, q subcomponent-like 1 (C1QL1) | Yes, 3.91e-70 | Downregulated (-1.456) | Yes, 4.61e-30 | Downregulated (-1.480) |

### 3.5. Conclusion

Colorectal cancer is a multifactorial disorder whose incidence and mortality are increasing daily due to a lack of panels for detecting CRC early in individuals. In conclusion, our study demonstrates the effectiveness of a univariate correlation statistics-based approach in identifying high-priority methylation sites for early colorectal cancer detection. We identified a panel of 15 upregulated methylation sites on EVC, FRMD6, FRMD6-AS2, LHFPL6, LIFR, and ZFPM2, along with three intergenic regions, as optimal classifiers. Using a random forest (RF)-based machine learning model, our approach achieved a specificity of 86.21% and sensitivity of 88.46%, comparable to the widely used methylated SEPTIN9 assay (88% specificity, 90% sensitivity). These findings underscore the potential of cfDNA methylation markers in enhancing non-invasive CRC detection and warrant further validation in clinical settings.

### Funding Source

The current work has been supported by the Department of Biotechnology (DBT) grant BT/PR40158/BTIS/137/24/2021.

### Conflict of interest

The authors declare no competing financial and non-financial interests.

### Authors' contributions

KM collected the dataset. KM, SJ, and NK processed the dataset. KM, NK, and GPSR implemented the algorithms and developed the prediction models. NK, SJ, KM, and GPSR analysed the results. NB, SJ, KM, NK, and GPSR performed the writing, reviewing, and draft preparation of the manuscript. GPSR conceived and coordinated the project. All authors have read and approved the final manuscript.

### Acknowledgments

Authors are thankful to the University Grants Commission (UGC), Council of Scientific & Industrial Research (CSIR), and IIIT-Delhi for fellowships and financial support, and the Department of Computational Biology, IIITD New Delhi, for infrastructure and facilities.

## References

[1] Baojun Duan 1 2 , Yaning Zhao 1 , Jun Bai 2 , Jianhua Wang 3 , Xianglong Duan 3 4 , Xiaohui Luo 1 , Rong Zhang 5 , Yansong Pu 3 , Mingqing Kou 6 , Jianyuan Lei 7 , Shangzhen Yang 8 Jose Andres Morgado-Diaz 1 , editors., Colorectal Cancer: An Overview, Gastrointestinal Cancers [Internet]., Brisbane (AU): Exon Publications, 2022. https://doi.org/10.36255/exon-publications-gastrointestinal-cancers-colorectal-cancer.

[2] H. Song, C. Ruan, Y. Xu, T. Xu, R. Fan, T. Jiang, M. Cao, J. Song, Survival stratification for colorectal cancer via multi-omics integration using an autoencoder-based model, Exp. Biol. Med. (Maywood) 247 (2022) 898–909. https://doi.org/10.1177/15353702211065010.

[3] S.T. Krishnan, D. Winkler, D. Creek, D. Anderson, C. Kirana, G.J. Maddern, K. Fenix, E. Hauben, D. Rudd, N.H. Voelcker, Staging of colorectal cancer using lipid biomarkers and machine learning, Metabolomics 19 (2023) 84. https://doi.org/10.1007/s11306-023-02049-z.

[4] R.L. Siegel, A.N. Giaquinto, A. Jemal, Cancer statistics, 2024, CA Cancer J. Clin. 74 (2024) 12–49. https://doi.org/10.3322/caac.21820.

[5] E.P. Whitlock, J. Lin, E. Liles, T. Beil, R. Fu, E. O'Connor, R.N. Thompson, T. Cardenas, Screening for colorectal cancer: An updated systematic review, Agency for Healthcare Research and Quality (US), Rockville (MD), 2008. https://www.ncbi.nlm.nih.gov/pubmed/20722162.

[6] W. Atkin, Options for screening for colorectal cancer, Scand. J. Gastroenterol. Suppl. (2003) 13–16. https://doi.org/10.1080/00855910310001421.

[7] R.S. Bresalier, C. Senore, G.P. Young, J. Allison, R. Benamouzig, S. Benton, P.M.M. Bossuyt, L. Caro, B. Carvalho, H.-M. Chiu, V.M.H. Coupé, W. de Klaver, C.M. de Klerk, E. Dekker, S. Dolwani, C.G. Fraser, W. Grady, L. Guittet, S. Gupta, S.P. Halloran, U. Haug, G. Hoff, S. Itzkowitz, T. Kortlever, A. Koulaouzidis, U. Ladabaum, B. Lauby-Secretan, M. Leja, B. Levin, T.R. Levin, F. Macrae, G.A. Meijer, J. Melson, C. O'Morain, S. Parry, L. Rabeneck, D.F. Ransohoff, R. Sáenz, H. Saito, S. Sanduleanu-Dascalescu, R.E. Schoen, K.

Selby, H. Singh, R.J.C. Steele, J.J.Y. Sung, E.L. Symonds, S.J. Winawer, Members of the World Endoscopy Colorectal Cancer Screening New Test Evaluation Expert Working Group, An efficient strategy for evaluating new non-invasive screening tests for colorectal cancer: the guiding principles, Gut 72 (2023) 1904–1918. https://doi.org/10.1136/gutjnl-2023-329701.

[8] K. Garborg, Ø. Holme, M. Løberg, M. Kalager, H.O. Adami, M. Bretthauer, Current status of screening for colorectal cancer, Ann. Oncol. 24 (2013) 1963–1972. https://doi.org/10.1093/annonc/mdt157.

[9] C. Senore, C. Doubeni, L. Guittet, FIT as a comparator for evaluating the effectiveness of new non-invasive CRC screening test, Dig. Dis. Sci. (2024). https://doi.org/10.1007/s10620-024-08718-w.

[10] R. Gómez-Molina, M. Suárez, R. Martínez, M. Chilet, J.M. Bauça, J. Mateo, Utility of stool-based tests for colorectal cancer detection: A comprehensive review, Healthcare (Basel) 12 (2024) 1645. https://doi.org/10.3390/healthcare12161645.

[11] S. Grego, C.M. Welling, G.H. Miller, P.F. Coggan, K.L. Sellgren, B.T. Hawkins, G.S. Ginsburg, J.R. Ruiz, D.A. Fisher, B.R. Stoner, A hands-free stool sampling system for monitoring intestinal health and disease, Sci. Rep. 12 (2022) 10859. https://doi.org/10.1038/s41598-022-14803-9.

[12] A.Z. Gimeno-García, E. Quintero, Role of colonoscopy in colorectal cancer screening: Available evidence, Best Pract. Res. Clin. Gastroenterol. 66 (2023) 101838. https://doi.org/10.1016/j.bpg.2023.101838.

[13] P.S. Liang, J.A. Dominitz, Colorectal cancer screening: Is colonoscopy the best option?, Med. Clin. North Am. 103 (2019) 111–123. https://doi.org/10.1016/j.mcna.2018.08.010.

[14] M.D. Knudsen, K. Wang, L. Wang, G. Polychronidis, P. Berstad, A. Hjartåker, Z. Fang, S. Ogino, A.T. Chan, M. Song, Colorectal cancer incidence and mortality after negative colonoscopy screening results, JAMA Oncol. 11 (2025) 46–54. https://doi.org/10.1001/jamaoncol.2024.5227.

[15] C.M. Spiceland, N. Lodhia, Endoscopy in inflammatory bowel disease: Role in diagnosis, management, and treatment, World J. Gastroenterol. 24 (2018) 4014–4020. https://doi.org/10.3748/wjg.v24.i35.4014.

[16] P.W.W. Chan, J.H. Ngu, Z. Poh, R. Soetikno, Colorectal cancer screening, Singapore Med. J. 58 (2017) 24–28. https://doi.org/10.11622/smedj.2017004.

[17] S.W. Chung, S. Hakim, S. Siddiqui, B.D. Cash, Update on flexible sigmoidoscopy, computed tomographic colonography, and capsule colonoscopy, Gastrointest. Endosc. Clin. N. Am. 30 (2020) 569–583. https://doi.org/10.1016/j.giec.2020.02.009.

[18] Q.-N. Liu, Y. Ye, X.-Q. Jia, Role of different examination methods in colorectal cancer screening: Insights and future directions, World J. Gastroenterol. 30 (2024) 4741–4744. https://doi.org/10.3748/wjg.v30.i44.4741.

[19] E. Halilovic, I. Rasic, A. Sofic, A. Mujic, A. Rovcanin, E. Hodzic, E. Kulovic, The importance of determining preoperative serum concentration of Carbohydrate antigen 19-9 and Carcinoembryonic antigen in assessing the progression of colorectal cancer, Med. Arch. 74 (2020) 346–349. https://doi.org/10.5455/medarh.2020.74.346-349.

[20] R.A. Rittgers, G. Steele Jr, N. Zamcheck, M.S. Loewenstein, P.H. Sugarbaker, R.J. Mayer, J.J. Lokich, J. Maltz, R.E. Wilso, Transient carcinoembryonic antigen (CEA) elevations following resection of colorectal cancer: a limitation in the use of serial CEA levels as an indicator for second-look surgery, J. Natl. Cancer Inst. 61 (1978) 315–318. https://www.ncbi.nlm.nih.gov/pubmed/277718.

[21] X. Li, L. Stassen, P. Schrotz-King, Z. Zhao, R. Cardoso, J.R. Raut, M. Bhardwaj, H. Brenner, Potential of fecal carcinoembryonic antigen for noninvasive detection of colorectal cancer: A systematic review, Cancers (Basel) 15 (2023) 5656. https://doi.org/10.3390/cancers15235656.

[22] M.G. Fakih, A. Padmanabhan, CEA monitoring in colorectal cancer. What you should know, Oncology (Williston Park) 20 (2006) 579–87; discussion 588, 594, 596 passim. https://www.ncbi.nlm.nih.gov/pubmed/16773844.

[23] J. Stiksma, D.C. Grootendorst, P.W.G. van der Linden, CA 19-9 as a marker in addition to CEA to monitor colorectal cancer, Clin. Colorectal Cancer 13 (2014) 239–244. https://doi.org/10.1016/j.clcc.2014.09.004.

[24] L. Lakemeyer, S. Sander, M. Wittau, D. Henne-Bruns, M. Kornmann, J. Lemke, Diagnostic and prognostic value of CEA and CA19-9 in colorectal cancer, Diseases 9 (2021) 21. https://doi.org/10.3390/diseases9010021.

[25] J. Hu, B. Hu, Y.-C. Gui, Z.-B. Tan, J.-W. Xu, Diagnostic value and clinical significance of methylated SEPT9 for colorectal cancer: A meta-analysis, Med. Sci. Monit. 25 (2019) 5813–5822. https://doi.org/10.12659/MSM.915472.

[26] J.D. Warren, W. Xiong, A.M. Bunker, C.P. Vaughn, L.V. Furtado, W.L. Roberts, J.C. Fang, W.S. Samowitz, K.A. Heichman, Septin 9 methylated DNA is a sensitive and specific blood test for colorectal cancer, BMC Med. 9 (2011) 133. https://doi.org/10.1186/1741-7015-9-133.

[27] K. Ashouri, A. Wong, P. Mittal, L. Torres-Gonzalez, J.H. Lo, S. Soni, S. Algaze, T. Khoukaz, W. Zhang, Y. Yang, J. Millstein, H.-J. Lenz, F. Battaglin, Exploring predictive and prognostic biomarkers in colorectal cancer: A comprehensive review, Cancers (Basel) 16 (2024) 2796. https://doi.org/10.3390/cancers16162796.

[28] L. Dong, H. Ren, Blood-based DNA methylation biomarkers for early detection of colorectal cancer, J. Proteomics Bioinform. 11 (2018) 120–126. https://doi.org/10.4172/jpb.1000477.

[29] A.G. Vaiopoulos, K.C. Athanasoula, A.G. Papavassiliou, Epigenetic modifications in colorectal cancer: molecular insights and therapeutic challenges, Biochim. Biophys. Acta 1842 (2014) 971–980. https://doi.org/10.1016/j.bbadis.2014.02.006.

[30] H.Y.S. Essa, G. Kusaf, O. Yuruker, R. Kalkan, Epigenetic alteration in colorectal cancer: A biomarker for diagnostic and therapeutic application, Glob. Med. Genet. 9 (2022) 258–262. https://doi.org/10.1055/s-0042-1757404.

[31] Y. Li, J. Xiao, T. Zhang, Y. Zheng, H. Jin, Analysis of KRAS, NRAS, and BRAF mutations, microsatellite instability, and relevant prognosis effects in patients with early colorectal cancer: A cohort study in east Asia, Front. Oncol. 12 (2022) 897548. https://doi.org/10.3389/fonc.2022.897548.

[32] J. Taieb, F.A. Sinicrope, L. Pederson, S. Lonardi, S.R. Alberts, T.J. George, G. Yothers, E. Van Cutsem, L. Saltz, S. Ogino, R. Kerr, T. Yoshino, R.M. Goldberg, T. André, P. Laurent-Puig, Q. Shi, Different prognostic values of KRAS exon 2 submutations and BRAF V600E mutation in microsatellite stable (MSS) and unstable (MSI) stage III colon cancer: an ACCENT/IDEA pooled analysis of seven trials, Ann. Oncol. 34 (2023) 1025–1034. https://doi.org/10.1016/j.annonc.2023.08.006.

[33] J. González-Montero, C.I. Rojas, M. Burotto, Predictors of response to immunotherapy in colorectal cancer, Oncologist 29 (2024) 824–832. https://doi.org/10.1093/oncolo/oyae152.

[34] Z. Tang, B. Kang, C. Li, T. Chen, Z. Zhang, GEPIA2: an enhanced web server for large-scale expression profiling and interactive analysis, Nucleic Acids Res. 47 (2019) W556–W560. https://doi.org/10.1093/nar/gkz430.

[35] B. Chen, X. Ding, A. Wan, X. Qi, X. Lin, H. Wang, W. Mu, G. Wang, J. Zheng, Author Correction: Comprehensive analysis of TLX2 in pan cancer as a prognostic and immunologic biomarker and validation in ovarian cancer, Sci. Rep. 13 (2023) 17678. https://doi.org/10.1038/s41598-023-44831-y.

[36] A. Kowal, A. Mostowska, D. Mydlak, B. Eberdt-Gołąbek, M. Misztal, P.P. Jagodziński, K.K. Hozyasz, EVC gene polymorphisms and risks of isolated hypospadias - a preliminary study, Cent. European J. Urol. 68 (2015) 257–262. https://doi.org/10.5173/ceju.2015.493.

[37] H.J. Cho, H. Kim, Y.-S. Lee, S.A. Moon, J.-M. Kim, H. Kim, M.J. Kim, J. Yu, K. Kim, I.-J. Baek, S.H. Lee, K.H. Ahn, S. Kim, J.-S. Kang, J.-M. Koh, SLIT3 promotes myogenic differentiation as a novel therapeutic factor against muscle loss, J. Cachexia Sarcopenia Muscle 12 (2021) 1724–1740. https://doi.org/10.1002/jcsm.12769.

[38] L. Ng, A.K.M. Chow, J.H.W. Man, T.C.C. Yau, T.M.H. Wan, D.N. Iyer, V.H.T. Kwan, R.T.P. Poon, R.W.C. Pang, W.-L. Law, Suppression of Slit3 induces tumor proliferation and chemoresistance in hepatocellular carcinoma through activation of GSK3β/β-catenin pathway, BMC Cancer 18 (2018) 621. https://doi.org/10.1186/s12885-018-4326-5.

[39] Y.-N. Wang, Y. Tang, Z. He, H. Ma, L. Wang, Y. Liu, Q. Yang, D. Pan, C. Zhu, S. Qian, Q.-Q. Tang, Slit3 secreted from M2-like macrophages increases sympathetic activity and thermogenesis in adipose tissue, Nat. Metab. 3 (2021) 1536–1551. https://doi.org/10.1038/s42255-021-00482-9.

[40] M. Yamagata, Structure and functions of sidekicks, Front. Mol. Neurosci. 13 (2020) 139. https://doi.org/10.3389/fnmol.2020.00139.

[41] K.M. Goodman, M. Yamagata, X. Jin, S. Mannepalli, P.S. Katsamba, G. Ahlsén, A.P. Sergeeva, B. Honig, J.R. Sanes, L. Shapiro, Molecular basis of sidekick-mediated cell-cell adhesion and specificity, Elife 5 (2016). https://doi.org/10.7554/eLife.19058.

[42] R.R. Parag, T. Yamamoto, K. Saito, D. Zhu, L. Yang, E.G. Van Meir, Novel isoforms of adhesion G protein-coupled receptor B1 (ADGRB1/BAI1) generated from an alternative

promoter in intron 17, Mol. Neurobiol. 62 (2025) 900–917. https://doi.org/10.1007/s12035-024-04293-3.

[43] G. Aust, D. Zhu, E.G. Van Meir, L. Xu, Adhesion GPCRs in tumorigenesis, Handb. Exp. Pharmacol. 234 (2016) 369–396. https://doi.org/10.1007/978-3-319-41523-9_17.

[44] Y. Fukushima, Y. Oshika, T. Tsuchida, T. Tokunaga, H. Hatanaka, H. Kijima, H. Yamazaki, Y. Ueyama, N. Tamaoki, M. Nakamura, Brain-specific angiogenesis inhibitor 1 expression is inversely correlated with vascularity and distant metastasis of colorectal cancer, Int. J. Oncol. 13 (1998) 967–970. https://doi.org/10.3892/ijo.13.5.967.

[45] A.J. Pellatt, L.E. Mullany, J.S. Herrick, L.C. Sakoda, R.K. Wolff, W.S. Samowitz, M.L. Slattery, The TGFβ-signaling pathway and colorectal cancer: associations between dysregulated genes and miRNAs, J. Transl. Med. 16 (2018) 191. https://doi.org/10.1186/s12967-018-1566-8.

[46] S. Krispin, A.N. Stratman, C.H. Melick, R.V. Stan, M. Malinverno, J. Gleklen, D. Castranova, E. Dejana, B.M. Weinstein, Growth differentiation factor 6 promotes vascular stability by restraining vascular endothelial growth factor signaling, Arterioscler. Thromb. Vasc. Biol. 38 (2018) 353–362. https://doi.org/10.1161/ATVBAHA.117.309571.

[47] H. Cui, J. Zhang, Z. Li, F. Chen, H. Cui, X. Du, H. Liu, J. Wang, A.D. Diwan, Z. Zheng, Growth differentiation factor-6 attenuates inflammatory and pain-related factors and degenerated disc-induced pain behaviors in rat model, J. Orthop. Res. 39 (2021) 959–970. https://doi.org/10.1002/jor.24793.

[48] J. Tang, M. Tan, Y. Deng, H. Tang, H. Shi, M. Li, W. Ma, J. Li, H. Dai, J. Li, S. Zhou, X. Li, F. Wei, X. Ma, L. Luo, Two novel pathogenic variants of TJP2 gene and the underlying molecular mechanisms in progressive familial intrahepatic cholestasis type 4 patients, Front. Cell Dev. Biol. 9 (2021) 661599. https://doi.org/10.3389/fcell.2021.661599.

[49] H. Heikelä, S.T. Ruohonen, M. Adam, R. Viitanen, H. Liljenbäck, O. Eskola, M. Gabriel, L. Mairinoja, A. Pessia, V. Velagapudi, A. Roivainen, F.-P. Zhang, L. Strauss, M. Poutanen, Hydroxysteroid (17β) dehydrogenase 12 is essential for metabolic homeostasis in adult mice, Am. J. Physiol. Endocrinol. Metab. 319 (2020) E494–E508. https://doi.org/10.1152/ajpendo.00042.2020.

[50] H. Kemiläinen, M. Adam, J. Mäki-Jouppila, P. Damdimopoulou, A.E. Damdimopoulos, J. Kere, O. Hovatta, T.D. Laajala, T. Aittokallio, J. Adamski, H. Ryberg, C. Ohlsson, L.

Strauss, M. Poutanen, The hydroxysteroid (17β) dehydrogenase family gene HSD17B12 is involved in the prostaglandin synthesis pathway, the ovarian function, and regulation of fertility, Endocrinology 157 (2016) 3719–3730. https://doi.org/10.1210/en.2016-1252.

[51] Y. Lin, Y. Meng, J. Zhang, L. Ma, L. Jiang, Y. Zhang, M. Yuan, A. Ren, W. Zhu, S. Li, Y. Shu, M. Du, L. Zhu, Functional genetic variant of HSD17B12 in the fatty acid biosynthesis pathway predicts the outcome of colorectal cancer, J. Cell. Mol. Med. 24 (2020) 14160–14170. https://doi.org/10.1111/jcmm.16026.

[52] A. Beenken, M. Mohammadi, The FGF family: biology, pathophysiology and therapy, Nat. Rev. Drug Discov. 8 (2009) 235–253. https://doi.org/10.1038/nrd2792.

[53] C.-S. Li, S.-X. Zhang, H.-J. Liu, Y.-L. Shi, L.-P. Li, X.-B. Guo, Z.-H. Zhang, Fibroblast growth factor receptor 4 as a potential prognostic and therapeutic marker in colorectal cancer, Biomarkers 19 (2014) 81–85. https://doi.org/10.3109/1354750X.2013.876555.

[54] S. Moghimyfiroozabad, M.A. Paul, S.M. Sigoillot, F. Selimi, Mapping and targeting of C1ql1-expressing cells in the mouse, Sci. Rep. 13 (2023) 17563. https://doi.org/10.1038/s41598-023-42924-2.

[55] X. Qiu, J.-R. Feng, F. Wang, P.-F. Chen, X.-X. Chen, R. Zhou, Y. Chang, J. Liu, Q. Zhao, Profiles of differentially expressed genes and overexpression of NEBL indicates a positive prognosis in patients with colorectal cancer, Mol. Med. Rep. 17 (2018) 3028–3034. https://doi.org/10.3892/mmr.2017.8210.

# SUPPLEMENTARY TABLES

**Supplementary Table 1: Unique ID representation of the top 100 positive and negative significantly methylated CpG sites.**

| Site ID | Chromosome no. | Start | End | Strand | Gene Symbol |
|---|---|---|---|---|---|
| 181489 | chr7 | 155164740 | 155164741 | - | NA |
| 196741 | chr8 | 73163886 | 73163887 | - | NA |
| 196742 | chr8 | 73163878 | 73163879 | - | NA |
| 181490 | chr7 | 155164736 | 155164737 | - | NA |
| 181491 | chr7 | 155164721 | 155164722 | - | NA |
| 93487 | chr1 | 181287827 | 181287828 | + | NA |
| 38139 | chr14 | 51560392 | 51560393 | - | TRIM9 |
| 86829 | chr19 | 58952130 | 58952131 | + | CTD-2619J13.19 |
| 189090 | chr8 | 105479178 | 105479179 | + | NA |
| 164637 | chr5 | 38556294 | 38556295 | + | LIFR |
| 197569 | chr8 | 97157394 | 97157395 | - | GDF6 |
| 33055 | chr13 | 39262101 | 39262102 | + | FREM2 |
| 19308 | chr11 | 69589200 | 69589201 | - | FGF4 |
| 155363 | chr4 | 5713260 | 5713261 | + | EVC |
| 93488 | chr1 | 181287836 | 181287837 | + | NA |
| 19307 | chr11 | 69589207 | 69589208 | - | FGF4 |
| 155366 | chr4 | 5713272 | 5713273 | + | EVC |
| 164638 | chr5 | 38556301 | 38556302 | + | LIFR |
| 109113 | chr20 | 21377300 | 21377301 | - | NKX2-4 |
| 155364 | chr4 | 5713266 | 5713267 | + | EVC |
| 22040 | chr12 | 108237684 | 108237685 | + | NA |
| 140674 | chr3 | 128274241 | 128274242 | + | NA |
| 155365 | chr4 | 5713270 | 5713271 | + | EVC |
| 140673 | chr3 | 128274237 | 128274238 | + | NA |
| 162111 | chr5 | 168727791 | 168727792 | + | NA |
| 14879 | chr11 | 31821271 | 31821272 | - | PAX6 |
| 197621 | chr8 | 97171940 | 97171941 | - | GDF6 |
| 53523 | chr16 | 55513384 | 55513385 | + | MMP2 |
| 22039 | chr12 | 108237667 | 108237668 | + | RP11-554D14.6 |
| 19310 | chr11 | 69589193 | 69589194 | - | FGF4 |
| 19309 | chr11 | 69589198 | 69589199 | - | FGF4 |
| 109111 | chr20 | 21377305 | 21377306 | - | NKX2-4 |
| 53525 | chr16 | 55513398 | 55513399 | + | MMP2 |
| 32305 | chr13 | 26042782 | 26042783 | + | ATP8A2 |
| 140675 | chr3 | 128274243 | 128274244 | + | NA |
| 32433 | chr13 | 27334449 | 27334450 | - | GPR12 |
| 14880 | chr11 | 31821265 | 31821266 | - | PAX6 |
| 142220 | chr3 | 156009151 | 156009152 | - | NA |
| 53524 | chr16 | 55513389 | 55513390 | + | MMP2 |
| 33056 | chr13 | 39262113 | 39262114 | + | FREM2 |
| 32303 | chr13 | 26042768 | 26042769 | + | ATP8A2 |
| 190943 | chr8 | 143533657 | 143533658 | + | NA |
| 162112 | chr5 | 168727793 | 168727794 | + | SLIT3 |
| 40169 | chr14 | 77737465 | 77737466 | + | NGB |
| 190944 | chr8 | 143533662 | 143533663 | + | NA |
| 190942 | chr8 | 143533649 | 143533650 | + | NA |
| 197570 | chr8 | 97157391 | 97157392 | - | GDF6 |
| 22768 | chr12 | 113909259 | 113909260 | + | NA |
| 140676 | chr3 | 128274247 | 128274248 | + | NA |
| 155367 | chr4 | 5713287 | 5713288 | + | EVC |
| 162113 | chr5 | 168727797 | 168727798 | + | NA |
| 14654 | chr11 | 2890925 | 2890926 | + | NA |
| 30599 | chr13 | 110960276 | 110960277 | + | NA |
| 32304 | chr13 | 26042777 | 26042778 | + | ATP8A2 |
| 140677 | chr3 | 128274252 | 128274253 | + | NA |
| 150767 | chr4 | 147559372 | 147559373 | - | NA |
| 14878 | chr11 | 31821274 | 31821275 | - | PAX6 |
| 109107 | chr20 | 21377312 | 21377313 | - | NKX2-4 |
| 163975 | chr5 | 1883206 | 1883207 | - | IRX4 |
| 140678 | chr3 | 128274261 | 128274262 | + | NA |
| 30600 | chr13 | 110960289 | 110960290 | + | NA |
| 15646 | chr11 | 43602856 | 43602857 | + | HSD17B12 |
| 150771 | chr4 | 147559354 | 147559355 | - | NA |
| 150770 | chr4 | 147559358 | 147559359 | - | NA |
| 62691 | chr17 | 43045319 | 43045320 | + | C1QL1 |
| 15643 | chr11 | 43602844 | 43602845 | + | HSD17B12 |
| 15645 | chr11 | 43602854 | 43602855 | + | HSD17B12 |
| 35227 | chr14 | 102030060 | 102030061 | - | NA |
| 155368 | chr4 | 5713291 | 5713292 | + | EVC |
| 206115 | chr9 | 71789398 | 71789399 | + | TJP2 |
| 162114 | chr5 | 168727809 | 168727810 | + | NA |
| 30601 | chr13 | 110960292 | 110960293 | + | NA |
| 145032 | chr3 | 32858836 | 32858837 | + | NA |
| 145033 | chr3 | 32858838 | 32858839 | + | NA |
| 197622 | chr8 | 97171925 | 97171926 | - | GDF6 |
| 155369 | chr4 | 5713302 | 5713303 | + | EVC |
| 15644 | chr11 | 43602846 | 43602847 | + | HSD17B12 |
| 140679 | chr3 | 128274264 | 128274265 | + | NA |
| 150768 | chr4 | 147559368 | 147559369 | - | NA |
| 19311 | chr11 | 69589170 | 69589171 | - | FGF4 |
| 137612 | chr2 | 74742649 | 74742650 | - | TLX2 |
| 162115 | chr5 | 168727813 | 168727814 | + | NA |
| 14336 | chr11 | 20618762 | 20618763 | + | NA |
| 40170 | chr14 | 77737494 | 77737495 | + | NGB |
| 155370 | chr4 | 5713305 | 5713306 | + | EVC |
| 184473 | chr7 | 3341549 | 3341550 | + | SDK1, AC073316.1 |
| 155374 | chr4 | 5713227 | 5713228 | - | EVC |
| 150773 | chr4 | 147559337 | 147559338 | - | NA |
| 1154 | chr10 | 110226219 | 110226220 | - | NA |
| 137625 | chr2 | 74743101 | 74743102 | - | TLX2 |
| 162118 | chr5 | 168727827 | 168727828 | + | NA |
| 184474 | chr7 | 3341554 | 3341555 | + | SDK1, AC073316.1 |
| 198323 | chr9 | 101706252 | 101706253 | - | NA |
| 198324 | chr9 | 101706249 | 101706250 | - | NA |
| 184476 | chr7 | 3341561 | 3341562 | + | SDK1, AC073316.1 |
| 184475 | chr7 | 3341559 | 3341560 | + | SDK1, AC073316.1 |
| 40171 | chr14 | 77737503 | 77737504 | + | NGB |
| 1155 | chr10 | 110226204 | 110226205 | - | NA |
| 1156 | chr10 | 110226202 | 110226203 | - | NA |
| 5779 | chr10 | 23462279 | 23462280 | - | NA |
| 3658 | chr10 | 134527979 | 134527980 | + | INPP5A |
| 100955 | chr1 | 2980307 | 2980308 | - | LINC00982 |
| 24643 | chr12 | 132837553 | 132837554 | - | GALNT9 |
| 24644 | chr12 | 132837550 | 132837551 | - | NA |
| 75322 | chr19 | 1071393 | 1071394 | - | HMHA1 |
| 100957 | chr1 | 2980290 | 2980291 | - | LINC00982 |
| 75321 | chr19 | 1071397 | 1071398 | - | HMHA1 |
| 189313 | chr8 | 1105774 | 1105775 | + | CTD-2281E23.2 |
| 100956 | chr1 | 2980300 | 2980301 | - | LINC00982 |
| 174999 | chr6 | 42227350 | 42227351 | + | TRERF1 |
| 9374 | chr10 | 75492828 | 75492829 | - | GLUD1P3, DUSP8P5 |
| 24639 | chr12 | 132837590 | 132837591 | - | GALNT9 |
| 180703 | chr7 | 150095096 | 150095097 | - | ZNF775 |
| 100954 | chr1 | 2980316 | 2980317 | - | LINC00982 |
| 125190 | chr2 | 107459700 | 107459701 | + | NA |
| 54573 | chr16 | 67686958 | 67686959 | - | RLTPR |
| 9375 | chr10 | 75492825 | 75492826 | - | GLUD1P3, DUSP8P5 |
| 114431 | chr20 | 62403000 | 62403001 | - | ZBTB46 |
| 180704 | chr7 | 150095094 | 150095095 | - | ZNF775 |
| 180705 | chr7 | 150095090 | 150095091 | - | NA |
| 75318 | chr19 | 1071414 | 1071415 | - | HMHA1 |
| 7577 | chr10 | 48429234 | 48429235 | - | GDF10 |
| 116026 | chr21 | 36164307 | 36164308 | - | RUNX1 |
| 125191 | chr2 | 107459708 | 107459709 | + | ST6GAL2 |
| 75320 | chr19 | 1071405 | 1071406 | - | HMHA1 |
| 17992 | chr11 | 65307223 | 65307224 | + | LTBP3 |
| 7575 | chr10 | 48429245 | 48429246 | - | GDF10 |
| 24642 | chr12 | 132837571 | 132837572 | - | GALNT9 |
| 100953 | chr1 | 2980333 | 2980334 | - | LINC00982 |
| 75319 | chr19 | 1071410 | 1071411 | - | HMHA1 |
| 126874 | chr2 | 129077403 | 129077404 | + | NA |
| 203525 | chr9 | 140173334 | 140173335 | - | TOR4A |
| 24640 | chr12 | 132837583 | 132837584 | - | GALNT9 |
| 78188 | chr19 | 17952552 | 17952553 | + | JAK3 |
| 24641 | chr12 | 132837574 | 132837575 | - | NA |
| 27300 | chr12 | 52473709 | 52473710 | + | RP11-1100L3.7, OR7E47P |
| 36304 | chr14 | 105609046 | 105609047 | - | JAG2 |
| 75616 | chr19 | 1122710 | 1122711 | - | SBNO2 |
| 171893 | chr6 | 167315154 | 167315155 | - | RPS6KA2, RP11-514O12.4 |
| 126871 | chr2 | 129077345 | 129077346 | + | NA |
| 97256 | chr1 | 228399680 | 228399681 | - | NA |
| 54574 | chr16 | 67686950 | 67686951 | - | RLTPR |
| 36303 | chr14 | 105609051 | 105609052 | - | NA |
| 202311 | chr9 | 138663007 | 138663008 | + | NA |
| 99171 | chr1 | 243637779 | 243637780 | + | SDCCAG8 |
| 126873 | chr2 | 129077362 | 129077363 | + | NA |
| 7576 | chr10 | 48429242 | 48429243 | - | GDF10 |
| 7701 | chr10 | 49659121 | 49659122 | + | ARHGAP22 |
| 7574 | chr10 | 48429250 | 48429251 | - | GDF10 |
| 149453 | chr4 | 11401606 | 11401607 | - | HS3ST1 |
| 7573 | chr10 | 48429258 | 48429259 | - | GDF10 |
| 78187 | chr19 | 17952537 | 17952538 | + | JAK3 |
| 99172 | chr1 | 243637788 | 243637789 | + | SDCCAG8 |
| 75129 | chr19 | 1046977 | 1046978 | + | ABCA7 |
| 126872 | chr2 | 129077352 | 129077353 | + | NA |
| 99173 | chr1 | 243637791 | 243637792 | + | SDCCAG8 |
| 90819 | chr1 | 151104697 | 151104698 | + | SEMA6C |
| 74073 | chr18 | 76782750 | 76782751 | - | NA |
| 62822 | chr17 | 43473133 | 43473134 | + | ARHGAP27 |
| 7700 | chr10 | 49659115 | 49659116 | + | ARHGAP22 |
| 89080 | chr1 | 112525331 | 112525332 | - | KCND3 |
| 36305 | chr14 | 105609043 | 105609044 | - | JAG2 |
| 160627 | chr5 | 140052645 | 140052646 | + | NA |
| 9376 | chr10 | 75492819 | 75492820 | - | GLUD1P3, DUSP8P5 |
| 7699 | chr10 | 49659109 | 49659110 | + | ARHGAP22 |
| 123337 | chr22 | 49935076 | 49935077 | + | C22orf34 |
| 7703 | chr10 | 49659153 | 49659154 | + | ARHGAP22 |
| 57404 | chr16 | 88721882 | 88721883 | + | MVD |
| 35413 | chr14 | 103373909 | 103373910 | - | NA |
| 90817 | chr1 | 151104661 | 151104662 | + | SEMA6C |
| 133220 | chr2 | 241989342 | 241989343 | + | NA |
| 127235 | chr2 | 132415683 | 132415684 | + | NA |
| 18002 | chr11 | 65307164 | 65307165 | - | LTBP3 |
| 17680 | chr11 | 64533706 | 64533707 | + | SF1 |
| 202310 | chr9 | 138662999 | 138663000 | + | KCNT1 |
| 113054 | chr20 | 58514147 | 58514148 | + | FAM217B, PPP1R3D |
| 58323 | chr17 | 10541763 | 10541764 | + | MYH3 |
| 75317 | chr19 | 1071419 | 1071420 | - | HMHA1 |
| 90224 | chr1 | 1375230 | 1375231 | - | VWA1 |
| 78181 | chr19 | 17942591 | 17942592 | + | JAK3 |
| 30580 | chr13 | 110918389 | 110918390 | - | NA |
| 116025 | chr21 | 36164318 | 36164319 | - | RUNX1 |
| 35412 | chr14 | 103373932 | 103373933 | - | TRAF3 |
| 35411 | chr14 | 103373938 | 103373939 | - | TRAF3 |
| 35414 | chr14 | 103373907 | 103373908 | - | NA |
| 202309 | chr9 | 138662987 | 138662988 | + | KCNT1 |
| 123336 | chr22 | 49935066 | 49935067 | + | C22orf34 |
| 24632 | chr12 | 132837654 | 132837655 | + | GALNT9 |
| 189312 | chr8 | 1105760 | 1105761 | + | CTD-2281E23.2 |
| 24631 | chr12 | 132837641 | 132837642 | + | NA |
| 41489 | chr15 | 101099302 | 101099303 | + | NA |
| 54571 | chr16 | 67686982 | 67686983 | - | RLTPR |
| 24476 | chr12 | 132393497 | 132393498 | + | ULK1 |
| 54570 | chr16 | 67686988 | 67686989 | - | RLTPR |
| 84787 | chr19 | 49183853 | 49183854 | - | SEC1P |
| 41491 | chr15 | 101099317 | 101099318 | + | RP11-526I2.1 |
| 204581 | chr9 | 34372181 | 34372182 | - | KIAA1161 |
| 41490 | chr15 | 101099305 | 101099306 | + | RP11-526I2.1 |
| 54572 | chr16 | 67686964 | 67686965 | - | RLTPR |
| 30581 | chr13 | 110918379 | 110918380 | - | COL4A1 |

**Supplementary Table 2: Statistical analysis of the top 100 positive and negative significantly methylated CpG sites**

| Site ID | Mean_ CRC | StdDev_ CRC | Variance_ CRC | Mean_ NOR | StdDev_ NOR | Variance_ NOR | Abs_Mean _Diff | Univariate score | t-test score | p-value |
|---|---|---|---|---|---|---|---|---|---|---|
| 181489 | 5.27 | 9.44 | 89.05 | 0.63 | 1.46 | 2.13 | 4.64 | 0.43 | 5.59 | 5.47E-08 |
| 196741 | 6.48 | 13.55 | 183.73 | 0.50 | 1.33 | 1.77 | 5.98 | 0.40 | 5.05 | 0.000000824 |
| 196742 | 6.15 | 13.11 | 171.79 | 0.37 | 1.03 | 1.07 | 5.78 | 0.41 | 5.05 | 0.000000805 |
| 181490 | 4.38 | 8.49 | 72.06 | 0.38 | 1.17 | 1.36 | 4.00 | 0.41 | 5.37 | 0.000000169 |
| 181491 | 3.33 | 7.24 | 52.40 | 0.12 | 0.52 | 0.27 | 3.21 | 0.41 | 5.08 | 0.000000688 |
| 93487 | 3.33 | 7.17 | 51.43 | 0.15 | 0.52 | 0.27 | 3.19 | 0.41 | 5.09 | 0.000000664 |
| 38139 | 7.06 | 14.88 | 221.27 | 0.61 | 1.32 | 1.74 | 6.44 | 0.40 | 4.96 | 0.00000125 |
| 86829 | 3.96 | 7.02 | 49.23 | 0.61 | 1.32 | 1.75 | 3.35 | 0.40 | 5.40 | 0.000000146 |
| 189090 | 3.00 | 4.57 | 20.91 | 0.61 | 1.32 | 1.75 | 2.40 | 0.41 | 5.80 | 1.82E-08 |
| 164637 | 3.00 | 6.71 | 45.01 | 0.12 | 0.52 | 0.27 | 2.87 | 0.40 | 4.91 | 0.00000158 |
| 197569 | 2.47 | 5.20 | 27.01 | 0.13 | 0.48 | 0.23 | 2.34 | 0.41 | 5.14 | 0.000000522 |
| 33055 | 10.08 | 14.67 | 215.31 | 2.46 | 2.99 | 8.92 | 7.62 | 0.43 | 5.85 | 0.000000014 |
| 19308 | 3.39 | 7.60 | 57.79 | 0.09 | 0.34 | 0.11 | 3.30 | 0.42 | 4.99 | 0.0000011 |

| | | | | | | | | | |
|---|---|---|---|---|---|---|---|---|---|
| 155363 | 2.85 | 4.97 | 24.72 | 0.34 | 1.11 | 1.24 | 2.50 | 0.41 | 5.65 | 4.05E-08 |
| 93488 | 2.98 | 6.85 | 46.96 | 0.08 | 0.35 | 0.12 | 2.90 | 0.40 | 4.86 | 0.00000203 |
| 19307 | 4.02 | 8.09 | 65.38 | 0.38 | 0.92 | 0.85 | 3.64 | 0.40 | 5.14 | 0.000000515 |
| 155366 | 2.45 | 4.69 | 22.01 | 0.10 | 0.40 | 0.16 | 2.35 | 0.46 | 5.74 | 2.57E-08 |
| 164638 | 2.70 | 6.05 | 36.60 | 0.11 | 0.44 | 0.20 | 2.59 | 0.40 | 4.91 | 0.00000158 |
| 109113 | 3.11 | 7.04 | 49.53 | 0.11 | 0.48 | 0.23 | 3.00 | 0.40 | 4.88 | 0.0000018 |
| 155364 | 2.70 | 4.90 | 23.99 | 0.28 | 1.01 | 1.03 | 2.42 | 0.41 | 5.56 | 0.000000065 |
| 22040 | 3.11 | 5.65 | 31.96 | 0.37 | 1.13 | 1.28 | 2.73 | 0.40 | 5.45 | 0.000000113 |
| 140674 | 2.32 | 4.62 | 21.38 | 0.10 | 0.49 | 0.24 | 2.22 | 0.43 | 5.48 | 9.51E-08 |
| 155365 | 2.49 | 4.73 | 22.40 | 0.11 | 0.40 | 0.16 | 2.38 | 0.46 | 5.77 | 2.17E-08 |
| 140673 | 2.53 | 4.80 | 23.08 | 0.18 | 0.59 | 0.35 | 2.35 | 0.44 | 5.59 | 5.47E-08 |
| 162111 | 1.63 | 2.79 | 7.81 | 0.14 | 0.48 | 0.23 | 1.48 | 0.45 | 6.02 | 5.64E-09 |
| 14879 | 6.02 | 11.59 | 134.44 | 0.75 | 1.35 | 1.83 | 5.27 | 0.41 | 5.19 | 0.000000412 |
| 197621 | 2.47 | 5.29 | 28.03 | 0.02 | 0.13 | 0.02 | 2.45 | 0.45 | 5.32 | 0.000000218 |
| 53523 | 5.11 | 10.52 | 110.73 | 0.47 | 1.00 | 1.00 | 4.64 | 0.40 | 5.05 | 0.00000082 |
| 22039 | 4.62 | 7.91 | 62.58 | 0.60 | 1.27 | 1.62 | 4.02 | 0.44 | 5.77 | 2.14E-08 |
| 19310 | 2.89 | 6.74 | 45.37 | 0.05 | 0.26 | 0.07 | 2.84 | 0.41 | 4.83 | 0.00000224 |
| 19309 | 3.20 | 7.32 | 53.65 | 0.09 | 0.34 | 0.11 | 3.11 | 0.41 | 4.88 | 0.00000184 |
| 109111 | 3.43 | 7.47 | 55.76 | 0.18 | 0.66 | 0.43 | 3.25 | 0.40 | 4.98 | 0.00000112 |
| 53525 | 3.22 | 6.67 | 44.55 | 0.17 | 0.56 | 0.31 | 3.05 | 0.42 | 5.23 | 0.000000334 |
| 32305 | 6.06 | 12.22 | 149.41 | 0.54 | 1.16 | 1.34 | 5.51 | 0.41 | 5.16 | 0.000000475 |
| 140675 | 2.21 | 4.35 | 18.96 | 0.09 | 0.46 | 0.21 | 2.11 | 0.44 | 5.54 | 6.96E-08 |
| 32433 | 2.25 | 4.88 | 23.85 | 0.06 | 0.30 | 0.09 | 2.19 | 0.42 | 5.14 | 0.000000515 |
| 14880 | 5.18 | 10.09 | 101.88 | 0.61 | 1.16 | 1.34 | 4.57 | 0.41 | 5.17 | 0.000000446 |
| 142220 | 2.23 | 4.56 | 20.77 | 0.15 | 0.56 | 0.32 | 2.08 | 0.41 | 5.20 | 0.000000392 |

| | | | | | | | | | |
|---|---|---|---|---|---|---|---|---|---|
| 53524 | 4.61 | 9.66 | 93.27 | 0.39 | 0.96 | 0.91 | 4.22 | 0.40 | 5.00 | 0.00000104 |
| 33056 | 6.39 | 9.67 | 93.48 | 1.25 | 2.19 | 4.78 | 5.14 | 0.43 | 5.97 | 7.55E-09 |
| 32303 | 7.15 | 14.13 | 199.77 | 0.90 | 1.53 | 2.34 | 6.25 | 0.40 | 5.05 | 0.000000803 |
| 190943 | 3.78 | 8.71 | 75.82 | 0.03 | 0.23 | 0.05 | 3.75 | 0.42 | 4.95 | 0.0000013 |
| 162112 | 1.42 | 2.48 | 6.13 | 0.09 | 0.40 | 0.16 | 1.33 | 0.46 | 6.08 | 4.02E-09 |
| 40169 | 2.50 | 4.09 | 16.72 | 0.37 | 0.95 | 0.90 | 2.12 | 0.42 | 5.82 | 1.66E-08 |
| 190944 | 3.60 | 8.32 | 69.16 | 0.03 | 0.23 | 0.05 | 3.57 | 0.42 | 4.93 | 0.0000014 |
| 190942 | 3.91 | 9.03 | 81.63 | 0.07 | 0.38 | 0.15 | 3.84 | 0.41 | 4.88 | 0.00000184 |
| 197570 | 2.23 | 4.91 | 24.06 | 0.09 | 0.33 | 0.11 | 2.14 | 0.41 | 5.00 | 0.00000102 |
| 22768 | 2.94 | 5.67 | 32.12 | 0.34 | 0.82 | 0.67 | 2.60 | 0.40 | 5.21 | 0.000000365 |
| 140676 | 2.05 | 4.12 | 16.99 | 0.06 | 0.34 | 0.11 | 1.99 | 0.45 | 5.53 | 7.52E-08 |
| 155367 | 1.63 | 3.42 | 11.73 | 0.05 | 0.27 | 0.07 | 1.58 | 0.43 | 5.28 | 0.000000263 |
| 162113 | 1.35 | 2.42 | 5.84 | 0.09 | 0.40 | 0.16 | 1.26 | 0.45 | 5.90 | 0.000000011 |
| 14654 | 4.84 | 5.17 | 26.70 | 1.75 | 2.59 | 6.72 | 3.09 | 0.40 | 6.18 | 2.37E-09 |
| 30599 | 2.93 | 5.34 | 28.57 | 0.38 | 0.89 | 0.80 | 2.55 | 0.41 | 5.42 | 0.000000132 |
| 32304 | 6.65 | 13.30 | 176.76 | 0.74 | 1.33 | 1.77 | 5.91 | 0.40 | 5.08 | 0.000000696 |
| 140677 | 1.96 | 3.98 | 15.83 | 0.06 | 0.34 | 0.11 | 1.90 | 0.44 | 5.46 | 0.000000106 |
| 150767 | 4.63 | 7.35 | 54.01 | 0.87 | 1.50 | 2.25 | 3.76 | 0.43 | 5.77 | 2.15E-08 |
| 14878 | 9.01 | 15.32 | 234.83 | 1.59 | 2.44 | 5.98 | 7.42 | 0.42 | 5.50 | 8.98E-08 |
| 109107 | 3.67 | 7.52 | 56.49 | 0.35 | 0.83 | 0.69 | 3.32 | 0.40 | 5.05 | 0.000000828 |
| 163975 | 2.50 | 5.13 | 26.35 | 0.17 | 0.57 | 0.33 | 2.33 | 0.41 | 5.18 | 0.000000429 |
| 140678 | 1.88 | 3.87 | 15.01 | 0.06 | 0.34 | 0.11 | 1.82 | 0.43 | 5.38 | 0.000000158 |
| 30600 | 2.74 | 5.04 | 25.37 | 0.33 | 0.86 | 0.74 | 2.41 | 0.41 | 5.42 | 0.000000129 |
| 15646 | 2.38 | 4.97 | 24.72 | 0.10 | 0.42 | 0.18 | 2.28 | 0.42 | 5.25 | 0.000000312 |
| 150771 | 2.06 | 4.32 | 18.62 | 0.11 | 0.40 | 0.16 | 1.94 | 0.41 | 5.15 | 0.000000506 |

| | | | | | | | | | |
|---|---|---|---|---|---|---|---|---|---|
| 150770 | 2.13 | 4.45 | 19.81 | 0.13 | 0.41 | 0.17 | 2.00 | 0.41 | 5.14 | 0.000000516 |
| 62691 | 0.85 | 1.74 | 3.04 | 0.02 | 0.13 | 0.02 | 0.83 | 0.44 | 5.48 | 9.69E-08 |
| 15643 | 2.84 | 5.58 | 31.18 | 0.24 | 0.68 | 0.46 | 2.59 | 0.41 | 5.30 | 0.000000244 |
| 15645 | 2.44 | 5.02 | 25.22 | 0.11 | 0.43 | 0.18 | 2.32 | 0.43 | 5.30 | 0.000000244 |
| 35227 | 1.17 | 2.16 | 4.66 | 0.12 | 0.43 | 0.18 | 1.06 | 0.41 | 5.52 | 7.82E-08 |
| 155368 | 1.49 | 3.31 | 10.97 | 0.03 | 0.22 | 0.05 | 1.46 | 0.41 | 5.06 | 0.000000779 |
| 206115 | 1.86 | 4.34 | 18.85 | 0.04 | 0.26 | 0.07 | 1.82 | 0.40 | 4.82 | 0.00000239 |
| 162114 | 1.17 | 2.20 | 4.85 | 0.07 | 0.36 | 0.13 | 1.10 | 0.43 | 5.65 | 4.12E-08 |
| 30601 | 2.45 | 4.65 | 21.66 | 0.24 | 0.68 | 0.46 | 2.21 | 0.41 | 5.40 | 0.000000147 |
| 145032 | 1.32 | 2.85 | 8.11 | 0.06 | 0.27 | 0.07 | 1.26 | 0.40 | 5.07 | 0.000000747 |
| 145033 | 1.32 | 2.85 | 8.11 | 0.06 | 0.27 | 0.07 | 1.26 | 0.40 | 5.07 | 0.000000747 |
| 197622 | 1.30 | 3.10 | 9.62 | 0.00 | 0.00 | 0.00 | 1.30 | 0.42 | 4.83 | 0.00000231 |
| 155369 | 1.20 | 2.75 | 7.56 | 0.01 | 0.09 | 0.01 | 1.19 | 0.42 | 4.98 | 0.00000116 |
| 15644 | 2.66 | 5.42 | 29.33 | 0.20 | 0.61 | 0.37 | 2.46 | 0.41 | 5.18 | 0.000000438 |
| 140679 | 1.59 | 3.52 | 12.38 | 0.03 | 0.21 | 0.04 | 1.55 | 0.42 | 5.06 | 0.000000778 |
| 150768 | 3.67 | 6.40 | 40.93 | 0.59 | 1.10 | 1.21 | 3.08 | 0.41 | 5.45 | 0.000000114 |
| 19311 | 1.21 | 2.83 | 8.01 | 0.01 | 0.08 | 0.01 | 1.20 | 0.41 | 4.89 | 0.00000177 |
| 137612 | 1.00 | 2.25 | 5.05 | 0.03 | 0.17 | 0.03 | 0.97 | 0.40 | 4.93 | 0.00000141 |
| 162115 | 0.98 | 1.91 | 3.63 | 0.06 | 0.35 | 0.12 | 0.92 | 0.41 | 5.45 | 0.000000111 |
| 14336 | 1.00 | 2.47 | 6.08 | 0.00 | 0.05 | 0.00 | 1.00 | 0.40 | 4.66 | 0.00000501 |
| 40170 | 1.32 | 2.93 | 8.58 | 0.04 | 0.21 | 0.04 | 1.28 | 0.41 | 5.02 | 0.000000954 |
| 155370 | 0.92 | 2.30 | 5.29 | 0.00 | 0.00 | 0.00 | 0.92 | 0.40 | 4.61 | 0.00000617 |
| 184473 | 1.09 | 2.33 | 5.42 | 0.01 | 0.12 | 0.01 | 1.08 | 0.44 | 5.31 | 0.000000232 |
| 155374 | 1.00 | 2.28 | 5.18 | 0.03 | 0.17 | 0.03 | 0.97 | 0.40 | 4.89 | 0.00000169 |
| 150773 | 1.01 | 2.27 | 5.17 | 0.03 | 0.16 | 0.03 | 0.98 | 0.40 | 4.93 | 0.00000146 |

| | | | | | | | | | |
|---|---|---|---|---|---|---|---|---|---|
| 1154 | 0.51 | 1.04 | 1.08 | 0.01 | 0.12 | 0.02 | 0.50 | 0.43 | 5.44 | 0.000000118 |
| 137625 | 1.12 | 2.44 | 5.96 | 0.03 | 0.23 | 0.05 | 1.08 | 0.41 | 5.08 | 0.000000713 |
| 162118 | 0.65 | 1.42 | 2.02 | 0.02 | 0.13 | 0.02 | 0.64 | 0.41 | 5.12 | 0.000000576 |
| 184474 | 1.04 | 2.33 | 5.44 | 0.01 | 0.12 | 0.01 | 1.03 | 0.42 | 5.08 | 0.000000698 |
| 198323 | 0.61 | 1.34 | 1.79 | 0.01 | 0.12 | 0.01 | 0.60 | 0.41 | 5.10 | 0.000000627 |
| 198324 | 0.59 | 1.32 | 1.75 | 0.01 | 0.12 | 0.01 | 0.58 | 0.40 | 5.02 | 0.000000953 |
| 184476 | 0.93 | 2.07 | 4.30 | 0.00 | 0.00 | 0.00 | 0.93 | 0.45 | 5.13 | 0.000000562 |
| 184475 | 0.93 | 2.07 | 4.30 | 0.00 | 0.00 | 0.00 | 0.93 | 0.45 | 5.13 | 0.000000562 |
| 40171 | 0.88 | 2.21 | 4.89 | 0.00 | 0.00 | 0.00 | 0.88 | 0.40 | 4.56 | 0.00000777 |
| 1155 | 0.46 | 0.99 | 0.98 | 0.01 | 0.12 | 0.02 | 0.45 | 0.40 | 5.12 | 0.000000578 |
| 1156 | 0.43 | 0.98 | 0.96 | 0.00 | 0.00 | 0.00 | 0.43 | 0.44 | 5.00 | 0.00000102 |
| 5779 | 0.51 | 1.22 | 1.49 | 0.00 | 0.00 | 0.00 | 0.51 | 0.42 | 4.81 | 0.00000247 |
| 3658 | 53.29 | 20.69 | 428.12 | 69.25 | 29.20 | 852.48 | 15.96 | -0.32 | -5.25 | 0.000000307 |
| 100955 | 56.67 | 20.93 | 438.03 | 71.94 | 23.45 | 549.88 | 15.27 | -0.34 | -5.69 | 0.000000032 |
| 24643 | 44.55 | 18.36 | 337.21 | 58.01 | 20.99 | 440.39 | 13.46 | -0.34 | -5.66 | 3.86E-08 |
| 24644 | 38.41 | 16.60 | 275.56 | 49.85 | 18.83 | 354.40 | 11.44 | -0.32 | -5.35 | 0.000000191 |
| 75322 | 26.92 | 11.93 | 142.35 | 35.68 | 14.41 | 207.75 | 8.76 | -0.33 | -5.49 | 9.04E-08 |
| 100957 | 50.67 | 19.48 | 379.37 | 64.09 | 21.46 | 460.40 | 13.42 | -0.33 | -5.43 | 0.000000127 |
| 75321 | 31.06 | 13.85 | 191.80 | 41.76 | 16.50 | 272.27 | 10.70 | -0.35 | -5.83 | 1.59E-08 |
| 189313 | 46.04 | 18.20 | 331.30 | 58.67 | 22.25 | 495.28 | 12.63 | -0.31 | -5.16 | 0.000000481 |
| 100956 | 53.52 | 19.99 | 399.59 | 67.64 | 22.02 | 484.81 | 14.13 | -0.34 | -5.57 | 6.23E-08 |
| 174999 | 44.89 | 21.78 | 474.29 | 59.67 | 24.47 | 598.63 | 14.78 | -0.32 | -5.29 | 0.000000253 |
| 9374 | 46.19 | 23.41 | 547.98 | 63.85 | 30.59 | 935.98 | 17.66 | -0.33 | -5.39 | 0.000000155 |
| 24639 | 90.58 | 30.57 | 934.59 | 114.08 | 36.08 | 1301.54 | 23.50 | -0.35 | -5.83 | 1.56E-08 |
| 180703 | 113.51 | 30.34 | 920.29 | 136.50 | 36.19 | 1309.92 | 22.99 | -0.35 | -5.71 | 2.92E-08 |
| 100954 | 63.74 | 21.14 | 447.02 | 80.53 | 26.57 | 705.87 | 16.78 | -0.35 | -5.81 | 1.78E-08 |
| 125190 | 63.03 | 20.51 | 420.81 | 77.88 | 24.07 | 579.44 | 14.85 | -0.33 | -5.51 | 8.39E-08 |
| 54573 | 27.32 | 11.84 | 140.27 | 37.02 | 18.73 | 350.98 | 9.70 | -0.32 | -5.16 | 0.000000478 |
| 9375 | 40.96 | 20.80 | 432.82 | 56.32 | 27.76 | 770.80 | 15.36 | -0.32 | -5.21 | 0.00000038 |
| 114431 | 79.97 | 36.88 | 1360.41 | 104.13 | 42.48 | 1804.56 | 24.16 | -0.30 | -5.04 | 0.000000863 |
| 180704 | 109.89 | 30.10 | 906.19 | 131.53 | 35.10 | 1232.17 | 21.64 | -0.33 | -5.49 | 9.28E-08 |
| 180705 | 100.65 | 27.37 | 749.21 | 121.40 | 32.82 | 1076.95 | 20.75 | -0.34 | -5.70 | 3.12E-08 |
| 75318 | 40.28 | 15.74 | 247.59 | 53.41 | 19.82 | 392.81 | 13.12 | -0.37 | -6.09 | 3.8E-09 |
| 7577 | 35.25 | 15.09 | 227.74 | 49.22 | 25.23 | 636.68 | 13.97 | -0.35 | -5.61 | 5.05E-08 |
| 116026 | 18.14 | 9.36 | 87.66 | 25.61 | 14.01 | 196.37 | 7.47 | -0.32 | -5.22 | 0.000000348 |
| 125191 | 54.37 | 17.66 | 311.87 | 65.92 | 20.52 | 421.27 | 11.55 | -0.30 | -5.00 | 0.00000102 |
| 75320 | 33.38 | 14.31 | 204.76 | 45.00 | 17.91 | 320.61 | 11.62 | -0.36 | -5.96 | 7.96E-09 |
| 17992 | 35.92 | 13.87 | 192.37 | 45.56 | 17.31 | 299.63 | 9.64 | -0.31 | -5.11 | 0.00000062 |
| 7575 | 74.01 | 22.56 | 509.15 | 97.82 | 41.19 | 1696.52 | 23.81 | -0.37 | -5.99 | 6.64E-09 |
| 24642 | 56.23 | 21.82 | 476.14 | 72.91 | 26.22 | 687.25 | 16.68 | -0.35 | -5.74 | 2.53E-08 |
| 100953 | 67.69 | 21.74 | 472.41 | 84.82 | 28.43 | 808.43 | 17.13 | -0.34 | -5.62 | 4.62E-08 |
| 75319 | 36.83 | 15.58 | 242.78 | 49.73 | 19.26 | 370.79 | 12.89 | -0.37 | -6.11 | 3.39E-09 |
| 126874 | 20.55 | 11.69 | 136.55 | 28.76 | 15.73 | 247.45 | 8.21 | -0.30 | -4.93 | 0.00000145 |
| 203525 | 98.26 | 33.40 | 1115.60 | 120.27 | 35.66 | 1271.57 | 22.01 | -0.32 | -5.28 | 0.00000027 |
| 24640 | 86.67 | 28.94 | 837.78 | 109.02 | 34.69 | 1203.51 | 22.35 | -0.35 | -5.81 | 1.78E-08 |
| 78188 | 120.67 | 41.09 | 1688.18 | 150.03 | 47.41 | 2247.34 | 29.36 | -0.33 | -5.49 | 9.24E-08 |
| 24641 | 61.55 | 22.93 | 525.76 | 78.99 | 27.43 | 752.55 | 17.44 | -0.35 | -5.72 | 2.73E-08 |
| 27300 | 25.23 | 14.09 | 198.45 | 35.05 | 17.21 | 296.23 | 9.82 | -0.31 | -5.18 | 0.000000427 |
| 36304 | 204.85 | 71.46 | 5106.14 | 252.14 | 81.36 | 6620.09 | 47.29 | -0.31 | -5.12 | 0.000000577 |
| 75616 | 57.27 | 22.57 | 509.22 | 71.12 | 22.96 | 527.34 | 13.85 | -0.30 | -5.03 | 0.000000878 |
| 171893 | 36.57 | 20.32 | 413.02 | 53.67 | 32.34 | 1045.99 | 17.09 | -0.32 | -5.28 | 0.000000269 |
| 126871 | 126.99 | 46.97 | 2205.83 | 161.17 | 66.11 | 4370.85 | 34.19 | -0.30 | -4.96 | 0.00000124 |
| 97256 | 29.19 | 10.41 | 108.39 | 36.85 | 14.78 | 218.56 | 7.66 | -0.30 | -4.99 | 0.00000109 |
| 54574 | 16.83 | 7.85 | 61.70 | 23.53 | 14.22 | 202.29 | 6.70 | -0.30 | -4.87 | 0.00000188 |
| 36303 | 215.85 | 75.75 | 5738.04 | 265.03 | 86.63 | 7504.95 | 49.17 | -0.30 | -5.01 | 0.00000098 |
| 202311 | 137.84 | 55.34 | 3062.27 | 176.45 | 69.77 | 4867.47 | 38.61 | -0.31 | -5.09 | 0.000000658 |
| 99171 | 115.43 | 36.89 | 1360.67 | 142.46 | 47.42 | 2248.90 | 27.03 | -0.32 | -5.29 | 0.000000256 |
| 126873 | 102.93 | 39.41 | 1553.00 | 132.18 | 57.30 | 3282.84 | 29.25 | -0.30 | -4.95 | 0.00000129 |
| 7576 | 68.95 | 21.59 | 466.10 | 90.84 | 37.33 | 1393.48 | 21.89 | -0.37 | -5.99 | 6.53E-09 |
| 7701 | 207.57 | 78.54 | 6168.45 | 261.10 | 89.26 | 7966.58 | 53.52 | -0.32 | -5.28 | 0.000000267 |
| 7574 | 84.47 | 24.29 | 589.99 | 110.74 | 42.92 | 1842.44 | 26.27 | -0.39 | -6.29 | 1.26E-09 |
| 149453 | 88.94 | 41.07 | 1687.05 | 120.11 | 62.01 | 3845.52 | 31.18 | -0.30 | -4.94 | 0.00000138 |
| 7573 | 124.14 | 35.50 | 1260.17 | 155.61 | 57.84 | 3345.17 | 31.47 | -0.34 | -5.47 | 0.000000102 |
| 78187 | 136.58 | 48.10 | 2313.91 | 168.10 | 55.32 | 3060.54 | 31.52 | -0.30 | -5.04 | 0.000000841 |
| 99172 | 112.99 | 36.18 | 1308.69 | 139.27 | 46.24 | 2138.56 | 26.29 | -0.32 | -5.26 | 0.000000292 |
| 75129 | 66.63 | 20.61 | 424.61 | 83.95 | 36.22 | 1311.88 | 17.31 | -0.30 | -4.91 | 0.0000016 |
| 126872 | 118.73 | 45.14 | 2037.74 | 151.79 | 64.11 | 4110.58 | 33.06 | -0.30 | -4.96 | 0.00000123 |
| 99173 | 110.07 | 35.55 | 1263.93 | 135.45 | 45.06 | 2030.20 | 25.37 | -0.31 | -5.19 | 0.000000405 |
| 90819 | 22.99 | 14.38 | 206.90 | 32.32 | 16.62 | 276.15 | 9.34 | -0.30 | -4.98 | 0.00000112 |
| 74073 | 56.83 | 22.93 | 525.81 | 74.39 | 35.46 | 1257.60 | 17.56 | -0.30 | -4.90 | 0.00000164 |
| 62822 | 130.58 | 42.92 | 1842.46 | 162.93 | 51.49 | 2651.32 | 32.34 | -0.34 | -5.66 | 3.79E-08 |
| 7700 | 221.23 | 83.49 | 6971.20 | 277.17 | 94.18 | 8868.98 | 55.94 | -0.31 | -5.21 | 0.000000372 |
| 89080 | 97.17 | 39.20 | 1536.28 | 123.58 | 48.16 | 2318.99 | 26.41 | -0.30 | -4.99 | 0.00000106 |
| 36305 | 196.88 | 69.33 | 4806.14 | 241.40 | 78.51 | 6163.68 | 44.52 | -0.30 | -4.98 | 0.00000112 |
| 160627 | 32.38 | 15.87 | 251.98 | 42.48 | 17.57 | 308.76 | 10.10 | -0.30 | -5.00 | 0.00000105 |
| 9376 | 12.32 | 8.75 | 76.58 | 19.37 | 14.50 | 210.30 | 7.05 | -0.30 | -4.91 | 0.00000155 |
| 7699 | 239.20 | 90.37 | 8165.86 | 298.27 | 101.59 | 10319.95 | 59.07 | -0.31 | -5.09 | 0.00000066 |
| 123337 | 67.43 | 21.73 | 472.40 | 81.75 | 25.45 | 647.94 | 14.32 | -0.30 | -5.02 | 0.000000942 |
| 7703 | 58.65 | 27.01 | 729.40 | 76.61 | 28.75 | 826.66 | 17.96 | -0.32 | -5.33 | 0.000000204 |
| 57404 | 15.38 | 8.31 | 69.00 | 21.99 | 12.31 | 151.58 | 6.60 | -0.32 | -5.24 | 0.000000327 |
| 35413 | 99.38 | 44.16 | 1950.22 | 128.07 | 46.68 | 2179.30 | 28.69 | -0.32 | -5.23 | 0.000000343 |
| 90817 | 126.62 | 40.55 | 1644.33 | 153.63 | 47.74 | 2279.41 | 27.02 | -0.31 | -5.06 | 0.000000772 |
| 133220 | 168.67 | 54.24 | 2942.33 | 205.88 | 67.56 | 4564.54 | 37.21 | -0.31 | -5.04 | 0.000000836 |
| 127235 | 71.38 | 26.43 | 698.47 | 87.54 | 26.98 | 728.05 | 16.16 | -0.30 | -5.01 | 0.000000999 |
| 18002 | 87.35 | 32.39 | 1049.01 | 110.10 | 40.66 | 1653.55 | 22.75 | -0.31 | -5.14 | 0.000000524 |
| 17680 | 34.38 | 11.44 | 130.85 | 43.56 | 17.65 | 311.68 | 9.19 | -0.32 | -5.15 | 0.000000508 |
| 202310 | 146.77 | 58.37 | 3406.96 | 187.92 | 74.60 | 5564.67 | 41.15 | -0.31 | -5.10 | 0.000000625 |
| 113054 | 50.69 | 23.12 | 534.45 | 65.46 | 24.63 | 606.52 | 14.78 | -0.31 | -5.12 | 0.00000057 |
| 58323 | 29.22 | 14.18 | 201.15 | 38.91 | 18.02 | 324.67 | 9.69 | -0.30 | -4.97 | 0.00000121 |
| 75317 | 46.30 | 17.75 | 315.05 | 59.71 | 22.80 | 519.78 | 13.41 | -0.33 | -5.45 | 0.000000112 |
| 90224 | 48.01 | 16.82 | 282.85 | 60.08 | 23.00 | 529.00 | 12.07 | -0.30 | -4.98 | 0.00000111 |
| 78181 | 92.58 | 26.43 | 698.64 | 111.35 | 33.60 | 1129.18 | 18.77 | -0.31 | -5.16 | 0.000000482 |
| 30580 | 58.66 | 23.94 | 573.34 | 77.68 | 36.40 | 1325.28 | 19.02 | -0.32 | -5.14 | 0.000000517 |
| 116025 | 46.43 | 17.72 | 313.98 | 59.85 | 25.65 | 657.74 | 13.42 | -0.31 | -5.07 | 0.000000743 |
| 35412 | 200.32 | 86.76 | 7527.10 | 259.22 | 91.15 | 8308.53 | 58.90 | -0.33 | -5.48 | 9.69E-08 |
| 35411 | 232.14 | 98.92 | 9784.26 | 298.92 | 112.34 | 12619.23 | 66.78 | -0.32 | -5.23 | 0.000000338 |
| 35414 | 97.62 | 43.56 | 1897.58 | 125.24 | 45.83 | 2100.49 | 27.63 | -0.31 | -5.12 | 0.000000593 |
| 202309 | 172.12 | 67.28 | 4526.42 | 218.42 | 86.71 | 7519.47 | 46.30 | -0.30 | -4.96 | 0.00000126 |
| 123336 | 89.67 | 26.30 | 691.84 | 107.95 | 34.19 | 1169.23 | 18.28 | -0.30 | -4.98 | 0.00000112 |
| 24632 | 70.50 | 27.40 | 751.01 | 87.30 | 28.59 | 817.67 | 16.80 | -0.30 | -4.97 | 0.00000121 |
| 189312 | 101.85 | 31.50 | 992.23 | 123.62 | 37.10 | 1376.40 | 21.76 | -0.32 | -5.25 | 0.000000312 |
| 24631 | 96.48 | 37.11 | 1376.89 | 120.51 | 40.74 | 1659.38 | 24.03 | -0.31 | -5.11 | 0.000000607 |
| 41489 | 146.28 | 44.66 | 1994.81 | 176.32 | 51.76 | 2679.09 | 30.04 | -0.31 | -5.15 | 0.000000492 |
| 54571 | 40.44 | 16.11 | 259.45 | 52.38 | 22.94 | 526.42 | 11.94 | -0.31 | -5.01 | 0.000000972 |
| 24476 | 92.17 | 30.90 | 954.83 | 116.80 | 45.61 | 2080.37 | 24.62 | -0.32 | -5.26 | 0.000000287 |
| 54570 | 42.23 | 16.65 | 277.34 | 54.50 | 23.87 | 569.58 | 12.26 | -0.30 | -4.96 | 0.00000124 |
| 84787 | 59.56 | 20.54 | 422.09 | 73.64 | 25.56 | 653.18 | 14.08 | -0.31 | -5.04 | 0.000000838 |
| 41491 | 139.96 | 42.97 | 1846.28 | 167.72 | 49.43 | 2443.74 | 27.76 | -0.30 | -4.97 | 0.00000118 |
| 204581 | 26.89 | 11.63 | 135.33 | 36.01 | 18.69 | 349.35 | 9.13 | -0.30 | -4.89 | 0.00000174 |
| 41490 | 142.69 | 43.54 | 1895.43 | 171.00 | 50.69 | 2569.58 | 28.31 | -0.30 | -4.97 | 0.00000119 |
| 54572 | 33.59 | 13.40 | 179.59 | 44.31 | 20.83 | 434.02 | 10.72 | -0.31 | -5.10 | 0.000000631 |
| 30581 | 51.49 | 21.87 | 478.33 | 68.48 | 31.24 | 975.66 | 17.00 | -0.32 | -5.25 | 0.000000311 |

**Supplementary Table 3: Performance of single features of Top 100 positively and bottom 100 negatively correlated Methylated sites(from 30791 MS)**

| Top 100 Positively correlated methylation sites | | | | | | | | | | | |
|---|---|---|---|---|---|---|---|---|---|---|---|
| Site ID | Mean_CRC | Mean_NOR | Threshold | True Positive | True Negative | False Positive | False Negative | Sensitivity | Specificity | FPR | AUC-ROC |
| 181489 | 5.27 | 0.63 | 0.63 | 103 | 102 | 30 | 39 | 0.73 | 0.77 | 0.23 | 0.75 |
| 196741 | 6.48 | 0.50 | 0.50 | 93 | 109 | 23 | 49 | 0.65 | 0.83 | 0.17 | 0.74 |
| 196742 | 6.15 | 0.37 | 0.37 | 88 | 113 | 19 | 54 | 0.62 | 0.86 | 0.14 | 0.74 |
| 181490 | 4.38 | 0.38 | 0.38 | 90 | 111 | 21 | 52 | 0.63 | 0.84 | 0.16 | 0.74 |
| 181491 | 3.33 | 0.12 | 0.48 | 74 | 124 | 8 | 68 | 0.52 | 0.94 | 0.06 | 0.73 |
| 93487 | 3.33 | 0.15 | 0.15 | 75 | 120 | 12 | 67 | 0.53 | 0.91 | 0.09 | 0.72 |
| 38139 | 7.06 | 0.61 | 0.61 | 100 | 96 | 36 | 42 | 0.70 | 0.73 | 0.27 | 0.72 |
| 86829 | 3.96 | 0.61 | 0.98 | 92 | 103 | 29 | 50 | 0.65 | 0.78 | 0.22 | 0.71 |
| 189090 | 3.00 | 0.61 | 0.61 | 93 | 101 | 31 | 49 | 0.65 | 0.77 | 0.23 | 0.71 |
| 164637 | 3.00 | 0.12 | 0.44 | 68 | 124 | 8 | 74 | 0.48 | 0.94 | 0.06 | 0.71 |
| 197569 | 2.47 | 0.13 | 0.39 | 71 | 120 | 12 | 71 | 0.50 | 0.91 | 0.09 | 0.70 |
| 33055 | 10.08 | 2.46 | 3.31 | 95 | 97 | 35 | 47 | 0.67 | 0.73 | 0.27 | 0.70 |
| 19308 | 3.39 | 0.09 | 0.46 | 67 | 123 | 9 | 75 | 0.47 | 0.93 | 0.07 | 0.70 |
| 155363 | 2.85 | 0.34 | 0.62 | 76 | 114 | 18 | 66 | 0.54 | 0.86 | 0.14 | 0.70 |
| 93488 | 2.98 | 0.08 | 0.72 | 63 | 126 | 6 | 79 | 0.44 | 0.95 | 0.05 | 0.70 |
| 19307 | 4.02 | 0.38 | 0.38 | 85 | 105 | 27 | 57 | 0.60 | 0.80 | 0.20 | 0.70 |
| 155366 | 2.45 | 0.10 | 0.62 | 64 | 124 | 8 | 78 | 0.45 | 0.94 | 0.06 | 0.70 |
| 164638 | 2.70 | 0.11 | 0.40 | 64 | 124 | 8 | 78 | 0.45 | 0.94 | 0.06 | 0.70 |
| 109113 | 3.11 | 0.11 | 0.44 | 64 | 124 | 8 | 78 | 0.45 | 0.94 | 0.06 | 0.70 |
| 155364 | 2.70 | 0.28 | 0.28 | 72 | 116 | 16 | 70 | 0.51 | 0.88 | 0.12 | 0.69 |
| 22040 | 3.10 | 0.37 | 0.37 | 75 | 113 | 19 | 67 | 0.53 | 0.86 | 0.14 | 0.69 |
| 140674 | 2.32 | 0.10 | 0.35 | 61 | 126 | 6 | 81 | 0.43 | 0.95 | 0.05 | 0.69 |
| 155365 | 2.49 | 0.11 | 0.64 | 64 | 123 | 9 | 78 | 0.45 | 0.93 | 0.07 | 0.69 |
| 140673 | 2.53 | 0.18 | 0.44 | 67 | 120 | 12 | 75 | 0.47 | 0.91 | 0.09 | 0.69 |
| 162111 | 1.63 | 0.14 | 0.47 | 68 | 119 | 13 | 74 | 0.48 | 0.90 | 0.10 | 0.69 |
| 14879 | 6.02 | 0.75 | 0.75 | 98 | 91 | 41 | 44 | 0.69 | 0.69 | 0.31 | 0.69 |
| 197621 | 2.47 | 0.02 | 0.29 | 56 | 130 | 2 | 86 | 0.39 | 0.98 | 0.02 | 0.69 |
| 53523 | 5.11 | 0.47 | 0.47 | 87 | 101 | 31 | 55 | 0.61 | 0.77 | 0.23 | 0.69 |
| 22039 | 4.62 | 0.60 | 0.60 | 88 | 100 | 32 | 54 | 0.62 | 0.76 | 0.24 | 0.69 |
| 19310 | 2.89 | 0.05 | 0.68 | 60 | 126 | 6 | 82 | 0.42 | 0.95 | 0.05 | 0.69 |
| 19309 | 3.20 | 0.09 | 0.44 | 63 | 123 | 9 | 79 | 0.44 | 0.93 | 0.07 | 0.69 |
| 109111 | 3.43 | 0.18 | 0.54 | 65 | 121 | 11 | 77 | 0.46 | 0.92 | 0.08 | 0.69 |
| 53525 | 3.22 | 0.17 | 0.17 | 67 | 119 | 13 | 75 | 0.47 | 0.90 | 0.10 | 0.69 |
| 32305 | 6.06 | 0.54 | 0.54 | 84 | 103 | 29 | 58 | 0.59 | 0.78 | 0.22 | 0.69 |
| 140675 | 2.21 | 0.09 | 0.33 | 59 | 126 | 6 | 83 | 0.42 | 0.95 | 0.05 | 0.69 |
| 32433 | 2.25 | 0.06 | 0.55 | 59 | 126 | 6 | 83 | 0.42 | 0.95 | 0.05 | 0.69 |
| 14880 | 5.18 | 0.61 | 0.61 | 92 | 95 | 37 | 50 | 0.65 | 0.72 | 0.28 | 0.68 |
| 142220 | 2.23 | 0.15 | 0.38 | 64 | 121 | 11 | 78 | 0.45 | 0.92 | 0.08 | 0.68 |
| 53524 | 4.61 | 0.39 | 0.39 | 79 | 107 | 25 | 63 | 0.56 | 0.81 | 0.19 | 0.68 |
| 33056 | 6.39 | 1.25 | 1.25 | 95 | 92 | 40 | 47 | 0.67 | 0.70 | 0.30 | 0.68 |
| 32303 | 7.15 | 0.90 | 0.90 | 95 | 92 | 40 | 47 | 0.67 | 0.70 | 0.30 | 0.68 |
| 190943 | 3.78 | 0.03 | 0.45 | 55 | 129 | 3 | 87 | 0.39 | 0.98 | 0.02 | 0.68 |
| 162112 | 1.42 | 0.09 | 0.39 | 60 | 124 | 8 | 82 | 0.42 | 0.94 | 0.06 | 0.68 |
| 40169 | 2.50 | 0.37 | 0.61 | 76 | 109 | 23 | 66 | 0.54 | 0.83 | 0.17 | 0.68 |
| 190944 | 3.60 | 0.03 | 0.82 | 53 | 130 | 2 | 89 | 0.37 | 0.98 | 0.02 | 0.68 |
| 190942 | 3.91 | 0.07 | 0.50 | 56 | 127 | 5 | 86 | 0.39 | 0.96 | 0.04 | 0.68 |
| 197570 | 2.23 | 0.09 | 0.57 | 61 | 122 | 10 | 81 | 0.43 | 0.92 | 0.08 | 0.68 |
| 22768 | 2.94 | 0.34 | 0.34 | 77 | 107 | 25 | 65 | 0.54 | 0.81 | 0.19 | 0.68 |
| 140676 | 2.05 | 0.06 | 0.50 | 55 | 127 | 5 | 87 | 0.39 | 0.96 | 0.04 | 0.67 |
| 155367 | 1.63 | 0.05 | 0.40 | 55 | 127 | 5 | 87 | 0.39 | 0.96 | 0.04 | 0.67 |
| 162113 | 1.35 | 0.09 | 0.37 | 58 | 124 | 8 | 84 | 0.41 | 0.94 | 0.06 | 0.67 |
| 14654 | 4.84 | 1.75 | 3.47 | 74 | 109 | 23 | 68 | 0.52 | 0.83 | 0.17 | 0.67 |
| 30599 | 2.93 | 0.38 | 0.38 | 77 | 106 | 26 | 65 | 0.54 | 0.80 | 0.20 | 0.67 |
| 32304 | 6.65 | 0.74 | 0.74 | 93 | 91 | 41 | 49 | 0.65 | 0.69 | 0.31 | 0.67 |
| 140677 | 1.96 | 0.06 | 0.48 | 54 | 127 | 5 | 88 | 0.38 | 0.96 | 0.04 | 0.67 |
| 150767 | 4.63 | 0.87 | 0.87 | 98 | 86 | 46 | 44 | 0.69 | 0.65 | 0.35 | 0.67 |
| 14878 | 9.01 | 1.59 | 3.24 | 70 | 112 | 20 | 72 | 0.49 | 0.85 | 0.15 | 0.67 |
| 109107 | 3.67 | 0.35 | 0.72 | 73 | 109 | 23 | 69 | 0.51 | 0.83 | 0.17 | 0.67 |
| 163975 | 2.50 | 0.17 | 0.43 | 62 | 118 | 14 | 80 | 0.44 | 0.89 | 0.11 | 0.67 |
| 140678 | 1.88 | 0.06 | 0.46 | 52 | 127 | 5 | 90 | 0.37 | 0.96 | 0.04 | 0.66 |
| 30600 | 2.74 | 0.33 | 0.33 | 69 | 111 | 21 | 73 | 0.49 | 0.84 | 0.16 | 0.66 |
| 15646 | 2.38 | 0.10 | 0.35 | 56 | 123 | 9 | 86 | 0.39 | 0.93 | 0.07 | 0.66 |
| 150771 | 2.06 | 0.11 | 0.33 | 60 | 119 | 13 | 82 | 0.42 | 0.90 | 0.10 | 0.66 |
| 150770 | 2.13 | 0.13 | 0.35 | 61 | 118 | 14 | 81 | 0.43 | 0.89 | 0.11 | 0.66 |
| 62691 | 0.85 | 0.02 | 0.48 | 49 | 129 | 3 | 93 | 0.35 | 0.98 | 0.02 | 0.66 |
| 15643 | 2.84 | 0.24 | 0.53 | 65 | 114 | 18 | 77 | 0.46 | 0.86 | 0.14 | 0.66 |
| 15645 | 2.44 | 0.11 | 0.37 | 56 | 122 | 10 | 86 | 0.39 | 0.92 | 0.08 | 0.66 |
| 35227 | 1.17 | 0.12 | 0.47 | 56 | 122 | 10 | 86 | 0.39 | 0.92 | 0.08 | 0.66 |
| 155368 | 1.49 | 0.03 | 0.35 | 48 | 129 | 3 | 94 | 0.34 | 0.98 | 0.02 | 0.66 |
| 206115 | 1.86 | 0.04 | 0.44 | 48 | 129 | 3 | 94 | 0.34 | 0.98 | 0.02 | 0.66 |
| 162114 | 1.17 | 0.07 | 0.44 | 52 | 125 | 7 | 90 | 0.37 | 0.95 | 0.05 | 0.66 |
| 30601 | 2.45 | 0.24 | 0.24 | 61 | 116 | 16 | 81 | 0.43 | 0.88 | 0.12 | 0.65 |
| 145032 | 1.32 | 0.06 | 0.62 | 50 | 126 | 6 | 92 | 0.35 | 0.95 | 0.05 | 0.65 |
| 145033 | 1.32 | 0.06 | 0.62 | 50 | 126 | 6 | 92 | 0.35 | 0.95 | 0.05 | 0.65 |
| 197622 | 1.30 | 0.00 | 0.43 | 43 | 132 | 0 | 99 | 0.30 | 1.00 | 0.00 | 0.65 |
| 155369 | 1.20 | 0.01 | 0.41 | 44 | 131 | 1 | 98 | 0.31 | 0.99 | 0.01 | 0.65 |
| 15644 | 2.66 | 0.20 | 0.47 | 61 | 115 | 17 | 81 | 0.43 | 0.87 | 0.13 | 0.65 |
| 140679 | 1.59 | 0.03 | 0.55 | 47 | 128 | 4 | 95 | 0.33 | 0.97 | 0.03 | 0.65 |
| 150768 | 3.67 | 0.59 | 0.93 | 80 | 97 | 35 | 62 | 0.56 | 0.73 | 0.27 | 0.65 |
| 19311 | 1.21 | 0.01 | 0.54 | 43 | 131 | 1 | 99 | 0.30 | 0.99 | 0.01 | 0.65 |
| 137612 | 1.00 | 0.03 | 0.57 | 45 | 129 | 3 | 97 | 0.32 | 0.98 | 0.02 | 0.65 |
| 162115 | 0.98 | 0.06 | 0.47 | 46 | 127 | 5 | 96 | 0.32 | 0.96 | 0.04 | 0.64 |
| 14336 | 1.00 | 0.00 | 0.44 | 41 | 131 | 1 | 101 | 0.29 | 0.99 | 0.01 | 0.64 |
| 40170 | 1.32 | 0.04 | 0.47 | 44 | 128 | 4 | 98 | 0.31 | 0.97 | 0.03 | 0.64 |
| 155370 | 0.92 | 0.00 | 0.41 | 39 | 132 | 0 | 103 | 0.27 | 1.00 | 0.00 | 0.64 |

| Site ID | | | | | | | | | | |
|---|---|---|---|---|---|---|---|---|---|---|
| 184473 | 1.09 | 0.01 | 0.37 | 40 | 131 | 1 | 102 | 0.28 | 0.99 | 0.01 | 0.64 |
| 155374 | 1.00 | 0.03 | 0.78 | 41 | 130 | 2 | 101 | 0.29 | 0.98 | 0.02 | 0.64 |
| 150773 | 1.01 | 0.03 | 0.47 | 43 | 127 | 5 | 99 | 0.30 | 0.96 | 0.04 | 0.63 |
| 1154 | 0.51 | 0.01 | 0.45 | 37 | 131 | 1 | 105 | 0.26 | 0.99 | 0.01 | 0.63 |
| 137625 | 1.12 | 0.03 | 0.51 | 39 | 129 | 3 | 103 | 0.27 | 0.98 | 0.02 | 0.63 |
| 162118 | 0.65 | 0.02 | 0.58 | 37 | 130 | 2 | 105 | 0.26 | 0.98 | 0.02 | 0.62 |
| 184474 | 1.04 | 0.01 | 0.47 | 35 | 131 | 1 | 107 | 0.25 | 0.99 | 0.01 | 0.62 |
| 198323 | 0.61 | 0.01 | 0.61 | 36 | 130 | 2 | 106 | 0.25 | 0.98 | 0.02 | 0.62 |
| 198324 | 0.59 | 0.01 | 0.59 | 36 | 130 | 2 | 106 | 0.25 | 0.98 | 0.02 | 0.62 |
| 184476 | 0.93 | 0.00 | 0.52 | 33 | 132 | 0 | 109 | 0.23 | 1.00 | 0.00 | 0.62 |
| 184475 | 0.93 | 0.00 | 0.52 | 33 | 132 | 0 | 109 | 0.23 | 1.00 | 0.00 | 0.62 |
| 40171 | 0.88 | 0.00 | 0.49 | 33 | 132 | 0 | 109 | 0.23 | 1.00 | 0.00 | 0.62 |
| 1155 | 0.46 | 0.01 | 0.46 | 34 | 131 | 1 | 108 | 0.24 | 0.99 | 0.01 | 0.62 |
| 1156 | 0.43 | 0.00 | 0.43 | 31 | 132 | 0 | 111 | 0.22 | 1.00 | 0.00 | 0.61 |
| 5779 | 0.51 | 0.00 | 0.45 | 31 | 132 | 0 | 111 | 0.22 | 1.00 | 0.00 | 0.61 |

| Top 100 Negatively correlated methylation sites | | | | | | | | | | | |
|---|---|---|---|---|---|---|---|---|---|---|---|
| Site ID | Mean_CRC | Mean_NOR | Threshold | True Positive | True Negative | False Positive | False Negative | Sensitivity | Specificity | FPR | AUC-ROC |
| 3658 | 53.29 | 69.25 | 61.27 | 106 | 79 | 53 | 36 | 0.75 | 0.60 | 0.40 | 0.67 |
| 100955 | 56.67 | 71.94 | 64.31 | 105 | 79 | 53 | 37 | 0.74 | 0.60 | 0.40 | 0.67 |
| 24643 | 44.55 | 58.01 | 51.28 | 102 | 81 | 51 | 40 | 0.72 | 0.61 | 0.39 | 0.67 |
| 24644 | 38.41 | 49.85 | 44.13 | 99 | 83 | 49 | 43 | 0.70 | 0.63 | 0.37 | 0.66 |
| 75322 | 26.92 | 35.68 | 31.30 | 103 | 79 | 53 | 39 | 0.73 | 0.60 | 0.40 | 0.66 |
| 100957 | 50.67 | 64.09 | 57.38 | 103 | 79 | 53 | 39 | 0.73 | 0.60 | 0.40 | 0.66 |

| ID | | | | | | | | | | |
|---|---|---|---|---|---|---|---|---|---|---|
| 75321 | 31.06 | 41.76 | 36.41 | 106 | 76 | 56 | 36 | 0.75 | 0.58 | 0.42 | 0.66 |
| 189313 | 46.04 | 58.67 | 52.36 | 102 | 79 | 53 | 40 | 0.72 | 0.60 | 0.40 | 0.66 |
| 100956 | 53.52 | 67.64 | 60.58 | 103 | 78 | 54 | 39 | 0.73 | 0.59 | 0.41 | 0.66 |
| 174999 | 44.89 | 59.67 | 52.28 | 104 | 77 | 55 | 38 | 0.73 | 0.58 | 0.42 | 0.66 |
| 9374 | 46.19 | 63.85 | 55.02 | 110 | 71 | 61 | 32 | 0.77 | 0.54 | 0.46 | 0.66 |
| 24639 | 90.58 | 114.08 | 102.33 | 100 | 80 | 52 | 42 | 0.70 | 0.61 | 0.39 | 0.66 |
| 180703 | 113.51 | 136.50 | 125.00 | 100 | 80 | 52 | 42 | 0.70 | 0.61 | 0.39 | 0.66 |
| 100954 | 63.74 | 80.53 | 72.14 | 102 | 78 | 54 | 40 | 0.72 | 0.59 | 0.41 | 0.65 |
| 125190 | 63.03 | 77.88 | 70.46 | 103 | 77 | 55 | 39 | 0.73 | 0.58 | 0.42 | 0.65 |
| 54573 | 27.32 | 37.02 | 32.17 | 103 | 77 | 55 | 39 | 0.73 | 0.58 | 0.42 | 0.65 |
| 9375 | 40.96 | 56.32 | 48.64 | 110 | 70 | 62 | 32 | 0.77 | 0.53 | 0.47 | 0.65 |
| 114431 | 79.97 | 104.13 | 92.05 | 98 | 81 | 51 | 44 | 0.69 | 0.61 | 0.39 | 0.65 |
| 180704 | 109.89 | 131.53 | 120.71 | 100 | 79 | 53 | 42 | 0.70 | 0.60 | 0.40 | 0.65 |
| 180705 | 100.65 | 121.40 | 111.02 | 101 | 78 | 54 | 41 | 0.71 | 0.59 | 0.41 | 0.65 |
| 75318 | 40.28 | 53.41 | 46.84 | 104 | 75 | 57 | 38 | 0.73 | 0.57 | 0.43 | 0.65 |
| 7577 | 35.25 | 49.22 | 42.24 | 105 | 74 | 58 | 37 | 0.74 | 0.56 | 0.44 | 0.65 |
| 116026 | 18.14 | 25.61 | 21.88 | 103 | 75 | 57 | 39 | 0.73 | 0.57 | 0.43 | 0.65 |
| 125191 | 54.37 | 65.92 | 60.14 | 103 | 75 | 57 | 39 | 0.73 | 0.57 | 0.43 | 0.65 |
| 75320 | 33.38 | 45.00 | 39.19 | 104 | 74 | 58 | 38 | 0.73 | 0.56 | 0.44 | 0.65 |
| 17992 | 35.92 | 45.56 | 40.74 | 94 | 83 | 49 | 48 | 0.66 | 0.63 | 0.37 | 0.65 |
| 7575 | 74.01 | 97.82 | 85.92 | 97 | 80 | 52 | 45 | 0.68 | 0.61 | 0.39 | 0.64 |
| 24642 | 56.23 | 72.91 | 64.57 | 98 | 79 | 53 | 44 | 0.69 | 0.60 | 0.40 | 0.64 |
| 100953 | 67.69 | 84.82 | 76.26 | 101 | 76 | 56 | 41 | 0.71 | 0.58 | 0.42 | 0.64 |
| 75319 | 36.83 | 49.73 | 43.28 | 102 | 75 | 57 | 40 | 0.72 | 0.57 | 0.43 | 0.64 |
| 126874 | 20.55 | 28.76 | 24.66 | 104 | 73 | 59 | 38 | 0.73 | 0.55 | 0.45 | 0.64 |
| 203525 | 98.26 | 120.27 | 109.26 | 95 | 81 | 51 | 47 | 0.67 | 0.61 | 0.39 | 0.64 |

| | | | | | | | | | | |
|---|---|---|---|---|---|---|---|---|---|---|
| 24640 | 86.67 | 109.02 | 97.84 | 98 | 78 | 54 | 44 | 0.69 | 0.59 | 0.41 | 0.64 |
| 78188 | 120.67 | 150.03 | 135.35 | 99 | 77 | 55 | 43 | 0.70 | 0.58 | 0.42 | 0.64 |
| 24641 | 61.55 | 78.99 | 70.27 | 100 | 76 | 56 | 42 | 0.70 | 0.58 | 0.42 | 0.64 |
| 27300 | 25.23 | 35.05 | 30.14 | 105 | 71 | 61 | 37 | 0.74 | 0.54 | 0.46 | 0.64 |
| 36304 | 204.85 | 252.14 | 228.50 | 96 | 79 | 53 | 46 | 0.68 | 0.60 | 0.40 | 0.64 |
| 75616 | 57.27 | 71.12 | 64.20 | 96 | 79 | 53 | 46 | 0.68 | 0.60 | 0.40 | 0.64 |
| 171893 | 36.57 | 53.67 | 45.12 | 100 | 75 | 57 | 42 | 0.70 | 0.57 | 0.43 | 0.64 |
| 126871 | 126.99 | 161.17 | 144.08 | 100 | 75 | 57 | 42 | 0.70 | 0.57 | 0.43 | 0.64 |
| 97256 | 29.19 | 36.85 | 33.02 | 94 | 80 | 52 | 48 | 0.66 | 0.61 | 0.39 | 0.63 |
| 54574 | 16.83 | 23.53 | 20.18 | 103 | 71 | 61 | 39 | 0.73 | 0.54 | 0.46 | 0.63 |
| 36303 | 215.85 | 265.03 | 240.44 | 96 | 77 | 55 | 46 | 0.68 | 0.58 | 0.42 | 0.63 |
| 202311 | 137.84 | 176.45 | 157.14 | 99 | 74 | 58 | 43 | 0.70 | 0.56 | 0.44 | 0.63 |
| 99171 | 115.43 | 142.46 | 128.94 | 100 | 73 | 59 | 42 | 0.70 | 0.55 | 0.45 | 0.63 |
| 126873 | 102.93 | 132.18 | 117.56 | 100 | 73 | 59 | 42 | 0.70 | 0.55 | 0.45 | 0.63 |
| 7576 | 68.95 | 90.84 | 79.90 | 91 | 81 | 51 | 51 | 0.64 | 0.61 | 0.39 | 0.63 |
| 7701 | 207.57 | 261.10 | 234.34 | 92 | 80 | 52 | 50 | 0.65 | 0.61 | 0.39 | 0.63 |
| 7574 | 84.47 | 110.74 | 97.60 | 93 | 79 | 53 | 49 | 0.65 | 0.60 | 0.40 | 0.63 |
| 149453 | 88.94 | 120.11 | 104.52 | 93 | 79 | 53 | 49 | 0.65 | 0.60 | 0.40 | 0.63 |
| 7573 | 124.14 | 155.61 | 139.88 | 96 | 76 | 56 | 46 | 0.68 | 0.58 | 0.42 | 0.63 |
| 78187 | 136.58 | 168.10 | 152.34 | 97 | 75 | 57 | 45 | 0.68 | 0.57 | 0.43 | 0.63 |
| 99172 | 112.99 | 139.27 | 126.13 | 99 | 73 | 59 | 43 | 0.70 | 0.55 | 0.45 | 0.63 |
| 75129 | 66.63 | 83.94 | 75.28 | 99 | 73 | 59 | 43 | 0.70 | 0.55 | 0.45 | 0.63 |
| 126872 | 118.73 | 151.79 | 135.26 | 99 | 73 | 59 | 43 | 0.70 | 0.55 | 0.45 | 0.63 |
| 99173 | 110.07 | 135.45 | 122.76 | 100 | 72 | 60 | 42 | 0.70 | 0.55 | 0.45 | 0.62 |
| 90819 | 22.99 | 32.32 | 27.66 | 100 | 72 | 60 | 42 | 0.70 | 0.55 | 0.45 | 0.62 |
| 74073 | 56.83 | 74.39 | 65.61 | 103 | 69 | 63 | 39 | 0.73 | 0.52 | 0.48 | 0.62 |
| 62822 | 130.58 | 162.93 | 146.76 | 93 | 78 | 54 | 49 | 0.65 | 0.59 | 0.41 | 0.62 |
| 7700 | 221.23 | 277.17 | 249.20 | 93 | 78 | 54 | 49 | 0.65 | 0.59 | 0.41 | 0.62 |
| 89080 | 97.17 | 123.58 | 110.38 | 93 | 78 | 54 | 49 | 0.65 | 0.59 | 0.41 | 0.62 |
| 36305 | 196.88 | 241.40 | 219.14 | 97 | 74 | 58 | 45 | 0.68 | 0.56 | 0.44 | 0.62 |
| 160627 | 32.38 | 42.48 | 37.43 | 99 | 72 | 60 | 43 | 0.70 | 0.55 | 0.45 | 0.62 |
| 9376 | 12.32 | 19.37 | 15.84 | 103 | 68 | 64 | 39 | 0.73 | 0.52 | 0.48 | 0.62 |
| 7699 | 239.20 | 298.27 | 268.74 | 94 | 76 | 56 | 48 | 0.66 | 0.58 | 0.42 | 0.62 |
| 123337 | 67.43 | 81.75 | 74.59 | 94 | 76 | 56 | 48 | 0.66 | 0.58 | 0.42 | 0.62 |
| 7703 | 58.65 | 76.61 | 67.63 | 95 | 75 | 57 | 47 | 0.67 | 0.57 | 0.43 | 0.62 |
| 57404 | 15.38 | 21.99 | 18.68 | 96 | 74 | 58 | 46 | 0.68 | 0.56 | 0.44 | 0.62 |
| 35413 | 99.38 | 128.07 | 113.72 | 96 | 74 | 58 | 46 | 0.68 | 0.56 | 0.44 | 0.62 |
| 90817 | 126.62 | 153.63 | 140.12 | 96 | 74 | 58 | 46 | 0.68 | 0.56 | 0.44 | 0.62 |
| 133220 | 168.67 | 205.88 | 187.27 | 97 | 73 | 59 | 45 | 0.68 | 0.55 | 0.45 | 0.62 |
| 127235 | 71.38 | 87.54 | 79.46 | 97 | 73 | 59 | 45 | 0.68 | 0.55 | 0.45 | 0.62 |
| 18002 | 87.35 | 110.10 | 98.72 | 98 | 72 | 60 | 44 | 0.69 | 0.55 | 0.45 | 0.62 |
| 17680 | 34.38 | 43.56 | 38.97 | 94 | 75 | 57 | 48 | 0.66 | 0.57 | 0.43 | 0.62 |
| 202310 | 146.77 | 187.92 | 167.34 | 94 | 75 | 57 | 48 | 0.66 | 0.57 | 0.43 | 0.62 |
| 113054 | 50.69 | 65.46 | 58.08 | 95 | 74 | 58 | 47 | 0.67 | 0.56 | 0.44 | 0.61 |
| 58323 | 29.22 | 38.91 | 34.06 | 95 | 74 | 58 | 47 | 0.67 | 0.56 | 0.44 | 0.61 |
| 75317 | 46.30 | 59.71 | 53.00 | 97 | 72 | 60 | 45 | 0.68 | 0.55 | 0.45 | 0.61 |
| 90224 | 48.01 | 60.08 | 54.04 | 97 | 72 | 60 | 45 | 0.68 | 0.55 | 0.45 | 0.61 |
| 78181 | 92.58 | 111.35 | 101.96 | 98 | 71 | 61 | 44 | 0.69 | 0.54 | 0.46 | 0.61 |
| 30580 | 58.66 | 77.68 | 68.17 | 100 | 69 | 63 | 42 | 0.70 | 0.52 | 0.48 | 0.61 |
| 116025 | 46.43 | 59.85 | 53.14 | 93 | 75 | 57 | 49 | 0.65 | 0.57 | 0.43 | 0.61 |
| 35412 | 200.32 | 259.22 | 229.77 | 95 | 73 | 59 | 47 | 0.67 | 0.55 | 0.45 | 0.61 |
| 35411 | 232.14 | 298.92 | 265.53 | 95 | 73 | 59 | 47 | 0.67 | 0.55 | 0.45 | 0.61 |

| | | | | | | | | | | |
|---|---|---|---|---|---|---|---|---|---|---|
| 35414 | 97.62 | 125.24 | 111.43 | 95 | 73 | 59 | 47 | 0.67 | 0.55 | 0.45 | 0.61 |
| 202309 | 172.12 | 218.42 | 195.27 | 95 | 73 | 59 | 47 | 0.67 | 0.55 | 0.45 | 0.61 |
| 123336 | 89.67 | 107.95 | 98.81 | 95 | 72 | 60 | 47 | 0.67 | 0.55 | 0.45 | 0.61 |
| 24632 | 70.50 | 87.30 | 78.90 | 88 | 78 | 54 | 54 | 0.62 | 0.59 | 0.41 | 0.61 |
| 189312 | 101.85 | 123.62 | 112.74 | 89 | 77 | 55 | 53 | 0.63 | 0.58 | 0.42 | 0.61 |
| 24631 | 96.48 | 120.51 | 108.50 | 91 | 75 | 57 | 51 | 0.64 | 0.57 | 0.43 | 0.60 |
| 41489 | 146.28 | 176.32 | 161.30 | 94 | 71 | 61 | 48 | 0.66 | 0.54 | 0.46 | 0.60 |
| 54571 | 40.44 | 52.38 | 46.41 | 95 | 70 | 62 | 47 | 0.67 | 0.53 | 0.47 | 0.60 |
| 24476 | 92.17 | 116.80 | 104.48 | 92 | 72 | 60 | 50 | 0.65 | 0.55 | 0.45 | 0.60 |
| 54570 | 42.23 | 54.50 | 48.36 | 92 | 72 | 60 | 50 | 0.65 | 0.55 | 0.45 | 0.60 |
| 84787 | 59.56 | 73.64 | 66.60 | 93 | 71 | 61 | 49 | 0.65 | 0.54 | 0.46 | 0.60 |
| 41491 | 139.96 | 167.72 | 153.84 | 94 | 70 | 62 | 48 | 0.66 | 0.53 | 0.47 | 0.60 |
| 204581 | 26.89 | 36.01 | 31.45 | 97 | 67 | 65 | 45 | 0.68 | 0.51 | 0.49 | 0.60 |
| 41490 | 142.69 | 171.00 | 156.84 | 92 | 71 | 61 | 50 | 0.65 | 0.54 | 0.46 | 0.59 |
| 54572 | 33.59 | 44.31 | 38.95 | 95 | 68 | 64 | 47 | 0.67 | 0.52 | 0.48 | 0.59 |
| 30581 | 51.49 | 68.48 | 59.98 | 101 | 62 | 70 | 41 | 0.71 | 0.47 | 0.53 | 0.59 |

**Supplementary Table 4: Performance of the Machine learning model built on 30791 methylated sites.**

| | | | | | Training | | | | | | | Validation Set | | | | | | |
|---|---|---|---|---|---|---|---|---|---|---|---|---|---|---|---|---|---|---|
| Model Name | Thr | TPR | FPR | YC | Sp | Sn | AUC | Acc | Pre | MC | K | Sp | Sn | AUC | Acc | Pre | MC | K |
| SVC | 0.66 | 0.48 | 0.07 | 0.41 | 71.43 | 69.57 | 0.71 | 70.45 | 72.73 | 0.41 | 0.41 | 92.86 | 48.15 | 0.71 | 70.91 | 86.67 | 0.46 | 0.41 |
| Logistic Regression | 0.61 | 0.85 | 0.18 | 0.67 | 80.95 | 65.22 | 0.73 | 72.73 | 78.95 | 0.47 | 0.46 | 82.14 | 85.19 | 0.84 | 83.64 | 82.14 | 0.67 | 0.67 |
| Naive Bayes | 1.00 | 1.00 | 0.86 | 0.14 | 30.00 | 95.65 | 0.63 | 65.12 | 61.11 | 0.35 | 0.27 | 14.29 | 100.00 | 0.57 | 56.36 | 52.94 | 0.28 | 0.14 |
| Decision Tree | 1.00 | 0.78 | 0.39 | 0.38 | 71.43 | 65.22 | 0.68 | 68.18 | 71.43 | 0.37 | 0.36 | 60.71 | 77.78 | 0.69 | 69.09 | 65.63 | 0.39 | 0.38 |

| | #Thr | YC | | | Sp | Sn | | Acc | Pre | | MC | | Sp | Sn | | Acc | Pre | MC | K |
|---|---|---|---|---|---|---|---|---|---|---|---|---|---|---|---|---|---|---|---|
| Random Forest | 0.54 | 0.89 | 0.18 | 0.71 | 76.19 | 82.61 | 0.79 | 79.55 | 79.17 | 0.59 | 0.59 | 82.14 | 88.89 | 0.86 | 85.45 | 82.76 | 0.71 | 0.71 |
| Extra Trees | 0.51 | 0.93 | 0.18 | 0.75 | 70.00 | 69.57 | 0.70 | 69.77 | 72.73 | 0.39 | 0.39 | 82.14 | 92.59 | 0.87 | 87.27 | 83.33 | 0.75 | 0.75 |
| AdaBoost | 0.50 | 0.93 | 0.18 | 0.75 | 85.71 | 86.96 | 0.86 | 86.36 | 86.96 | 0.73 | 0.73 | 82.14 | 92.59 | 0.87 | 87.27 | 83.33 | 0.75 | 0.75 |
| Gradient Boosting | 0.57 | 0.85 | 0.11 | 0.74 | 80.95 | 82.61 | 0.82 | 81.82 | 82.61 | 0.64 | 0.64 | 89.29 | 85.19 | 0.87 | 87.27 | 88.46 | 0.75 | 0.75 |
| Multiple Layer Perceptron | 1.00 | 0.70 | 0.04 | 0.67 | 70.00 | 69.57 | 0.70 | 69.77 | 72.73 | 0.39 | 0.39 | 96.43 | 70.37 | 0.83 | 83.64 | 95.00 | 0.69 | 0.67 |
| XGBoost | 0.17 | 0.96 | 0.36 | 0.61 | 75.00 | 69.57 | 0.72 | 72.09 | 76.19 | 0.44 | 0.44 | 64.29 | 96.30 | 0.80 | 80.00 | 72.22 | 0.64 | 0.60 |

#Thr: Threshold; YC: Youden Coeff; Sp: Specificity; Sn: Sensitivity; Acc: Accuracy; Pre: Precision; MC: Mathews Coeff; K :Kappa

**Supplementary Table 5.1: List of selected features based on Univariate analysis for positievly and negatively correlated Methylated sites**

| Positive_Top_25_MS | Positive_Top_15_MS | Positive_Top_5_MS | Negative_Top_5_MS | Negative_Top_15_MS | Negative_Top_25_MS |
|---|---|---|---|---|---|
| 19307 | 19308 | 181489 | 24642 | 3658 | 3658 |
| 19308 | 33055 | 181490 | 24643 | 9374 | 7576 |
| 22040 | 38139 | 181491 | 100955 | 24641 | 9374 |
| 33055 | 86829 | 196741 | 100956 | 24642 | 9375 |
| 38139 | 93487 | 196742 | 100957 | 24643 | 17992 |
| 86829 | 93488 | | | 36303 | 24639 |
| 93487 | 155363 | | | 36304 | 24641 |
| 93488 | 164637 | | | 75319 | 24642 |
| 109113 | 181489 | | | 100953 | 24643 |
| 140673 | 181490 | | | 100954 | 24644 |
| 140674 | 181491 | | | 100955 | 36303 |
| 155363 | 189090 | | | 100956 | 36304 |

| | | | | | | |
|---|---|---|---|---|---|---|
| 155364 | 196741 | | | 100957 | 36305 | |
| 155365 | 196742 | | | 180703 | 54573 | |
| 155366 | 197569 | | | 189313 | 75318 | |
| 162111 | | | | | 75319 | |
| 164637 | | | | | 100953 | |
| 164638 | | | | | 100954 | |
| 181489 | | | | | 100955 | |
| 181490 | | | | | 100956 | |
| 181491 | | | | | 100957 | |
| 189090 | | | | | 180703 | |
| 196741 | | | | | 180704 | |
| 196742 | | | | | 180705 | |
| 197569 | | | | | 189313 | |

**Supplementary Table 5.2: ML Performance of Univariate selected feature sets on Top 100 positively and negatively Correlated Methylated sites**

| Positively_Corr_25_performance_metrics | | | | | | | | | | | |
|---|---|---|---|---|---|---|---|---|---|---|---|
| | | Training | | | | | Validation | | | | |
| Name | Thresh | Sens | Spec | Acc | AUC | MCC | Sens | Spec | Acc | AUC | MCC |
| RF | 0.51 | 75.68 | 75.93 | 75.80 | 0.86 | 0.52 | 80.65 | 70.83 | 76.36 | 0.87 | 0.52 |
| **LR** | **0.42** | **77.48** | **77.78** | **77.63** | **0.88** | **0.55** | **83.87** | **79.17** | **81.82** | **0.87** | **0.63** |
| MLP | 0.46 | 76.58 | 75.93 | 76.26 | 0.84 | 0.53 | 90.32 | 66.67 | 80.00 | 0.87 | 0.59 |
| XGB | 0.53 | 73.87 | 74.07 | 73.97 | 0.82 | 0.48 | 83.87 | 66.67 | 76.36 | 0.83 | 0.52 |
| KN | 0.60 | 71.17 | 84.26 | 77.63 | 0.84 | 0.56 | 64.52 | 70.83 | 67.27 | 0.75 | 0.35 |
| ET | 0.51 | 76.58 | 75.93 | 76.26 | 0.86 | 0.53 | 90.32 | 70.83 | 81.82 | 0.88 | 0.63 |

| Positively_Corr_15_performance_metrics | | | | | | | | | | | |
|---|---|---|---|---|---|---|---|---|---|---|---|
| | | Training | | | | | Validation | | | | |
| Name | Thresh | Sens | Spec | Acc | AUC | MCC | Sens | Spec | Acc | AUC | MCC |
| RF | 0.51 | 78.38 | 78.70 | 78.54 | 0.85 | 0.57 | 80.65 | 66.67 | 74.55 | 0.86 | 0.48 |
| **LR** | **0.42** | **79.28** | **79.63** | **79.45** | **0.88** | **0.59** | **83.87** | **79.17** | **81.82** | **0.88** | **0.63** |
| MLP | 0.46 | 76.58 | 75.93 | 76.26 | 0.83 | 0.53 | 87.10 | 70.83 | 80.00 | 0.86 | 0.59 |
| XGB | 0.35 | 74.78 | 75.00 | 74.89 | 0.84 | 0.50 | 87.10 | 62.50 | 76.36 | 0.84 | 0.52 |
| KN | 0.60 | 77.48 | 82.41 | 79.91 | 0.86 | 0.60 | 74.19 | 66.67 | 70.91 | 0.76 | 0.41 |
| ET | 0.52 | 77.48 | 75.93 | 76.71 | 0.85 | 0.53 | 87.10 | 58.33 | 74.55 | 0.88 | 0.48 |

| Positively_Corr_5_performance_metrics | | | | | | | | | | | |
|---|---|---|---|---|---|---|---|---|---|---|---|
| | | Training | | | | | Validation | | | | |
| Name | Thresh | Sens | Spec | Acc | AUC | MCC | Sens | Spec | Acc | AUC | MCC |
| RF | 0.35 | 80.18 | 80.56 | 80.37 | 0.83 | 0.61 | 80.65 | 79.17 | 80.00 | 0.76 | 0.60 |
| LR | 0.33 | 75.68 | 75.93 | 75.80 | 0.80 | 0.52 | 74.19 | 58.33 | 67.27 | 0.81 | 0.33 |
| MLP | 0.40 | 77.48 | 76.85 | 77.17 | 0.82 | 0.54 | 83.87 | 66.67 | 76.36 | 0.86 | 0.52 |
| XGB | 0.28 | 75.68 | 79.63 | 77.63 | 0.79 | 0.55 | 80.65 | 79.17 | 80.00 | 0.76 | 0.60 |
| **KN** | **0.60** | **73.87** | **80.56** | **77.17** | **0.83** | **0.55** | **67.74** | **79.17** | **72.73** | **0.80** | **0.47** |
| ET | 0.28 | 81.08 | 81.48 | 81.28 | 0.84 | 0.63 | 80.65 | 79.17 | 80.00 | 0.74 | 0.60 |

| Negatively_Corr_25_performance_metrics | | | | | | | | | | | |
|---|---|---|---|---|---|---|---|---|---|---|---|
| | | Training | | | | | Validation | | | | |
| Name | Thresh | Sens | Spec | Acc | AUC | MCC | Sens | Spec | Acc | AUC | MCC |
| RF | 0.49 | 71.17 | 69.44 | 70.32 | 0.78 | 0.41 | 80.65 | 50.00 | 67.27 | 0.69 | 0.32 |
| LR | 0.47 | 61.26 | 61.11 | 61.19 | 0.60 | 0.22 | 48.39 | 45.83 | 47.27 | 0.43 | -0.06 |

| Name | Thresh | Sens | Spec | Acc | AUC | MCC | Sens | Spec | Acc | AUC | MCC |
|---|---|---|---|---|---|---|---|---|---|---|---|
| MLP | 0.46 | 52.25 | 51.85 | 52.06 | 0.55 | 0.04 | 41.94 | 66.67 | 52.73 | 0.51 | 0.09 |
| XGB | 0.56 | 68.47 | 68.52 | 68.49 | 0.75 | 0.37 | 61.29 | 58.33 | 60.00 | 0.67 | 0.20 |
| KN | 0.60 | 76.58 | 72.22 | 74.43 | 0.76 | 0.49 | 80.65 | 58.33 | 70.91 | 0.70 | 0.40 |
| ET | **0.50** | **73.87** | **74.07** | **73.97** | **0.78** | **0.48** | **83.87** | **54.17** | **70.91** | **0.74** | **0.40** |

| Negatively_Corr_15_performance_metrics | | | | | | | | | | | |
|---|---|---|---|---|---|---|---|---|---|---|---|
| | | Training | | | | | Validation | | | | |
| Name | Thresh | Sens | Spec | Acc | AUC | MCC | Sens | Spec | Acc | AUC | MCC |
| **RF** | **0.48** | **66.67** | **66.67** | **66.67** | **0.75** | **0.33** | **80.65** | **58.33** | **70.91** | **0.69** | **0.40** |
| LR | 0.47 | 61.26 | 62.96 | 62.10 | 0.66 | 0.24 | 64.52 | 41.67 | 54.55 | 0.52 | 0.06 |
| MLP | 0.48 | 51.35 | 50.00 | 50.69 | 0.56 | 0.01 | 48.39 | 70.83 | 58.18 | 0.66 | 0.20 |
| XGB | 0.41 | 67.57 | 67.59 | 67.58 | 0.74 | 0.35 | 74.19 | 54.17 | 65.46 | 0.65 | 0.29 |
| KN | 0.60 | 71.17 | 67.59 | 69.41 | 0.74 | 0.39 | 70.97 | 54.17 | 63.64 | 0.64 | 0.25 |
| ET | 0.51 | 68.47 | 67.59 | 68.04 | 0.75 | 0.36 | 70.97 | 54.17 | 63.64 | 0.67 | 0.25 |

| Negatively_Corr_5_performance_metrics | | | | | | | | | | | |
|---|---|---|---|---|---|---|---|---|---|---|---|
| | | Training | | | | | Validation | | | | |
| Name | Thresh | Sens | Spec | Acc | AUC | MCC | Sens | Spec | Acc | AUC | MCC |
| RF | 0.52 | 63.96 | 62.96 | 63.47 | 0.66 | 0.27 | 58.07 | 54.17 | 56.36 | 0.59 | 0.12 |
| LR | 0.53 | 71.17 | 68.52 | 69.86 | 0.73 | 0.40 | 64.52 | 45.83 | 56.36 | 0.61 | 0.11 |
| MLP | 0.48 | 57.66 | 59.26 | 58.45 | 0.57 | 0.17 | 54.84 | 45.83 | 50.91 | 0.49 | 0.01 |
| XGB | 0.50 | 62.16 | 62.04 | 62.10 | 0.62 | 0.24 | 54.84 | 45.83 | 50.91 | 0.56 | 0.01 |
| KN | 0.60 | 63.06 | 68.52 | 65.75 | 0.65 | 0.32 | 70.97 | 62.50 | 67.27 | 0.62 | 0.34 |
| **ET** | **0.52** | **62.16** | **62.04** | **62.10** | **0.65** | **0.24** | **67.74** | **58.33** | **63.64** | **0.64** | **0.26** |

**Supplementary Table 6.1: List of selected features based on RFE for positively and negatively correlated Methylated sites**

| Positive Methylated Sites | | | | | | Negative Methylated Sites | | | | |
|---|---|---|---|---|---|---|---|---|---|---|
| RFE_25 | RFE_20 | RFE_15 | RFE_10 | RFE_5 | | RFE_25 | RFE_20 | RFE_15 | RFE_10 | RFE_5 |
| 1156 | 1156 | 1156 | 1156 | 1156 | | 7575 | 9374 | 9374 | 24639 | 24639 |
| 5779 | 5779 | 5779 | 5779 | 5779 | | 7577 | 24476 | 24639 | 24640 | 24640 |
| 15644 | 15644 | 15646 | 19310 | 19310 | | 9374 | 24639 | 24640 | 36304 | 54573 |
| 15646 | 15646 | 19310 | 62691 | 62691 | | 17680 | 24640 | 36304 | 36305 | 75317 |
| 19310 | 19310 | 62691 | 137612 | 197622 | | 24476 | 24642 | 36305 | 54573 | 75318 |
| 19311 | 62691 | 137612 | 155370 | | | 24632 | 35411 | 54573 | 75317 | |
| 40170 | 137612 | 150770 | 162118 | | | 24639 | 35412 | 180703 | 75318 | |
| 62691 | 150770 | 155370 | 184476 | | | 24640 | 36304 | 75317 | 75321 | |
| 137612 | 150771 | 155374 | 197622 | | | 24643 | 36305 | 75318 | 180703 | |
| 150770 | 155366 | 162118 | 206115 | | | 35411 | 54573 | 75321 | 180704 | |
| 150771 | 155370 | 184475 | | | | 35412 | 58323 | 75322 | | |
| 155366 | 155374 | 184476 | | | | 36304 | 75317 | 99171 | | |
| 155370 | 162118 | 190943 | | | | 36305 | 75318 | 99172 | | |
| 155374 | 184475 | 197622 | | | | 41490 | 75321 | 180703 | | |
| 162112 | 184476 | 206115 | | | | 41491 | 75322 | 180704 | | |
| 162115 | 190943 | | | | | 54571 | 99171 | | | |
| 162118 | 197569 | | | | | 54573 | 99172 | | | |
| 164637 | 197570 | | | | | 58323 | 180703 | | | |
| 184475 | 197622 | | | | | 75317 | 180704 | | | |
| 184476 | 206115 | | | | | 75318 | 203525 | | | |
| 190943 | | | | | | 126872 | | | | |
| 197569 | | | | | | 126873 | | | | |
| 197570 | | | | | | 180703 | | | | |
| 197622 | | | | | | 180704 | | | | |

| 206115 | | | | | | 203525 | | | | |
|---|---|---|---|---|---|---|---|---|---|---|

## Supplementary Table 6.2: ML Performance of RFE selected feature sets for positively and negatively correlated Methylated sites

| | RFE_positively_Corr_25_performance_metrics | | | | | | | | | | | | | | | | |
|---|---|---|---|---|---|---|---|---|---|---|---|---|---|---|---|---|---|
| | | | | | Training | | | | | | | Validation | | | | | |
| Model ID | Thr | TPR | FPR | YC | Sp | Sn | AUC | Acc | Pre | MC | K | Sp | Sn | AUC | Acc | Pre | MC | K |
| SVC | 0.64 | 0.88 | 0.17 | 0.71 | 100.00 | 65.22 | 0.83 | 81.82 | 100.00 | 0.69 | 0.64 | 82.76 | 88.46 | 0.86 | 85.45 | 82.14 | 0.71 | 0.71 |
| LR | 0.40 | 0.88 | 0.21 | 0.68 | 90.48 | 91.30 | 0.91 | 90.91 | 91.30 | 0.82 | 0.82 | 79.31 | 88.46 | 0.84 | 83.64 | 79.31 | 0.68 | 0.67 |
| Naive Bayes | 0.00 | 0.88 | 0.14 | 0.75 | 76.19 | 78.26 | 0.77 | 77.27 | 78.26 | 0.54 | 0.54 | 86.21 | 88.46 | 0.87 | 87.27 | 85.19 | 0.75 | 0.75 |
| DT | 1.00 | 0.73 | 0.28 | 0.45 | 66.67 | 69.57 | 0.68 | 68.18 | 69.57 | 0.36 | 0.36 | 72.41 | 73.08 | 0.73 | 72.73 | 70.37 | 0.45 | 0.45 |
| RF | 0.27 | 0.88 | 0.24 | 0.64 | 76.19 | 73.91 | 0.75 | 75.00 | 77.27 | 0.50 | 0.50 | 75.86 | 88.46 | 0.82 | 81.82 | 76.67 | 0.64 | 0.64 |
| ET | 0.61 | 0.85 | 0.17 | 0.67 | 80.95 | 86.96 | 0.84 | 84.09 | 83.33 | 0.68 | 0.68 | 82.76 | 84.62 | 0.84 | 83.64 | 81.48 | 0.67 | 0.67 |
| AdaBooSt | 0.50 | 0.85 | 0.17 | 0.67 | 80.00 | 79.17 | 0.80 | 79.55 | 82.61 | 0.59 | 0.59 | 82.76 | 84.62 | 0.84 | 83.64 | 81.48 | 0.67 | 0.67 |
| Gradient Boosting | 0.64 | 0.81 | 0.17 | 0.64 | 85.71 | 86.96 | 0.86 | 86.36 | 86.96 | 0.73 | 0.73 | 82.76 | 80.77 | 0.82 | 81.82 | 80.77 | 0.64 | 0.64 |
| MLP | 0.63 | 0.85 | 0.07 | 0.78 | 90.48 | 91.30 | 0.91 | 90.91 | 91.30 | 0.82 | 0.82 | 93.10 | 84.62 | 0.89 | 89.09 | 91.67 | 0.78 | 0.78 |
| XGBoost | 0.25 | 0.88 | 0.28 | 0.61 | 80.95 | 78.26 | 0.80 | 79.55 | 81.82 | 0.59 | 0.59 | 72.41 | 88.46 | 0.80 | 80.00 | 74.19 | 0.61 | 0.60 |

| | RFE_positively_Corr_20_performance_metrics | | | | | | | | | | | | | | | | |
|---|---|---|---|---|---|---|---|---|---|---|---|---|---|---|---|---|---|
| | | | | | Training | | | | | | | Validation | | | | | |
| Model ID | Thr | TPR | FPR | YC | Sp | Sn | AUC | Acc | Pre | MC | K | Sp | Sn | AUC | Acc | Pre | MC | K |
| SVC | 0.88 | 0.65 | 0.00 | 0.65 | 100.00 | 60.87 | 0.80 | 79.55 | 100.00 | 0.65 | 0.60 | 100.0 | 65.38 | 0.83 | 83.64 | 100.0 | 0.71 | 0.67 |
| LR | 0.34 | 0.88 | 0.21 | 0.68 | 85.71 | 86.96 | 0.86 | 86.36 | 86.96 | 0.73 | 0.73 | 79.31 | 88.46 | 0.84 | 83.64 | 79.31 | 0.68 | 0.67 |
| Naive | 0.00 | 0.88 | 0.07 | 0.82 | 85.71 | 78.26 | 0.82 | 81.82 | 85.71 | 0.64 | 0.64 | 93.10 | 88.46 | 0.91 | 90.91 | 92.00 | 0.82 | 0.82 |

| Model ID | Thr | TPR | FPR | YC | Sp | Sn | AUC | Acc | Pre | MC | K | Sp | Sn | AUC | Acc | Pre | MC | K |
|---|---|---|---|---|---|---|---|---|---|---|---|---|---|---|---|---|---|---|
| Bayes | | | | | | | | | | | | | | | | | | |
| DT | 1.00 | 0.77 | 0.17 | 0.60 | 80.00 | 73.91 | 0.77 | 76.74 | 80.95 | 0.54 | 0.54 | 82.76 | 76.92 | 0.80 | 80.00 | 80.00 | 0.60 | 0.60 |
| RF | 0.65 | 0.81 | 0.14 | 0.67 | 75.00 | 73.91 | 0.74 | 74.42 | 77.27 | 0.49 | 0.49 | 86.21 | 80.77 | 0.83 | 83.64 | 84.00 | 0.67 | 0.67 |
| ET | 0.31 | 0.92 | 0.24 | 0.68 | 80.95 | 82.61 | 0.82 | 81.82 | 82.61 | 0.64 | 0.64 | 75.86 | 92.31 | 0.84 | 83.64 | 77.42 | 0.69 | 0.68 |
| AdaBooSt | 0.54 | 0.85 | 0.21 | 0.64 | 85.71 | 91.30 | 0.89 | 88.64 | 87.50 | 0.77 | 0.77 | 79.31 | 84.62 | 0.82 | 81.82 | 78.57 | 0.64 | 0.64 |
| Gradient Boosting | 0.66 | 0.77 | 0.14 | 0.63 | 76.19 | 73.91 | 0.75 | 75.00 | 77.27 | 0.50 | 0.50 | 86.21 | 76.92 | 0.82 | 81.82 | 83.33 | 0.64 | 0.63 |
| MLP | 0.19 | 0.88 | 0.21 | 0.68 | 90.48 | 91.30 | 0.91 | 90.91 | 91.30 | 0.82 | 0.82 | 79.31 | 88.46 | 0.84 | 83.64 | 79.31 | 0.68 | 0.67 |
| XGBoost | 0.85 | 0.73 | 0.07 | 0.66 | 80.00 | 73.91 | 0.77 | 76.74 | 80.95 | 0.54 | 0.54 | 93.10 | 73.08 | 0.83 | 83.64 | 90.48 | 0.68 | 0.67 |

| | RFE_positively_Corr_15_performance_metrics | | | | | | | | | | | | | | | | | |
|---|---|---|---|---|---|---|---|---|---|---|---|---|---|---|---|---|---|---|
| | | | | | Training | | | | | | | Validation | | | | | | |
| Model ID | Thr | TPR | FPR | YC | Sp | Sn | AUC | Acc | Pre | MC | K | Sp | Sn | AUC | Acc | Pre | MC | K |
| SVC | 0.22 | 0.92 | 0.21 | 0.72 | 100.00 | 78.26 | 0.89 | 88.64 | 100.00 | 0.80 | 0.77 | 79.31 | 92.31 | 0.86 | 85.45 | 80.00 | 0.72 | 0.71 |
| LR | 0.29 | 0.92 | 0.21 | 0.72 | 90.48 | 91.30 | 0.91 | 90.91 | 91.30 | 0.82 | 0.82 | 79.31 | 92.31 | 0.86 | 85.45 | 80.00 | 0.72 | 0.71 |
| Naive Bayes | 0.00 | 0.92 | 0.21 | 0.72 | 80.95 | 91.30 | 0.86 | 86.36 | 84.00 | 0.73 | 0.73 | 79.31 | 92.31 | 0.86 | 85.45 | 80.00 | 0.72 | 0.71 |
| DT | 1.00 | 0.81 | 0.14 | 0.67 | 85.71 | 86.96 | 0.86 | 86.36 | 86.96 | 0.73 | 0.73 | 86.21 | 80.77 | 0.83 | 83.64 | 84.00 | 0.67 | 0.67 |
| RF | 0.41 | 0.88 | 0.14 | 0.75 | 80.95 | 82.61 | 0.82 | 81.82 | 82.61 | 0.64 | 0.64 | 86.21 | 88.46 | 0.87 | 87.27 | 85.19 | 0.75 | 0.75 |
| ET | 0.28 | 0.88 | 0.14 | 0.75 | 90.48 | 86.96 | 0.89 | 88.64 | 90.91 | 0.77 | 0.77 | 86.21 | 88.46 | 0.87 | 87.27 | 85.19 | 0.75 | 0.75 |
| AdaBooSt | 0.66 | 0.69 | 0.03 | 0.66 | 76.19 | 78.26 | 0.77 | 77.27 | 78.26 | 0.54 | 0.54 | 96.55 | 69.23 | 0.83 | 83.64 | 94.74 | 0.69 | 0.67 |
| Gradient Boosting | 0.15 | 0.85 | 0.21 | 0.64 | 85.71 | 82.61 | 0.84 | 84.09 | 86.36 | 0.68 | 0.68 | 79.31 | 84.62 | 0.82 | 81.82 | 78.57 | 0.64 | 0.64 |
| MLP | 0.25 | 0.92 | 0.21 | 0.72 | 90.48 | 86.96 | 0.89 | 88.64 | 90.91 | 0.77 | 0.77 | 79.31 | 92.31 | 0.86 | 85.45 | 80.00 | 0.72 | 0.71 |
| XGBoost | 0.17 | 0.85 | 0.17 | 0.67 | 66.67 | 69.57 | 0.68 | 68.18 | 69.57 | 0.36 | 0.36 | 82.76 | 84.62 | 0.84 | 83.64 | 81.48 | 0.67 | 0.67 |

| | RFE_positively_Corr_10_performance_metrics |
|---|---|

|  |  |  |  |  | Training |  |  |  |  |  | Validation |  |  |  |  |  |
| --- | --- | --- | --- | --- | --- | --- | --- | --- | --- | --- | --- | --- | --- | --- | --- | --- |
| Model ID | Thr | TPR | FPR | YC | Sp | Sn | AUC | Acc | Pre | MC | K | Sp | Sn | AUC | Acc | Pre | MC | K |
| SVC | 0.36 | 0.81 | 0.14 | 0.67 | 100.00 | 69.57 | 0.85 | 84.09 | 100.00 | 0.72 | 0.69 | 86.21 | 80.77 | 0.83 | 83.64 | 84.00 | 0.67 | 0.67 |
| LR | 0.42 | 0.81 | 0.14 | 0.67 | 90.48 | 91.30 | 0.91 | 90.91 | 91.30 | 0.82 | 0.82 | 86.21 | 80.77 | 0.83 | 83.64 | 84.00 | 0.67 | 0.67 |
| Naive Bayes | 0.00 | 0.81 | 0.14 | 0.67 | 90.48 | 82.61 | 0.87 | 86.36 | 90.48 | 0.73 | 0.73 | 86.21 | 80.77 | 0.83 | 83.64 | 84.00 | 0.67 | 0.67 |
| DT | 1.00 | 0.65 | 0.14 | 0.52 | 90.48 | 86.96 | 0.89 | 88.64 | 90.91 | 0.77 | 0.77 | 86.21 | 65.38 | 0.76 | 76.36 | 80.95 | 0.53 | 0.52 |
| RF | 0.38 | 0.81 | 0.14 | 0.67 | 90.48 | 86.96 | 0.89 | 88.64 | 90.91 | 0.77 | 0.77 | 86.21 | 80.77 | 0.83 | 83.64 | 84.00 | 0.67 | 0.67 |
| ET | 0.33 | 0.81 | 0.14 | 0.67 | 90.48 | 86.96 | 0.89 | 88.64 | 90.91 | 0.77 | 0.77 | 86.21 | 80.77 | 0.83 | 83.64 | 84.00 | 0.67 | 0.67 |
| AdaBooSt | 0.52 | 0.77 | 0.14 | 0.63 | 90.48 | 91.30 | 0.91 | 90.91 | 91.30 | 0.82 | 0.82 | 86.21 | 76.92 | 0.82 | 81.82 | 83.33 | 0.64 | 0.63 |
| Gradient Boosting | 0.22 | 0.77 | 0.14 | 0.63 | 90.48 | 91.30 | 0.91 | 90.91 | 91.30 | 0.82 | 0.82 | 86.21 | 76.92 | 0.82 | 81.82 | 83.33 | 0.64 | 0.63 |
| MLP | 0.54 | 0.81 | 0.14 | 0.67 | 90.48 | 91.30 | 0.91 | 90.91 | 91.30 | 0.82 | 0.82 | 86.21 | 80.77 | 0.83 | 83.64 | 84.00 | 0.67 | 0.67 |
| XGBoost | 0.44 | 0.81 | 0.14 | 0.67 | 90.48 | 91.30 | 0.91 | 90.91 | 91.30 | 0.82 | 0.82 | 86.21 | 80.77 | 0.83 | 83.64 | 84.00 | 0.67 | 0.67 |

| RFE_positively_Corr_5_performance_metrics | | | | | | | | | | | | | | | | | | |
| --- | --- | --- | --- | --- | --- | --- | --- | --- | --- | --- | --- | --- | --- | --- | --- | --- | --- | --- |
|  |  |  |  |  | Training |  |  |  |  |  | Validation |  |  |  |  |  |
| Model ID | Thr | TPR | FPR | YC | Sp | Sn | AUC | Acc | Pre | MC | K | Sp | Sn | AUC | Acc | Pre | MC | K |
| SVC | 0.82 | 0.69 | 0.10 | 0.59 | 95.24 | 56.52 | 0.76 | 75.00 | 92.86 | 0.56 | 0.51 | 89.66 | 69.23 | 0.79 | 80.00 | 85.71 | 0.61 | 0.59 |
| LR | 0.73 | 0.69 | 0.07 | 0.62 | 95.24 | 73.91 | 0.85 | 84.09 | 94.44 | 0.70 | 0.68 | 93.10 | 69.23 | 0.81 | 81.82 | 90.00 | 0.65 | 0.63 |
| Naive Bayes | 1.00 | 0.69 | 0.07 | 0.62 | 95.24 | 69.57 | 0.82 | 81.82 | 94.12 | 0.66 | 0.64 | 93.10 | 69.23 | 0.81 | 81.82 | 90.00 | 0.65 | 0.63 |
| DT | 1.00 | 0.65 | 0.10 | 0.55 | 95.24 | 73.91 | 0.85 | 84.09 | 94.44 | 0.70 | 0.68 | 89.66 | 65.38 | 0.78 | 78.18 | 85.00 | 0.57 | 0.56 |
| RF | 0.93 | 0.69 | 0.10 | 0.59 | 95.24 | 73.91 | 0.85 | 84.09 | 94.44 | 0.70 | 0.68 | 89.66 | 69.23 | 0.79 | 80.00 | 85.71 | 0.61 | 0.59 |
| ET | 0.91 | 0.69 | 0.10 | 0.59 | 95.24 | 73.91 | 0.85 | 84.09 | 94.44 | 0.70 | 0.68 | 89.66 | 69.23 | 0.79 | 80.00 | 85.71 | 0.61 | 0.59 |
| AdaBooSt | 0.60 | 0.69 | 0.07 | 0.62 | 95.24 | 73.91 | 0.85 | 84.09 | 94.44 | 0.70 | 0.68 | 93.10 | 69.23 | 0.81 | 81.82 | 90.00 | 0.65 | 0.63 |
| Gradient Boosting | 0.71 | 0.69 | 0.10 | 0.59 | 95.24 | 73.91 | 0.85 | 84.09 | 94.44 | 0.70 | 0.68 | 89.66 | 69.23 | 0.79 | 80.00 | 85.71 | 0.61 | 0.59 |

| Model ID | Thr | TPR | FPR | YC | Sp | Sn | AUC | Acc | Pre | MC | K | Sp | Sn | AUC | Acc | Pre | MC | K |
|---|---|---|---|---|---|---|---|---|---|---|---|---|---|---|---|---|---|---|
| MLP | 0.84 | 0.69 | 0.07 | 0.62 | 95.24 | 73.91 | 0.85 | 84.09 | 94.44 | 0.70 | 0.68 | 93.10 | 69.23 | 0.81 | 81.82 | 90.00 | 0.65 | 0.63 |
| XGBoost | 0.81 | 0.69 | 0.10 | 0.59 | 95.24 | 73.91 | 0.85 | 84.09 | 94.44 | 0.70 | 0.68 | 89.66 | 69.23 | 0.79 | 80.00 | 85.71 | 0.61 | 0.59 |

| | **RFE_negatively_Corr_25_performance_metrics** | | | | | | | | | | | | | | | | | |
|---|---|---|---|---|---|---|---|---|---|---|---|---|---|---|---|---|---|---|
| | | | | | **Training** | | | | | | | **Validation** | | | | | | |
| Model ID | Thr | TPR | FPR | YC | Sp | Sn | AUC | Acc | Pre | MC | K | Sp | Sn | AUC | Acc | Pre | MC | K |
| SVC | 0.36 | 0.92 | 0.52 | 0.41 | 66.67 | 65.22 | 0.66 | 65.91 | 68.18 | 0.32 | 0.32 | 48.28 | 92.31 | 0.70 | 69.09 | 61.54 | 0.45 | 0.40 |
| LR | 0.29 | 0.96 | 0.79 | 0.17 | 57.14 | 52.17 | 0.55 | 54.55 | 57.14 | 0.09 | 0.09 | 20.69 | 96.15 | 0.58 | 56.36 | 52.08 | 0.25 | 0.16 |
| Naive Bayes | 0.00 | 0.96 | 0.52 | 0.44 | 75.00 | 73.91 | 0.74 | 74.42 | 77.27 | 0.49 | 0.49 | 48.28 | 96.15 | 0.72 | 70.91 | 62.50 | 0.50 | 0.43 |
| DT | 1.00 | 0.50 | 0.45 | 0.05 | 66.67 | 65.22 | 0.66 | 65.91 | 68.18 | 0.32 | 0.32 | 55.17 | 50.00 | 0.53 | 52.73 | 50.00 | 0.05 | 0.05 |
| RF | 0.51 | 0.85 | 0.34 | 0.50 | 61.90 | 73.91 | 0.68 | 68.18 | 68.00 | 0.36 | 0.36 | 65.52 | 84.62 | 0.75 | 74.55 | 68.75 | 0.51 | 0.50 |
| ET | 0.53 | 0.81 | 0.24 | 0.57 | 65.00 | 69.57 | 0.67 | 67.44 | 69.57 | 0.35 | 0.35 | 75.86 | 80.77 | 0.78 | 78.18 | 75.00 | 0.57 | 0.56 |
| AdaBooSt | 0.50 | 0.81 | 0.34 | 0.46 | 61.90 | 65.22 | 0.64 | 63.64 | 65.22 | 0.27 | 0.27 | 65.52 | 80.77 | 0.73 | 72.73 | 67.74 | 0.47 | 0.46 |
| Gradient Boosting | 0.36 | 0.85 | 0.28 | 0.57 | 71.43 | 69.57 | 0.71 | 70.45 | 72.73 | 0.41 | 0.41 | 72.41 | 84.62 | 0.79 | 78.18 | 73.33 | 0.57 | 0.57 |
| MLP | 0.56 | 0.69 | 0.48 | 0.21 | 57.14 | 60.87 | 0.59 | 59.09 | 60.87 | 0.18 | 0.18 | 51.72 | 69.23 | 0.60 | 60.00 | 56.25 | 0.21 | 0.21 |
| XGBoost | 0.45 | 0.85 | 0.28 | 0.57 | 70.00 | 69.57 | 0.70 | 69.77 | 72.73 | 0.39 | 0.39 | 72.41 | 84.62 | 0.79 | 78.18 | 73.33 | 0.57 | 0.57 |

| | **RFE_negatively_Corr_20_performance_metrics** | | | | | | | | | | | | | | | | | |
|---|---|---|---|---|---|---|---|---|---|---|---|---|---|---|---|---|---|---|
| | | | | | **Training** | | | | | | | **Validation** | | | | | | |
| Model ID | Thr | TPR | FPR | YC | Sp | Sn | AUC | Acc | Pre | MC | K | Sp | Sn | AUC | Acc | Pre | MC | K |
| SVC | 0.42 | 0.88 | 0.52 | 0.37 | 61.90 | 60.87 | 0.61 | 61.36 | 63.64 | 0.23 | 0.23 | 48.28 | 88.46 | 0.68 | 67.27 | 60.53 | 0.40 | 0.36 |
| LR | 0.17 | 0.96 | 0.83 | 0.13 | 61.90 | 52.17 | 0.57 | 56.82 | 60.00 | 0.14 | 0.14 | 17.24 | 96.15 | 0.57 | 54.55 | 51.02 | 0.21 | 0.13 |
| Naive Bayes | 0.15 | 0.81 | 0.41 | 0.39 | 75.00 | 73.91 | 0.74 | 74.42 | 77.27 | 0.49 | 0.49 | 58.62 | 80.77 | 0.70 | 69.09 | 63.64 | 0.40 | 0.39 |
| DT | 1.00 | 0.69 | 0.38 | 0.31 | 70.00 | 69.57 | 0.70 | 69.77 | 72.73 | 0.39 | 0.39 | 62.07 | 69.23 | 0.66 | 65.45 | 62.07 | 0.31 | 0.31 |

| Model ID | Thr | TPR | FPR | YC | Sp | Sn | AUC | Acc | Pre | MC | K | Sp | Sn | AUC | Acc | Pre | MC | K |
|---|---|---|---|---|---|---|---|---|---|---|---|---|---|---|---|---|---|---|
| RF | 0.62 | 0.73 | 0.17 | 0.56 | 71.43 | 73.91 | 0.73 | 72.73 | 73.91 | 0.45 | 0.45 | 82.76 | 73.08 | 0.78 | 78.18 | 79.17 | 0.56 | 0.56 |
| ET | 0.47 | 0.92 | 0.45 | 0.47 | 70.00 | 65.22 | 0.68 | 67.44 | 71.43 | 0.35 | 0.35 | 55.17 | 92.31 | 0.74 | 72.73 | 64.86 | 0.51 | 0.46 |
| AdaBooSt | 0.50 | 0.73 | 0.34 | 0.39 | 60.00 | 60.87 | 0.60 | 60.47 | 63.64 | 0.21 | 0.21 | 65.52 | 73.08 | 0.69 | 69.09 | 65.52 | 0.39 | 0.38 |
| Gradient Boosting | 0.87 | 0.62 | 0.10 | 0.51 | 61.90 | 60.87 | 0.61 | 61.36 | 63.64 | 0.23 | 0.23 | 89.66 | 61.54 | 0.76 | 76.36 | 84.21 | 0.54 | 0.52 |
| MLP | 0.16 | 0.96 | 0.62 | 0.34 | 70.00 | 66.67 | 0.68 | 68.18 | 72.73 | 0.37 | 0.36 | 37.93 | 96.15 | 0.67 | 65.45 | 58.14 | 0.41 | 0.33 |
| XGBoost | 0.60 | 0.73 | 0.24 | 0.49 | 60.00 | 60.87 | 0.60 | 60.47 | 63.64 | 0.21 | 0.21 | 75.86 | 73.08 | 0.74 | 74.55 | 73.08 | 0.49 | 0.49 |

| | RFE_negatively_Corr_15_performance_metrics | | | | | | | | | | | | | | | | | |
|---|---|---|---|---|---|---|---|---|---|---|---|---|---|---|---|---|---|---|
| | | | | | Training | | | | | | | Validation | | | | | | |
| Model ID | Thr | TPR | FPR | YC | Sp | Sn | AUC | Acc | Pre | MC | K | Sp | Sn | AUC | Acc | Pre | MC | K |
| SVC | 0.49 | 0.73 | 0.34 | 0.39 | 61.90 | 82.61 | 0.72 | 72.73 | 70.37 | 0.46 | 0.45 | 65.52 | 73.08 | 0.69 | 69.09 | 65.52 | 0.39 | 0.38 |
| LR | 0.59 | 0.35 | 0.28 | 0.07 | 65.00 | 58.33 | 0.62 | 61.36 | 66.67 | 0.23 | 0.23 | 72.41 | 34.62 | 0.54 | 54.55 | 52.94 | 0.08 | 0.07 |
| Naive Bayes | 0.76 | 0.77 | 0.34 | 0.42 | 80.00 | 73.91 | 0.77 | 76.74 | 80.95 | 0.54 | 0.54 | 65.52 | 76.92 | 0.71 | 70.91 | 66.67 | 0.43 | 0.42 |
| DT | 1.00 | 0.62 | 0.45 | 0.17 | 57.14 | 65.22 | 0.61 | 61.36 | 62.50 | 0.22 | 0.22 | 55.17 | 61.54 | 0.58 | 58.18 | 55.17 | 0.17 | 0.17 |
| RF | 0.46 | 0.92 | 0.48 | 0.44 | 57.14 | 56.52 | 0.57 | 56.82 | 59.09 | 0.14 | 0.14 | 51.72 | 92.31 | 0.72 | 70.91 | 63.16 | 0.48 | 0.43 |
| ET | 0.44 | 0.88 | 0.45 | 0.44 | 60.00 | 60.87 | 0.60 | 60.47 | 63.64 | 0.21 | 0.21 | 55.17 | 88.46 | 0.72 | 70.91 | 63.89 | 0.46 | 0.43 |
| AdaBooSt | 0.49 | 0.85 | 0.48 | 0.36 | 60.00 | 60.87 | 0.60 | 60.47 | 63.64 | 0.21 | 0.21 | 51.72 | 84.62 | 0.68 | 67.27 | 61.11 | 0.38 | 0.36 |
| Gradient Boosting | 0.41 | 0.92 | 0.48 | 0.44 | 60.00 | 56.52 | 0.58 | 58.14 | 61.90 | 0.16 | 0.16 | 51.72 | 92.31 | 0.72 | 70.91 | 63.16 | 0.48 | 0.43 |
| MLP | 0.57 | 0.46 | 0.28 | 0.19 | 50.00 | 54.17 | 0.52 | 52.27 | 56.52 | 0.04 | 0.04 | 72.41 | 46.15 | 0.59 | 60.00 | 60.00 | 0.19 | 0.19 |
| XGBoost | 0.24 | 0.85 | 0.45 | 0.40 | 61.90 | 65.22 | 0.64 | 63.64 | 65.22 | 0.27 | 0.27 | 55.17 | 84.62 | 0.70 | 69.09 | 62.86 | 0.41 | 0.39 |

| | RFE_negatively_Corr_10_performance_metrics | | | | | | | | | | | | | | | | | |
|---|---|---|---|---|---|---|---|---|---|---|---|---|---|---|---|---|---|---|
| | | | | | Training | | | | | | | Validation | | | | | | |
| Model ID | Thr | TPR | FPR | YC | Sp | Sn | AUC | Acc | Pre | MC | K | Sp | Sn | AUC | Acc | Pre | MC | K |

| Model ID | Thr | TPR | FPR | YC | Sp | Sn | AUC | Acc | Pre | MC | K | Sp | Sn | AUC | Acc | Pre | MC | K |
|---|---|---|---|---|---|---|---|---|---|---|---|---|---|---|---|---|---|---|
| SVC | 0.61 | 0.65 | 0.28 | 0.38 | 61.90 | 82.61 | 0.72 | 72.73 | 70.37 | 0.46 | 0.45 | 72.41 | 65.38 | 0.69 | 69.09 | 68.00 | 0.38 | 0.38 |
| LR | 0.39 | 0.77 | 0.62 | 0.15 | 61.90 | 69.57 | 0.66 | 65.91 | 66.67 | 0.32 | 0.32 | 37.93 | 76.92 | 0.57 | 56.36 | 52.63 | 0.16 | 0.15 |
| Naive Bayes | 0.75 | 0.73 | 0.31 | 0.42 | 75.00 | 69.57 | 0.72 | 72.09 | 76.19 | 0.44 | 0.44 | 68.97 | 73.08 | 0.71 | 70.91 | 67.86 | 0.42 | 0.42 |
| DT | 1.00 | 0.62 | 0.41 | 0.20 | 66.67 | 65.22 | 0.66 | 65.91 | 68.18 | 0.32 | 0.32 | 58.62 | 61.54 | 0.60 | 60.00 | 57.14 | 0.20 | 0.20 |
| RF | 0.42 | 0.88 | 0.38 | 0.51 | 65.00 | 60.87 | 0.63 | 62.79 | 66.67 | 0.26 | 0.26 | 62.07 | 88.46 | 0.75 | 74.55 | 67.65 | 0.52 | 0.50 |
| ET | 0.61 | 0.65 | 0.17 | 0.48 | 61.90 | 65.22 | 0.64 | 63.64 | 65.22 | 0.27 | 0.27 | 82.76 | 65.38 | 0.74 | 74.55 | 77.27 | 0.49 | 0.49 |
| AdaBooSt | 0.49 | 0.92 | 0.45 | 0.47 | 71.43 | 65.22 | 0.68 | 68.18 | 71.43 | 0.37 | 0.36 | 55.17 | 92.31 | 0.74 | 72.73 | 64.86 | 0.51 | 0.46 |
| Gradient Boosting | 0.40 | 0.88 | 0.34 | 0.54 | 61.90 | 60.87 | 0.61 | 61.36 | 63.64 | 0.23 | 0.23 | 65.52 | 88.46 | 0.77 | 76.36 | 69.70 | 0.55 | 0.53 |
| MLP | 0.34 | 0.73 | 0.62 | 0.11 | 60.00 | 58.33 | 0.59 | 59.09 | 63.64 | 0.18 | 0.18 | 37.93 | 73.08 | 0.56 | 54.55 | 51.35 | 0.12 | 0.11 |
| XGBoost | 0.66 | 0.69 | 0.24 | 0.45 | 61.90 | 52.17 | 0.57 | 56.82 | 60.00 | 0.14 | 0.14 | 75.86 | 69.23 | 0.73 | 72.73 | 72.00 | 0.45 | 0.45 |

| | RFE_negatively_Corr_5_performance_metrics | | | | | | | | | | | | | | | | | |
|---|---|---|---|---|---|---|---|---|---|---|---|---|---|---|---|---|---|---|
| | | | | | Training | | | | | | | Validation | | | | | | |
| Model ID | Thr | TPR | FPR | YC | Sp | Sn | AUC | Acc | Pre | MC | K | Sp | Sn | AUC | Acc | Pre | MC | K |
| SVC | 0.45 | 0.81 | 0.38 | 0.43 | 65.00 | 73.91 | 0.69 | 69.77 | 70.83 | 0.39 | 0.39 | 62.07 | 80.77 | 0.71 | 70.91 | 65.63 | 0.43 | 0.42 |
| LR | 0.49 | 0.77 | 0.38 | 0.39 | 71.43 | 78.26 | 0.75 | 75.00 | 75.00 | 0.50 | 0.50 | 62.07 | 76.92 | 0.70 | 69.09 | 64.52 | 0.39 | 0.39 |
| Naive Bayes | 0.52 | 0.77 | 0.38 | 0.39 | 70.00 | 78.26 | 0.74 | 74.42 | 75.00 | 0.48 | 0.48 | 62.07 | 76.92 | 0.70 | 69.09 | 64.52 | 0.39 | 0.39 |
| DT | 1.00 | 0.58 | 0.52 | 0.06 | 65.00 | 65.22 | 0.65 | 65.12 | 68.18 | 0.30 | 0.30 | 48.28 | 57.69 | 0.53 | 52.73 | 50.00 | 0.06 | 0.06 |
| RF | 0.47 | 0.81 | 0.48 | 0.32 | 71.43 | 69.57 | 0.71 | 70.45 | 72.73 | 0.41 | 0.41 | 51.72 | 80.77 | 0.66 | 65.45 | 60.00 | 0.34 | 0.32 |
| ET | 0.57 | 0.73 | 0.28 | 0.45 | 71.43 | 73.91 | 0.73 | 72.73 | 73.91 | 0.45 | 0.45 | 72.41 | 73.08 | 0.73 | 72.73 | 70.37 | 0.45 | 0.45 |
| AdaBooSt | 0.50 | 0.85 | 0.48 | 0.36 | 70.00 | 60.87 | 0.65 | 65.12 | 70.00 | 0.31 | 0.31 | 51.72 | 84.62 | 0.68 | 67.27 | 61.11 | 0.38 | 0.36 |
| Gradient Boosting | 0.60 | 0.69 | 0.41 | 0.28 | 76.19 | 78.26 | 0.77 | 77.27 | 78.26 | 0.54 | 0.54 | 58.62 | 69.23 | 0.64 | 63.64 | 60.00 | 0.28 | 0.28 |
| MLP | 0.48 | 0.58 | 0.41 | 0.16 | 61.90 | 69.57 | 0.66 | 65.91 | 66.67 | 0.32 | 0.32 | 58.62 | 57.69 | 0.58 | 58.18 | 55.56 | 0.16 | 0.16 |

| | Thr | TPR | FPR | YC | Sp | Sn | AUC | Acc | Pre | MC | K | Sp | Sn | AUC | Acc | Pre | MC | K |
|---|---|---|---|---|---|---|---|---|---|---|---|---|---|---|---|---|---|---|
| XGBoost | 0.05 | 1.00 | 0.76 | 0.24 | 61.90 | 60.87 | 0.61 | 61.36 | 63.64 | 0.23 | 0.23 | 24.14 | 100.0 | 0.62 | 60.00 | 54.17 | 0.36 | 0.23 |

#Thr: Threshold; YC: Youden Coeff; Sp: Specificity; Sn: Sensitivity; Acc: Accuracy; Pre: Precision; MC: Mathews Coeff; K :Kappa

**Supplementary Table 7.1: List of selected features based on SFS for positively and negatively correlated Methylated sites**

| SFS_10_Pos_Met_sites | SFS_4_Neg_Met_sites |
|---|---|
| 5779 | 17680 |
| 19311 | 75319 |
| 33055 | 100953 |
| 62691 | 116025 |
| 93487 | |
| 155374 | |
| 162112 | |
| 181490 | |
| 181491 | |
| 190944 | |

**Supplementary Table 7.2: ML performance of selected features based on SFS for positively and negatively correlated Methylated sites**

| | | | | | SFS_Positively_Top_10 correlated MS using SFS | | | | | | | | | | | | | |
|---|---|---|---|---|---|---|---|---|---|---|---|---|---|---|---|---|---|---|
| | | | | | Training | | | | | | | Validation | | | | | | |
| Model ID | Thr | TPR | FPR | YC | Sp | Sn | AUC | Acc | Pre | MC | K | Sp | Sn | AUC | Acc | Pre | MC | K |
| SVC | 0.56 | 0.77 | 0.24 | 0.53 | 80.95 | 78.26 | 0.80 | 79.55 | 81.82 | 0.59 | 0.59 | 75.86 | 76.92 | 0.76 | 76.36 | 74.07 | 0.53 | 0.53 |
| LR | 0.65 | 0.81 | 0.10 | 0.70 | 76.19 | 73.91 | 0.75 | 75.00 | 77.27 | 0.50 | 0.50 | 89.66 | 80.77 | 0.85 | 85.45 | 87.50 | 0.71 | 0.71 |
| Naive Bayes | 1.00 | 0.77 | 0.14 | 0.63 | 71.43 | 73.91 | 0.73 | 72.73 | 73.91 | 0.45 | 0.45 | 86.21 | 76.92 | 0.82 | 81.82 | 83.33 | 0.64 | 0.63 |

| Model ID | Thr | TPR | FPR | YC | Sp | Sn | AUC | Acc | Pre | MC | K | Sp | Sn | AUC | Acc | Pre | MC | K |
|---|---|---|---|---|---|---|---|---|---|---|---|---|---|---|---|---|---|---|
| DT | 1.00 | 0.85 | 0.21 | 0.64 | 80.95 | 91.30 | 0.86 | 86.36 | 84.00 | 0.73 | 0.73 | 79.31 | 84.62 | 0.82 | 81.82 | 78.57 | 0.64 | 0.64 |
| RF | 0.46 | 0.85 | 0.28 | 0.57 | 95.24 | 82.61 | 0.89 | 88.64 | 95.00 | 0.78 | 0.77 | 72.41 | 84.62 | 0.79 | 78.18 | 73.33 | 0.57 | 0.57 |
| ET | 0.76 | 0.77 | 0.10 | 0.67 | 80.95 | 86.96 | 0.84 | 84.09 | 83.33 | 0.68 | 0.68 | 89.66 | 76.92 | 0.83 | 83.64 | 86.96 | 0.67 | 0.67 |
| AdaBoost | 0.51 | 0.77 | 0.24 | 0.53 | 76.19 | 78.26 | 0.77 | 77.27 | 78.26 | 0.54 | 0.54 | 75.86 | 76.92 | 0.76 | 76.36 | 74.07 | 0.53 | 0.53 |
| Gradient Boosting | 0.89 | 0.69 | 0.07 | 0.62 | 80.95 | 86.96 | 0.84 | 84.09 | 83.33 | 0.68 | 0.68 | 93.10 | 69.23 | 0.81 | 81.82 | 90.00 | 0.65 | 0.63 |
| MLP | 0.79 | 0.73 | 0.07 | 0.66 | 76.19 | 78.26 | 0.77 | 77.27 | 78.26 | 0.54 | 0.54 | 93.10 | 73.08 | 0.83 | 83.64 | 90.48 | 0.68 | 0.67 |
| XGBoost | 0.81 | 0.73 | 0.14 | 0.59 | 80.95 | 82.61 | 0.82 | 81.82 | 82.61 | 0.64 | 0.64 | 86.21 | 73.08 | 0.80 | 80.00 | 82.61 | 0.60 | 0.60 |

| SFS_Negatively_Top 4 correlated MS using SFS ||||||||||||||||||
|---|---|---|---|---|---|---|---|---|---|---|---|---|---|---|---|---|---|---|
| | | | | | Training |||||| | Validation |||||||
| Model ID | Thr | TPR | FPR | YC | Sp | Sn | AUC | Acc | Pre | MC | K | Sp | Sn | AUC | Acc | Pre | MC | K |
| SVC | 0.62 | 0.77 | 0.24 | 0.53 | 75.00 | 82.61 | 0.79 | 79.07 | 79.17 | 0.58 | 0.58 | 75.86 | 76.92 | **0.76** | 76.36 | 74.07 | 0.53 | 0.53 |
| LR | 0.48 | 0.81 | 0.38 | 0.43 | 71.43 | 73.91 | 0.73 | 72.73 | 73.91 | 0.45 | 0.45 | 62.07 | 80.77 | 0.71 | 70.91 | 65.63 | 0.43 | 0.42 |
| Naive Bayes | 0.70 | 0.81 | 0.34 | 0.46 | 80.00 | 86.96 | 0.83 | 83.72 | 83.33 | 0.67 | 0.67 | 65.52 | 80.77 | 0.73 | 72.73 | 67.74 | 0.47 | 0.46 |
| DT | 1.00 | 0.77 | 0.72 | 0.05 | 66.67 | 69.57 | 0.68 | 68.18 | 69.57 | 0.36 | 0.36 | 27.59 | 76.92 | 0.52 | 50.91 | 48.78 | 0.05 | 0.04 |
| RF | 0.83 | 0.31 | 0.14 | 0.17 | 61.90 | 73.91 | 0.68 | 68.18 | 68.00 | 0.36 | 0.36 | 86.21 | 30.77 | 0.58 | 60.00 | 66.67 | 0.21 | 0.17 |
| ET | 0.57 | 0.73 | 0.31 | 0.42 | 66.67 | 78.26 | 0.72 | 72.73 | 72.00 | 0.45 | 0.45 | 68.97 | 73.08 | 0.71 | 70.91 | 67.86 | 0.42 | 0.42 |
| AdaBoost | 0.50 | 0.85 | 0.59 | 0.26 | 76.19 | 78.26 | 0.77 | 77.27 | 78.26 | 0.54 | 0.54 | 41.38 | 84.62 | 0.63 | 61.82 | 56.41 | 0.29 | 0.25 |
| Gradient Boosting | 0.46 | 0.77 | 0.48 | 0.29 | 57.14 | 60.87 | 0.59 | 59.09 | 60.87 | 0.18 | 0.18 | 51.72 | 76.92 | 0.64 | 63.64 | 58.82 | 0.29 | 0.28 |
| MLP | 0.47 | 0.81 | 0.72 | 0.08 | 55.00 | 58.33 | 0.57 | 56.82 | 60.87 | 0.13 | 0.13 | 27.59 | 80.77 | 0.54 | 52.73 | 50.00 | 0.10 | 0.08 |
| XGBoost | 0.99 | 0.27 | 0.10 | 0.17 | 61.90 | 60.87 | 0.61 | 61.36 | 63.64 | 0.23 | 0.23 | 89.66 | 26.92 | 0.58 | 60.00 | 70.00 | 0.21 | 0.17 |

**Supplementary Table 8.1: List of selected features based on SVC-L1 for positively and negatively correlated Methylated sites**

| SVC-L1_9_Pos_Met_sites | | SVC-L1_50_Neg_Met_sites | | | | |
|---|---|---|---|---|---|---|
| 33055 | | 3658 | 54572 | 125191 | 24476 | 89080 |
| 93487 | | 7573 | 54573 | 149453 | 24632 | 90817 |
| 14879 | | 7574 | 57404 | 171893 | 24643 | 90819 |
| 22039 | | 7575 | 58323 | 180704 | 30581 | 99173 |
| 32433 | | 7699 | 75129 | 189312 | 35411 | 100954 |
| 38139 | | 7703 | 75317 | 189313 | 35412 | 100957 |
| 86829 | | 9376 | 75319 | 202309 | 36303 | 113054 |
| 181489 | | 17680 | 75322 | 202310 | 36304 | 114431 |
| 196742 | | 17992 | 78188 | 203525 | 36305 | 116026 |
| | | 18002 | 84787 | 204581 | 41490 | 123336 |

**Supplementary Table 8.2: ML performance of selected features based on SVC-L1 for positively and negatively correlated Methylated sites**

| | Top 9 positively correlated MS using SVC-L1 | | | | | | | | | | | | | | | | |
|---|---|---|---|---|---|---|---|---|---|---|---|---|---|---|---|---|---|
| | | | | | Training | | | | | | | Validation | | | | | |
| Model ID | Thr | TPR | FPR | YC | Sp | Sn | AUC | Acc | Pre | MC | K | Sp | Sn | AUC | Acc | Pre | MC | K |
| SVC | 0.78 | 0.65 | 0.07 | 0.58 | 85.71 | 78.26 | 0.82 | 81.82 | 85.71 | 0.64 | 0.64 | 93.10 | 65.38 | 0.79 | 80.00 | 89.47 | 0.61 | 0.59 |
| LR | 0.73 | 0.69 | 0.03 | 0.66 | 76.19 | 73.91 | 0.75 | 75.00 | 77.27 | 0.50 | 0.50 | 96.55 | 69.23 | 0.83 | 83.64 | 94.74 | 0.69 | 0.67 |
| Naive Bayes | 0.00 | 0.88 | 0.24 | 0.64 | 95.24 | 60.87 | 0.78 | 77.27 | 93.33 | 0.59 | 0.55 | 75.86 | 88.46 | 0.82 | 81.82 | 76.67 | 0.64 | 0.64 |
| DT | 1.00 | 0.81 | 0.34 | 0.46 | 90.48 | 86.96 | 0.89 | 88.64 | 90.91 | 0.77 | 0.77 | 65.52 | 80.77 | 0.73 | 72.73 | 67.74 | 0.47 | 0.46 |
| RF | 0.38 | 0.88 | 0.34 | 0.54 | 75.00 | 65.22 | 0.70 | 69.77 | 75.00 | 0.40 | 0.40 | 65.52 | 88.46 | 0.77 | 76.36 | 69.70 | 0.55 | 0.53 |

| Model ID | Thr | TPR | FPR | YC | Sp | Sn | AUC | Acc | Pre | MC | K | Sp | Sn | AUC | Acc | Pre | MC | K |
|---|---|---|---|---|---|---|---|---|---|---|---|---|---|---|---|---|---|---|
| ET | | 0.57 | 0.81 | 0.24 | 0.57 | 66.67 | 65.22 | 0.66 | 65.91 | 68.18 | 0.32 | 0.32 | 75.86 | 80.77 | 0.78 | 78.18 | 75.00 | 0.57 | 0.56 |
| AdaBooSt | | 0.51 | 0.88 | 0.14 | 0.75 | 71.43 | 69.57 | 0.71 | 70.45 | 72.73 | 0.41 | 0.41 | 86.21 | 88.46 | 0.87 | 87.27 | 85.19 | 0.75 | 0.75 |
| Gradient Boosting | | 0.62 | 0.85 | 0.28 | 0.57 | 80.95 | 86.96 | 0.84 | 84.09 | 83.33 | 0.68 | 0.68 | 72.41 | 84.62 | 0.79 | 78.18 | 73.33 | 0.57 | 0.57 |
| MLP | | 0.61 | 0.77 | 0.17 | 0.60 | 66.67 | 65.22 | 0.66 | 65.91 | 68.18 | 0.32 | 0.32 | 82.76 | 76.92 | 0.80 | 80.00 | 80.00 | 0.60 | 0.60 |
| XGBoost | | 0.89 | 0.73 | 0.24 | 0.49 | 76.19 | 86.96 | 0.82 | 81.82 | 80.00 | 0.64 | 0.63 | 75.86 | 73.08 | 0.74 | 74.55 | 73.08 | 0.49 | 0.49 |

| | | | | | Top 50 negatively correlated MS using SVC-L1 | | | | | | | | | | | | | |
|---|---|---|---|---|---|---|---|---|---|---|---|---|---|---|---|---|---|---|
| | | | | | Training | | | | | | | Validation | | | | | | |
| Model ID | Thr | TPR | FPR | YC | Sp | Sn | AUC | Acc | Pre | MC | K | Sp | Sn | AUC | Acc | Pre | MC | K |
| SVC | 0.40 | 0.85 | 0.48 | 0.36 | 71.43 | 65.22 | 0.68 | 68.18 | 71.43 | 0.37 | 0.36 | 51.72 | 84.62 | 0.68 | 67.27 | 61.11 | 0.38 | 0.36 |
| LR | 0.33 | 0.88 | 0.59 | 0.30 | 65.00 | 60.87 | 0.63 | 62.79 | 66.67 | 0.26 | 0.26 | 41.38 | 88.46 | 0.65 | 63.64 | 57.50 | 0.33 | 0.29 |
| Naive Bayes | 0.00 | 0.92 | 0.45 | 0.47 | 75.00 | 73.91 | 0.74 | 74.42 | 77.27 | 0.49 | 0.49 | 55.17 | 92.31 | 0.74 | 72.73 | 64.86 | 0.51 | 0.46 |
| DT | 1.00 | 0.77 | 0.38 | 0.39 | 61.90 | 56.52 | 0.59 | 59.09 | 61.90 | 0.18 | 0.18 | 62.07 | 76.92 | 0.70 | 69.09 | 64.52 | 0.39 | 0.39 |
| RF | 0.55 | 0.69 | 0.14 | 0.55 | 71.43 | 69.57 | 0.71 | 70.45 | 72.73 | 0.41 | 0.41 | 86.21 | 69.23 | 0.78 | 78.18 | 81.82 | 0.57 | 0.56 |
| ET | 0.53 | 0.77 | 0.21 | 0.56 | 71.43 | 69.57 | 0.71 | 70.45 | 72.73 | 0.41 | 0.41 | 79.31 | 76.92 | 0.78 | 78.18 | 76.92 | 0.56 | 0.56 |
| AdaBooSt | 0.51 | 0.69 | 0.28 | 0.42 | 71.43 | 73.91 | 0.73 | 72.73 | 73.91 | 0.45 | 0.45 | 72.41 | 69.23 | 0.71 | 70.91 | 69.23 | 0.42 | 0.42 |
| Gradient Boosting | 0.43 | 0.85 | 0.38 | 0.47 | 71.43 | 78.26 | 0.75 | 75.00 | 75.00 | 0.50 | 0.50 | 62.07 | 84.62 | 0.73 | 72.73 | 66.67 | 0.48 | 0.46 |
| MLP | 0.02 | 0.92 | 0.55 | 0.37 | 50.00 | 50.00 | 0.50 | 50.00 | 54.55 | 0.00 | 0.00 | 44.83 | 92.31 | 0.69 | 67.27 | 60.00 | 0.42 | 0.36 |
| XGBoost | 0.37 | 0.92 | 0.34 | 0.58 | 71.43 | 73.91 | 0.73 | 72.73 | 73.91 | 0.45 | 0.45 | 65.52 | 92.31 | 0.79 | 78.18 | 70.59 | 0.59 | 0.57 |

| | | | | | Top 146 MS using SVC-L1 on 30971 Methylation Sites. | | | | | | | | | | | | | |
|---|---|---|---|---|---|---|---|---|---|---|---|---|---|---|---|---|---|---|
| | | | | | Training | | | | | | | Validation | | | | | | |
| Model ID | Thr | TPR | FPR | YC | Sp | Sn | AUC | Acc | Pre | MC | K | Sp | Sn | AUC | Acc | Pre | MC | K |

| Model | | | | | | | | | | | | | | | | | |
|---|---|---|---|---|---|---|---|---|---|---|---|---|---|---|---|---|---|
| SVC | 0.63 | 0.59 | 0.32 | 0.27 | 57.14 | 56.52 | 0.57 | 56.82 | 59.09 | 0.14 | 0.14 | 67.86 | 59.26 | 0.64 | 63.64 | 64.00 | 0.27 | 0.27 |
| LR | 0.62 | 0.59 | 0.32 | 0.27 | 57.14 | 56.52 | 0.57 | 56.82 | 59.09 | 0.14 | 0.14 | 67.86 | 59.26 | 0.64 | 63.64 | 64.00 | 0.27 | 0.27 |
| Naive Bayes | 0.97 | 0.81 | 0.04 | 0.78 | 90.00 | 86.96 | 0.88 | 88.37 | 90.91 | 0.77 | 0.77 | 96.43 | 81.48 | 0.89 | 89.09 | 95.65 | 0.79 | 0.78 |
| DT | 1.00 | 0.56 | 0.32 | 0.23 | 71.43 | 69.57 | 0.71 | 70.45 | 72.73 | 0.41 | 0.41 | 67.86 | 55.56 | 0.62 | 61.82 | 62.50 | 0.24 | 0.23 |
| RF | 1.00 | 0.74 | 0.50 | 0.24 | 66.67 | 65.22 | 0.66 | 65.91 | 68.18 | 0.32 | 0.32 | 50.00 | 74.07 | 0.62 | 61.82 | 58.82 | 0.25 | 0.24 |
| ET | 0.58 | 0.67 | 0.11 | 0.56 | 66.67 | 69.57 | 0.68 | 68.18 | 69.57 | 0.36 | 0.36 | 89.29 | 66.67 | 0.78 | 78.18 | 85.71 | 0.58 | 0.56 |
| AdaBooSt | 0.55 | 0.67 | 0.11 | 0.56 | 76.19 | 65.22 | 0.71 | 70.45 | 75.00 | 0.42 | 0.41 | 89.29 | 66.67 | 0.78 | 78.18 | 85.71 | 0.58 | 0.56 |
| Gradient Boosting | 0.48 | 0.85 | 0.25 | 0.60 | 80.00 | 78.26 | 0.79 | 79.07 | 81.82 | 0.58 | 0.58 | 75.00 | 85.19 | 0.80 | 80.00 | 76.67 | 0.60 | 0.60 |
| MLP | 0.85 | 0.78 | 0.14 | 0.63 | 70.00 | 65.22 | 0.68 | 67.44 | 71.43 | 0.35 | 0.35 | 85.71 | 77.78 | 0.82 | 81.82 | 84.00 | 0.64 | 0.64 |
| XGBoost | 1.00 | 0.74 | 0.25 | 0.49 | 71.43 | 73.91 | 0.73 | 72.73 | 73.91 | 0.45 | 0.45 | 75.00 | 74.07 | 0.75 | 74.55 | 74.07 | 0.49 | 0.49 |
| XGBoost | 0.80 | 0.78 | 0.18 | 0.60 | 76.19 | 78.26 | 0.77 | 77.27 | 78.26 | 0.54 | 0.54 | 82.14 | 77.78 | 0.80 | 80.00 | 80.77 | 0.60 | 0.60 |